\documentclass[
  aps,
  twocolumn,
  superscriptaddress,
  floatfix 
]{revtex4-1}
\usepackage[dvipdfmx]{graphicx}
\usepackage{dcolumn}
\usepackage{bm}
\usepackage{ulem}
\usepackage{amsmath}
\usepackage{amssymb}
\usepackage{txfonts}
\usepackage{hyperref}
\usepackage{color} 
\usepackage{tikz}
\usepackage{pgfplots}
\pgfplotsset{compat=1.8}
\newcommand{\bs}   {\boldsymbol}
\newcommand{\mb}   {\mathbf}

\newcommand{\mcal} {\mathcal}
\newcommand{\imag} {\mathrm{i}}
\newcommand{\dd}   {\mathrm{d}}
\newcommand{\e}    {\mathrm{e}}
\newcommand{\bra}  {\langle}
\newcommand{\ket}  {\rangle}
\newcommand{\up}   {\uparrow}
\newcommand{\dn}   {\downarrow}
\newcommand{\s}    {\sigma}
\newcommand{\w}    {\omega}

\newcommand{\tr}   {\mathrm{tr}}
\newcommand{\Tr}   {\mathrm{Tr}}

\hypersetup{
  colorlinks=true,
  linkcolor=[rgb]{0.85,0.00,0.15},
  citecolor=[rgb]{0.0,0.15,0.95},
  urlcolor=[rgb]{0.0,0.15,0.95},
  setpagesize=false
}

\begin{document}

\title{  
  Topological interpretation of Luttinger theorem
}

\author{Kazuhiro~Seki}
\affiliation{Computational Condensed Matter Physics Laboratory, RIKEN, Saitama 351-0198, Japan}
\affiliation{Computational Materials Science Research Team, RIKEN Advanced Institute for Computational Science (AICS),  Hyogo 650-0047,  Japan}
\author{Seiji~Yunoki}
\affiliation{Computational Condensed Matter Physics Laboratory, RIKEN, Saitama 351-0198, Japan}
\affiliation{Computational Materials Science Research Team, RIKEN Advanced Institute for Computational Science (AICS),  Hyogo 650-0047,  Japan}
\affiliation{Computational Quantum Matter Research Team, RIKEN, Center for Emergent Matter Science (CEMS), Saitama 351-0198, Japan}

\begin{abstract}
  Based solely on the analytical properties of the single-particle Green's function of fermions at finite temperatures, 
  we show that the generalized Luttinger theorem inherently possesses topological aspects.
  The topological interpretation of the generalized
  Luttinger theorem can be introduced because i) the Luttinger 
  volume is represented as the winding number of the single-particle Green's function and 
  thus ii) the deviation of the theorem, expressed with a ratio between 
  the interacting and noninteracting single-particle Green's functions, is also represented as 
  the winding number of this ratio. 
  The formulation based on the winding number naturally leads to two types of 
  the generalized Luttinger theorem. Exploring two examples of single-band 
  translationally invariant interacting electrons, i.e., simple metal and Mott insulator, 
  we show that the first type falls into the original statement for Fermi liquids given by Luttinger, where 
  poles of the single-particle Green's function appear at the chemical potential, while the second type 
  corresponds to the extended one for non metallic cases with no Fermi surface such as insulators 
  and superconductors 
  generalized by Dzyaloshinskii, where zeros of the single-particle Green's function appear 
  at the chemical potential. 
  This formulation also allows us to derive a sufficient condition for the validity of the Luttinger theorem 
  of the first type by applying the Rouche's theorem in complex analysis as an inequality. 
  Moreover, we can rigorously prove in a non-perturbative manner, 
  without assuming any detail of a microscopic Hamiltonian, 
  that the generalized Luttinger theorem of both types 
  is valid for generic interacting fermions as long as the particle-hole symmetry is preserved. 
  Finally, we show that the winding number of the single-particle Green's function can also be associated with 
  the distribution function of quasiparticles, and therefore the number of quasiparticles is equal
  to the Luttinger volume. This implies that the fundamental hypothesis of the Landau's Fermi-liquid theory, 
  the number of fermions being equal to that of quasiparticles, is guaranteed if the Luttinger theorem is valid 
  since the theorem states that the number of fermions is equal to the Luttinger volume.
  All these general statements are made possible because of the finding that the Luttinger volume is expressed 
  as the winding number of the single-particle Green's function at finite temperatures, 
  for which the complex analysis can be readily exploited. 
\end{abstract}


\date{\today}

\maketitle

\section{Introduction}
The Luttinger theorem states that the particle density 
of interacting fermions is equal to the 
volume in the momentum space enclosed by the Fermi surface~\cite{Luttinger1960}.  
The theorem has been proved valid for normal Fermi liquids originally in perturbation expansion of the 
interacting single-particle Green's function  in 60's~\cite{Luttinger1960, Luttinger-Ward1960} 
and later in a non-perturbative way~\cite{Oshikawa2000}. 
In the Green's function language, 
the Luttinger volume is bounded by the surface, named Luttinger surface,
in the momentum ($\mb{k}$) space on which 
the single-particle Green's function $G(\mb{k},\w=0)$ at zero frequency ($\w$), 
i.e., at the chemical potential, changes its sign~\cite{AGD}.  
The fact that the single-particle Green's function changes its sign 
in going through either poles or zeros~\cite{AGD,Dzyaloshinskii2003} allows Luttinger theorem to be 
extended even to insulating states~\cite{Dzyaloshinskii2003,Essler2002} 
of interacting fermions, including multi-orbital systems~\cite{Rosch2007} and 
non-translational-invariant systems~\cite{Ortloff2007}. 
The extended versions of the Luttinger theorem are  
called the generalized Luttinger theorem, 
stating that the particle density of interacting fermions 
is equal to the Luttinger volume. 
The Luttinger theorem has also been shown valid for 
the Tomonaga-Luttinger liquid in one spatial dimension~\cite{Yamanaka1997}.

Although the generalization of the Luttinger theorem 
has significant advantages, e.g.,  being able to treat 
metal and insulator on an equal footing~\cite{Stanescu2006,Sakai2009,Imada2011,Eder2011,Stanescu2007}, 
its validity has been proved only for limited systems. 
For example, the validity of the generalized Luttinger theorem 
has been proved for a particle-hole symmetric single-band Hubbard model 
on the square lattice~\cite{Stanescu2007,Phillips2012}.  
The proof is based on the moment expansion of the single-particle Green's function, 
which involves the commutation relations of the Hamiltonian and electron creation/annihilation operators. 
Therefore, the proof depends on the microscopic Hamiltonian and thus it is difficult to generalize to 
other systems.

In this paper, we show that 
the Luttinger volume at zero temperature is expressed as 
the winding number of the determinant of the single-particle Green's function. 
Therefore, the winding number of a ratio between the determinants of 
the interacting and noninteracting single-particle Green's functions
provides the topological interpretation of the generalized Luttinger theorem. 
We prove rigorously that the generalized Luttinger theorem is valid for generic 
interacting fermions as long as the particle-hole symmetry is preserved. 
The formulation based on the winding number of the single-particle Green's function 
also allows us to naturally classify the condition for the validity 
of generalized Luttinger theorem into two types,
depending on whether poles or zeros of the single-particle Green's function exist 
at the chemical potential.  
The first type (type I) corresponds to the original statement 
for Fermi liquids given by Luttinger~\cite{Luttinger1960}, 
whereas the second type (type II) corresponds to the extended one 
for single-particle gapful systems 
given by Dzyaloshinskii~\cite{Dzyaloshinskii2003}. 
Moreover, a sufficient condition for the validity of the Luttinger theorem of type I can be derived 
from the topological interpretation of the theorem. 
Associating the winding number of the single-particle Green's function with 
the distribution function of quasiparticles, we can show that the number of quasiparticle is equal to the Luttinger 
volume. 
These general results are based solely on analytical properties of the single-particle 
Green's function at finite temperatures, for which the complex analysis can be exploited unambiguously, 
without any detail of a microscopic Hamiltonian, 
and can be applied in any spatial dimension to
metallic and insulating  
states, independently of the strength of interactions. 
Several specific examples of interacting electrons are also provided to demonstrate these results.

The rest of the paper is organized as follows. 
Section~\ref{sec:formulation} introduces the notation used in this paper and summarizes 
analytical aspects of the single-particle Green's function at finite temperatures which are essential 
for the analysis in the following sections. 
Giving the definition of the Luttinger volume in Sec.~\ref{sec:lv}, 
we show in Sec.~\ref{sec:lv_wn} that the Luttinger volume is represented as the winding number of 
the determinant of the single-particle Green's function in the zero-temperature limit. 
In Sec.~\ref{sec:ti}, 
the topological interpretation of the generalized Luttinger theorem is provided and 
the condition for the validity of the generalized Luttinger theorem 
is classified into two types (type I and type II). 
A sufficient condition for the validity of the generalized Luttinger theorem of type I 
is also derived. 
In Sec.~\ref{sec:VGLT}, the generalized Luttinger theorem is proved valid 
for generic interacting fermions as long as the particle-hole symmetry is preserved. 
To give specific examples for 
the generalized Luttinger theorem of types I and II, 
we examine a simple metal in Sec.~\ref{sec:typeI} 
and a one-dimensional Mott insulator using the cluster perturbation 
theory (CPT)~\cite{Senechal2000,Senechal2012} in Sec.~\ref{sec:typeII}, respectively. 
Finally, several remarks on the topological interpretation of  
the generalized Luttinger theorem 
are provided in Sec.~\ref{sec:remarks}
before summarizing the paper in Sec.~\ref{sec:summary}. 
Additional discussions on the Luttinger-Ward functional and 
the quasiparticle distribution function at finite temperatures 
are given in Appendices~\ref{sec:ad} and \ref{sec:QP}, 
respectively. 
The single-band 
Hubbard model on the honeycomb lattice is analyzed in Appendix~\ref{sec:hubbard-I}.

\section{Single-particle Green's function}\label{sec:formulation}

In this section, we shall derive useful analytical properties of the single-particle Green's function 
at finite temperatures. 
As shown in Sec.~\ref{sec:GLT}, the finite-temperature formulation introduced here significantly 
simplifies the analysis 
without encountering any ambiguity in treating the singularities of the single-particle Green's function 
at the chemical potential, which is often overlooked in the zero-temperature formulation.

\subsection{Notation}\label{sec:notation}

First, we introduce the notation for the single-particle Green's function used here. 
We set $\hbar = k_{\rm B} = 1$ and refer to $z$ ($\omega$) as complex (real) frequency.
Following the notation in Ref.~\cite{Aichhorn2006}, the Lehmann representation~\cite{FW} of 
the single-particle Green's function at temperature $T$ is  
\begin{equation}
  \label{lehmann}
  G_{\alpha \beta} (z) = \sum_{m=1}^{N_{\rm ex}} 
  \frac{Q_{\alpha m} Q_{\beta m}^*}{z - \w_{m}} 
\end{equation}
with 
\begin{equation}
  Q_{\alpha m} = \sqrt{\e^{(\Omega -E_r)/T} + \e^{(\Omega -E_s)/T}} 
  \bra r |c_\alpha |s \ket 
\end{equation}
and $\w_{m} = E_s - E_r$, 
where $m = (r, s) = 1, 2,\cdots, N_{\rm ex}$ represents all possible pairs of 
eigenstates $|r \ket$ and $|s \ket$ of Hamiltonian ${H}$ 
with their eigenvalues $E_r$ and $E_s$, respectively~\cite{note2}.  
We adopt the convention that the chemical potential term is included in $H$ and therefore 
$z=0$ in $G_{\alpha \beta} (z)$ corresponds to the chemical potential. 
$\Omega = -T\ln \sum_r \e^{-E_r/T}$ is the grand potential and 
$c_{\alpha}$ is a fermion-annihilation operator with single-particle state 
$\alpha\, (= 1, 2,\cdots, L_{\rm s})$. 
For example, $\alpha$ can be 
a set of spin $\sigma$, momentum $\mb{k}$, and orbital $\xi$ indices 
[$\alpha \equiv (\s, \mb{k}, \xi)$], 
or simply a site index $i$ ($\alpha \equiv i$). 
The Green's function $G_{\alpha \beta} (z) $ is analytical in the complex plane 
except for the excitation energies at $\w_m$ and thus poles of $G_{\alpha \beta} (z) $ 
appear only on the real-frequency axis.

The Green's function is now written in an $L_{\rm s}\times L_{\rm s}$ matrix form
\begin{equation}
  \label{Gmat}
  \bs{G}(z) = \bs{Q} \bs{g}(z) \bs{Q}^\dag,
\end{equation}
where $\bs{Q} = \left[ Q_{\alpha m} \right]$ is an $L_{\rm s} \times N_{\rm ex}$ 
rectangular matrix and 
\begin{equation}
  \label{eq.g}
  \bs{g}(z) = {\rm diag}\left[1/(z - \w_1),\cdots, 1/(z - \w_{N_{\rm ex}}) \right]  
\end{equation}
is an $N_{\rm ex} \times N_{\rm ex}$ diagonal matrix with $\w_1\leqslant \w_2\leqslant \cdots\leqslant \w_{N_{\rm ex}}$. 
The anti-commutation relation of the fermionic operators  
$\{c_\alpha ^\dag , c_\beta\} = \delta_{\alpha \beta}$ 
guarantees the spectral weight sum rule, which is now written as 
$\sum_{m=1}^{N_{\rm ex}} Q_{\alpha m} Q_{\beta m}^{*} = \delta_{\alpha \beta}$ 
or equivalently
\begin{equation}
  \label{sumrule}
  \bs{Q Q}^{\dag} = \bs{I}. 
\end{equation}
It should be noted that in general $\bs{Q}$ is {\it not} a unitary matrix, i.e., 
$\bs{QQ}^\dag_{(L_{\rm s} \times L_{\rm s})} 
\not= \bs{Q}^\dag \bs{Q}_{(N_{\rm ex} \times N_{\rm ex})}$, 
where the subscripts denote the size of the resulting matrices.

\subsection{Diagonal elements of single-particle Green's function}\label{diagG}

Next, we consider analytical properties of the diagonal 
element of the single-particle Green's function $G_{\alpha \alpha}(z)$~\cite{Luttinger1961,Eder2014,Seki2016_2,Karlsson2016}  
because the particle number is evaluated through 
the trace of the single-particle Green's function. 
It is apparent from Eq.~(\ref{lehmann}) that the imaginary part of $G_{\alpha \alpha}(z)$, 
${\rm Im} G_{\alpha \alpha}(z)$, is always finite 
when frequency $z$ is away from the real axis.  
Therefore, zeros of $G_{\alpha \alpha}(z)$ must lie on the real-frequency axis. 
The fact that the single-particle Green's function 
is a rational function with respect to $z$ and 
$G_{\alpha \alpha}(z) \sim 1/z$ for large $|z|$ [see Eqs.~(\ref{lehmann}) and (\ref{sumrule})]   
ensures that the diagonal element of the single-particle Green's function 
is in the following form: 
\begin{equation}
  \label{prod}
  G_{\alpha \alpha}(z) = \frac
  {\prod_{l=1}^{Z_{\alpha \alpha}}\left(z - \zeta_l^{(\alpha)}\right) }
  {\prod_{m=1}^{P_{\alpha \alpha}}\left(z - \w_m^{(\alpha)}\right)}  
\end{equation}
with 
\begin{equation}
  \label{eq:pz}
  P_{\alpha \alpha} - Z_{\alpha \alpha} = 1, 
\end{equation}
where $\zeta_l^{(\alpha)}$ 
is a real frequency (with $\zeta_1^{(\alpha)}<\zeta_2^{(\alpha)}<\cdots<\zeta_{Z_{\alpha\alpha}}^{(\alpha)}$) 
at which $G_{\alpha \alpha}(\zeta_l^{(\alpha)})=0$, 
$\w_m^{(\alpha)}\in\{\w_1,\w_2,\cdots,\w_{N_{\rm ex}}\}$ with 
$\w_1^{(\alpha)}<\w_2^{(\alpha)}<\cdots<\w_{P_{\alpha\alpha}}^{(\alpha)}$, and 
$Z_{\alpha \alpha}$ ($P_{\alpha \alpha}$) is the number of zeros (poles) 
of $G_{\alpha \alpha} (z)$~\cite{note}.   
Here, $P_{\alpha \alpha}$ is counted only when the 
corresponding spectral weight is non-zero, i.e., $|Q_{\alpha m}|> 0$. 
Thus, $P_{\alpha \alpha}$ can be smaller than $N_{\rm ex}$.

\begin{figure}
  \begin{center}  
    \includegraphics[width=7.5cm]{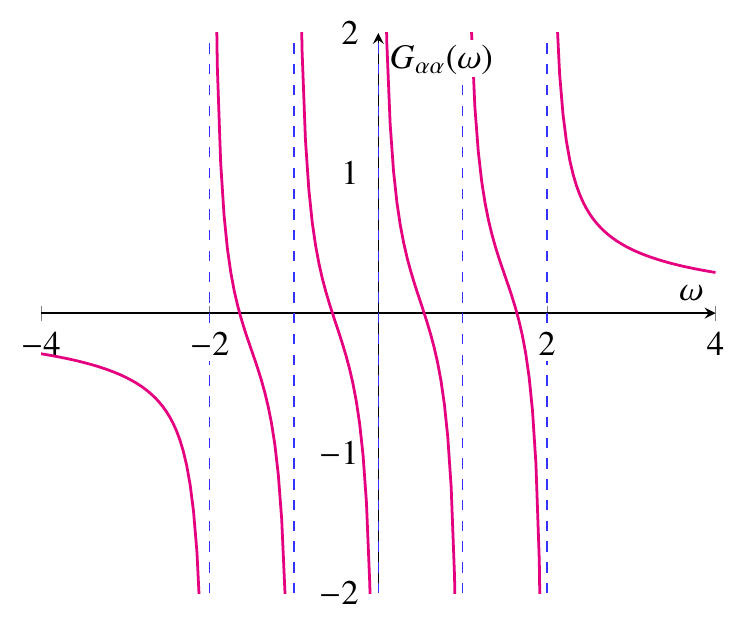}
    \caption{
      Schematic figure of the diagonal element of the single-particle Green's function 
      $G_{\alpha\alpha}(\w)=\sum_m |Q_{\alpha m}|^2/(\w-\w_m^{(\alpha)})$ 
      (thick solid lines) on the real-frequency axis  $\w$.  
      This toy Green's function $G_{\alpha\alpha}(\w)$ has five poles at 
      $\w_m^{(\alpha)} = -2, -1, 0, 1,$ and $2$ (indicated by dashed vertical lines) 
      with spectral weight 
      $|Q_{\alpha m}|^{2} = 1/5$ for $m=1, 2, \cdots, 5$. 
    }
    \label{toyReG}
  \end{center}
\end{figure}

Typical frequency dependence of $G_{\alpha \alpha}(\w)$ is 
shown in Fig.~\ref{toyReG}. 
The analytical properties of $G_{\alpha \alpha} (\w)$ is understood as follows. 
Since 
\begin{equation}\label{eq:gz_dag}
  \bs{G}(z)^\dag = \bs{Qg}(z)^\dag \bs{Q}^\dag = \bs{G}(z^*), 
\end{equation}
$\bs{G}(\w)$ is Hermitian and thus $G_{\alpha\alpha}(\w)$ is real for real frequency $\w$. 
In the vicinity of a pole at $\w_{m}^{(\alpha)}$, $G_{\alpha \alpha}(\w)$ is 
positive (negative) on the right (left) side of $\w_{m}^{(\alpha)}$ 
because 
\begin{equation}
  G_{\alpha \alpha}(\w \simeq \w_m^{(\alpha)}) \simeq \frac{|Q_{\alpha m}|^2}{\w - \w_m^{(\alpha)}}
\end{equation}
with the positive spectral weight, i.e., $|Q_{\alpha m}|^2 > 0$. 
On the other hand, its derivative 
\begin{equation}
  \frac{\partial G_{\alpha \alpha}(\w)}{\partial\w}
  = -\sum_{m=1}^{P_{\alpha \alpha}} \frac{|Q_{\alpha m}|^2}{ (\w - \w_{m}^{(\alpha)})^2}
\end{equation}
is always negative for $\w \not = \w_m^{(\alpha)}$, indicating that 
$G_{\alpha \alpha}(\w)$ is a decreasing function of $\w\,( \not = \w_m^{(\alpha)})$. 
This immediately concludes 
that there must exist only a single frequency at which $G_{\alpha\alpha}(\w)=0$ 
between two distinct successive real frequencies where $G_{\alpha\alpha}(\w)$ exhibits poles, i.e.,
\begin{equation}
\label{eq:wz}
  \w_1^{(\alpha)} < \zeta_1^{(\alpha)} < \w_2^{(\alpha)} < \cdots < \zeta_{Z_{\alpha \alpha}}^{(\alpha)} < \w_{P_{\alpha \alpha}}^{(\alpha)},
\end{equation}
with 
\begin{eqnarray}
  \left\{
  \begin{array}{l}
    G_{\alpha \alpha}(\w) > 0,   \ \ {\rm for}\,\, \w_{m}^{(\alpha)}  < \w < \zeta_{m}^{(\alpha)}   \\
    G_{\alpha \alpha}(\w) = 0,   \ \ {\rm for}\,\, \w = \zeta_{m}^{(\alpha)}   \\
    G_{\alpha \alpha}(\w) < 0,   \ \ {\rm for}\,\, \zeta_{m}^{(\alpha)} < \w < \w_{m+1}^{(\alpha)} 
  \end{array}
  \right. ,
  \label{signGaa}
\end{eqnarray}
as shown in Fig.~\ref{toyReG}.

\subsection{Determinant of single-particle Green's function}
Let us now examine analytical properties of the 
determinant of the single-particle Green's function, 
already analyzed to a certain extent in 
Refs.~\cite{Dzyaloshinskii2003}, \cite{Karlsson2016} and \cite{Eder2008}. 
Here, we shall show that the determinant of the single-particle 
Green's function can be expressed as a simple rational 
polynomial function as in Eq.~(\ref{detG}) with 
the numbers of zeros and poles satisfying Eq.~(\ref{NpNzdet}) 
(see also Appendix A of Ref.~\cite{Seki2016_2}).

From the Cauchy-Binet theorem, 
the determinant of the single-particle Green's function 
\begin{equation}
  \det \bs{G}(z) = \det \left[\bs{Qg}(z)\bs{Q}^\dag \right]
\end{equation}
is identically zero if $N_{\rm ex} < L_{\rm s}$. However, generally $N_{\rm ex} \geq L_{\rm s}$ and 
thus we can safely assume that $ \det \bs{G}(z)$ is not identically zero. 
From Eqs.~(\ref{eq.g}) and (\ref{sumrule}), the asymptotic behavior of the 
determinant for large $|z|$ is 
$\det \bs{G}(z) \sim (1/z)^{L_{\rm s}}.$
This already suggests that $\det \bs{G}(z)$ has a form shown in 
Eqs.~(\ref{detG}) and (\ref{NpNzdet}). 
In the following, we shall show that zeros of $\det \bs{G}(z)$ are all on the 
real-frequency axis. 

Let us first triangularize $\bs{G}(z)$ by a unitary transformation 
(Schur decomposition), 
\begin{eqnarray}
  \bs{R}(z) = \bs{U}(z) \bs{G}(z) \bs{U}(z)^\dag,
\end{eqnarray}
where 
$\bs{R}(z)$ is an upper triangle matrix and 
$\bs{U}(z)$ is a unitary matrix. 
From Eq.~(\ref{Gmat}), $\bs{R}(z)$ can be written as  
\begin{equation}
  \bs{R}(z) = \tilde {\bs{Q}}(z) \bs{g}(z) \tilde{\bs{Q}}(z)^\dag,
\end{equation}
where $\tilde{\bs{Q}}(z) = \bs{U}(z) \bs{Q}$ 
is an $L_{\rm s}\times N_{\rm ex}$ matrix with its matrix element 
\begin{equation}
\tilde{Q}_{\alpha m}(z) 
= \sqrt{\e^{(\Omega - E_r)/T}+ \e^{(\Omega - E_s)/T}} \bra r |\tilde{c}_{\alpha}(z) |s\ket
\end{equation} 
and 
$\tilde{c}_{\alpha}(z) = \sum_{\beta=1}^{L_{\rm s}} U_{\alpha \beta}(z) c_{\beta}.$
It is apparent from Eq.~(\ref{sumrule}) and the unitarity of $\bs{U}(z)$ 
that $\tilde{\bs{Q}}(z)$ fulfills the sum rule
\begin{equation}
  \tilde{\bs{Q}}(z) \tilde{\bs{Q}}(z)^\dag = \bs{I}
\end{equation}
as $\{\tilde{c}_{\alpha}^\dag(z), \tilde{c}_\beta(z)\} = \delta_{\alpha \beta}$.

The diagonal element of $\bs{R}(z)$ is now given as 
\begin{equation}
  R_{\alpha \alpha} (z) = \sum_{m=1}^{N_{\rm ex}} \frac{|\tilde{Q}_{\alpha m}(z)|^2} {z - \w_{m}}.
\end{equation}
Since the sum rule $\sum_{m=1}^{N_{\rm ex}} |\tilde{Q}_{\alpha m}(z)|^2 = 1$ must hold 
for arbitrary $z$, 
$\tilde{Q}_{\alpha m}(z)$ is bounded in the 
entire complex $z$ plane, and thus it must be constant, i.e., 
\begin{equation}
  \tilde{Q}_{\alpha m}(z) = \tilde{Q}_{\alpha m},
\end{equation}
known as Liouville's theorem~\cite{Ahlfors}. 
Therefore, $R_{\alpha \alpha}(z)$ 
has the same analytical properties as $G_{\alpha \alpha}(z)$ and 
it is written as 
\begin{equation}
  R_{\alpha \alpha} (z) = 
  \frac
      {\prod_{l=1}^{\tilde{Z}_{\alpha \alpha}} \left(z - \tilde{\zeta}_l^{(\alpha)}\right)} 
      {\prod_{m=1}^{\tilde{P}_{\alpha \alpha}} \left(z - \tilde\w_m^{(\alpha)}\right)} 
\end{equation}
with 
\begin{equation}
  \tilde{P}_{\alpha \alpha} - \tilde{Z}_{\alpha \alpha} = 1,
\end{equation} 
where $\tilde{\zeta}_z^{(\alpha)} $($\tilde{\zeta}_1^{(\alpha)}<\tilde{\zeta}_2^{(\alpha)}<\cdots<\tilde{\zeta}_{Z_{\alpha\alpha}}^{(\alpha)}$) 
is a real frequency 
at which $R_{\alpha \alpha}(\tilde{\zeta}_z^{(\alpha)})=0$, and 
$\tilde\w_m^{(\alpha)}\in\{\w_1,\w_2,\cdots,\w_{N_{\rm ex}}\}$ with 
$\tilde\w_1^{(\alpha)}<\tilde\w_2^{(\alpha)}<\cdots<\tilde\w_{\tilde P_{\alpha\alpha}}^{(\alpha)}$.   
$\tilde{Z}_{\alpha \alpha}$ ($\tilde{P}_{\alpha \alpha}$) is the number of zeros (poles) of 
$R_{\alpha \alpha} (z)$~\cite{note} and $\tilde{P}_{\alpha \alpha}$ is counted only when the 
corresponding spectral weight is non-zero, i.e., $|\tilde{Q}_{\alpha m}|>0$. 
Similarly to Eq.~(\ref{eq:wz}), we can also show that 
\begin{equation}\label{eq:pz2}
  \tilde\w_1^{(\alpha)} < \tilde\zeta_1^{(\alpha)} < \tilde\w_2^{(\alpha)} < \tilde\zeta_2^{(\alpha)} < \cdots < \tilde\zeta_{Z_{\alpha \alpha}}^{(\alpha)} < \tilde\w_{P_{\alpha \alpha}}^{(\alpha)}.
\end{equation}

Since $\det \bs{G}(z)=\det \bs{R}(z)$, $\det \bs{G}(z) $ is now readily evaluated as 
\begin{eqnarray}
  \det \bs{G}(z)
  &=&
  \prod_{\alpha =1}^{L_{\rm s}} 
  \left[
    \frac
    {\prod_{l' = 1}^{\tilde{Z}_{\alpha \alpha}} \left(z - \tilde{\zeta}_{l'}^{(\alpha)}\right)}
    {\prod_{m' = 1}^{\tilde{P}_{\alpha \alpha}} \left(z - \tilde\w_{m'}^{(\alpha)}\right) }
  \right] \nonumber \\
 & = &
  \frac
  {\prod_{l=1}^{Z_{\det}} \left(z - \tilde{\zeta}_{l}\right)}
  {\prod_{m=1}^{P_{\det}} \left(z - \tilde\w_{m}\right) } 
\label{detG}
\end{eqnarray}
with
\begin{equation}
  \tilde{\zeta}_{l} \in \{\tilde{\zeta}_{l'}^{(\alpha)} | \alpha=1,2,\dots,L_{\rm s};\, l'=1,2,\dots,\tilde{Z}_{\alpha\alpha}    \} 
  \label{det_zeta}
\end{equation}
and 
\begin{equation}
  \tilde{\w}_{m} \in \{ \tilde{\w}_{m'}^{(\alpha)}  | \alpha=1,2,\dots,L_{\rm s};\, m'=1,2,\dots,\tilde{P}_{\alpha\alpha}  \}, 
  \label{det_w}
\end{equation} 
where 
$Z_{\det} = \sum_{\alpha=1}^{L_{\rm s}} \tilde{Z}_{\alpha \alpha}$ is the number of zeros of $\det \bs{G}(z)$ 
and $P_{\det} = \sum_{\alpha=1}^{L_{\rm s}} \tilde{P}_{\alpha \alpha}$ 
is the number of poles of $\det \bs{G}(z)$. 
Here, each zero (pole) is counted in $Z_{\det}$ ($P_{\det}$) 
as many times as its order and thus some of $\tilde{\zeta}_{l}$  ($\tilde{\w}_{m}$) 
in Eq.~(\ref{det_zeta}) [Eq.~(\ref{det_w})] 
might be the same. We can now readily show that   
\begin{equation}
  P_{\det} - Z_{\det} = L_{\rm s}. 
  \label{NpNzdet}
\end{equation}
It is apparent above that zeros of $\det \bs{G}(z) $ are all on the real-frequency axis. 
Note however that generally there is no relation similar to Eq.~(\ref{eq:wz}) 
(i.e., only one zero between the two successive poles) 
for zeros and poles of $\det{\bs G}(z)$ in Eq.~(\ref{detG}).
Note also that $\ln \det\bs{G}(z)$ is analytical as long as $z$ is away from 
the real-frequency axis because $\tilde{\zeta}_{l}$ and $\tilde\w_{m}$ are both real.

\subsection{Particle number}

Using $\bs{G}(z)$, the average particle number $N$ is evaluated as
\begin{equation}
  \label{TrG}
  N = T \sum_{\nu = -\infty}^{\infty} \e^{\imag \w_\nu 0^+} \tr [\bs{G}(\imag \w_\nu)],
\end{equation}
where $\imag \w_\nu = (2\nu + 1)\imag \pi T$ with integer $\nu$ 
is the fermionic Matsubara frequency~\cite{Ezawa1957,Matsubara1955} and $0^+$ represents infinitesimally 
small positive real number.
The frequency sum in Eq.~(\ref{TrG}) 
can be converted to the contour integral, 
\begin{equation}
  \label{NLuttinger}
  N 
  = \oint_{\Gamma} \frac{\dd z}{2 \pi \imag} n_{\rm F} (z) \tr \left[\bs{G}(z)\right]
  = \sum_{\alpha=1}^{L_{\rm s}} \sum_{m=1}^{N_{\rm ex}} n_{\rm F} (\w_{m}) |Q_{\alpha m}|^{2}, 
\end{equation}
where 
\begin{equation}
  n_{\rm F} (z) = \frac{1}{\e^{z/T} + 1}
\end{equation}
is the Fermi-Dirac distribution function and contour $\Gamma$ 
encloses the singularities of $\tr \left[ \bs{G}(z) \right]$, 
not the ones of $n_{\rm F}(z)$, 
in the counter-clockwise direction, as shown in Fig.~\ref{contour0}(a).

\begin{center}
  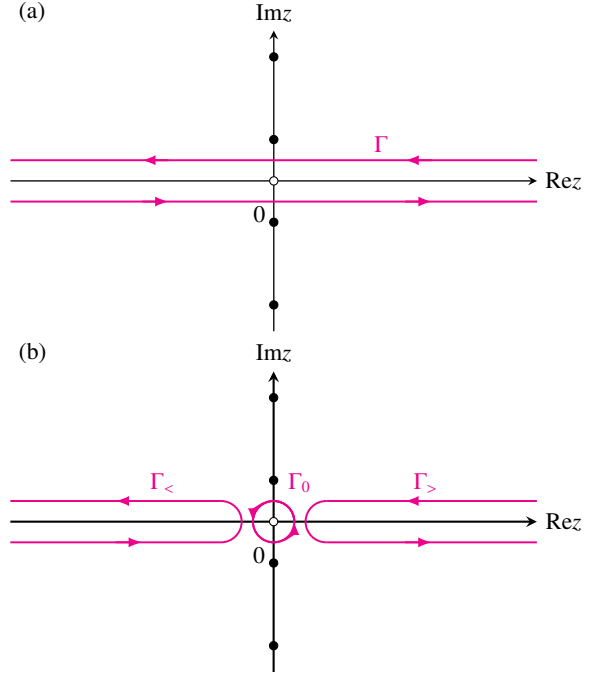
\begin{figure}
    \tikzstyle{contour1}=[magenta, line width=0.750pt]
    \def\xmin{-3.5}
    \def\xmax{3.5}
    \def\ymin{-2.0}
    \def\ymax{2.0}
    \def\angle{7.5}
    \def\dy{0.55}

    \begin{tikzpicture}
      \draw [line width=0.5pt, -stealth](\xmin, 0.0)--(\xmax,0.0) node [anchor=west]  { ${\rm Re}z$};
      \draw [line width=0.5pt, -stealth]( 0.0,\ymin)--(0.0,\ymax) node [anchor=south] { ${\rm Im}z$};
      \foreach \y in {-3,-1,1,3}{
        \filldraw[draw=black,fill=black] (0.0, \dy*\y) circle (0.055);
      }
      \filldraw[draw=black,fill=white] (0.0,0.0) circle (0.055);

      \draw[contour1,-latex] (\xmax*1.0,  \dy*0.5) -- (\xmax*0.5,  \dy*0.5);
      \draw[contour1,-latex] (\xmax*0.6,  \dy*0.5) -- (\xmin*0.5,  \dy*0.5);
      \draw[contour1]        (\xmin*0.4,  \dy*0.5) -- (\xmin*1.0,  \dy*0.5);
      \draw[contour1]        (\xmax*1.0, -\dy*0.5) -- (\xmax*0.5, -\dy*0.5);
      \draw[contour1,latex-] (\xmax*0.6, -\dy*0.5) -- (\xmin*0.5, -\dy*0.5);
      \draw[contour1,latex-] (\xmin*0.4, -\dy*0.5) -- (\xmin*1.0, -\dy*0.5);

      \node [magenta, anchor=south west] at (\xmax*0.35, \dy*0.5) { $\Gamma$};
      \node [black, anchor=north east] at   (0,-\dy*0.4) { $0$};
      \node [black, anchor=south west] at (\xmin,\ymax)  { (a)};

    \end{tikzpicture}

    \begin{tikzpicture}
      \draw [thick, -stealth](\xmin, 0.0)--(\xmax,0.0) node [anchor=west]  { ${\rm Re}z$};
      \draw [thick, -stealth]( 0.0,\ymin)--(0.0,\ymax) node [anchor=south] { ${\rm Im}z$};
      \foreach \y in {-3,-1,1,3}{
        \filldraw[draw=black,fill=black] (0.0, \dy*\y) circle (0.055);
      }
      \filldraw[draw=black,fill=white] (0.0,0.0) circle (0.055);


      \node [magenta, anchor=south west] at (\xmin*0.5,\dy*0.5) { $\Gamma_<$};
      \node [magenta, anchor=south west] at (\xmax*0.025,      \dy*0.5) { $\Gamma_0$};
      \node [magenta, anchor=south west] at (\xmax*0.5,\dy*0.5) { $\Gamma_>$};
      \node [black, anchor=north east] at (0,-\dy*0.4)  { $0$};
      \node [black, anchor=south west] at (\xmin,\ymax) { (b)};

      \draw[contour1,-latex] (\xmin*1.0,-\dy*0.5) -- (\xmin*0.5,-\dy*0.5);
      \draw[contour1]        (\xmin*0.6,-\dy*0.5) -- (\xmin*0.2,-\dy*0.5);
      \draw[contour1]        (\xmin*0.2,-\dy*0.5) arc  (-90:90:\dy*0.5);
      \draw[contour1,-latex] (\xmin*0.2,\dy*0.5) -- (\xmin*0.6,\dy*0.5);
      \draw[contour1]        (\xmin*0.5,\dy*0.5) -- (\xmin*1.0,\dy*0.5);

      \draw [contour1,latex-] (0:\dy*0.5) arc (360:0:\dy*0.5);
      \draw [contour1,latex-] (180:\dy*0.5) arc (180:0:\dy*0.5);
      \draw [contour1] (0:\dy*0.5) arc (360:0:\dy*0.5);

      \draw[contour1]        (\xmax*0.5,-\dy*0.5) -- (\xmax*1.0,-\dy*0.5);
      \draw[contour1,-latex]  (\xmax*0.2,-\dy*0.5) -- (\xmax*0.6,-\dy*0.5);
      \draw[contour1]        (\xmax*0.2,-\dy*0.5) arc  (270:90:\dy*0.5);
      \draw[contour1] (\xmax*0.6,\dy*0.5) -- (\xmax*0.2,\dy*0.5);
      \draw[contour1,latex-]        (\xmax*0.5,\dy*0.5) -- (\xmax*1.0,\dy*0.5);

    \end{tikzpicture}
    
    \caption{
      (a) Contour $\Gamma$ in complex $z$ plane. 
      (b) Contours $\Gamma_<$, $\Gamma_0$, and $\Gamma_>$ in complex $z$ plane. 
      Filled dots on the imaginary axis represent the 
      Matsubara frequencies $\imag \w_\nu = (2\nu + 1) \imag \pi T$ with $\nu$ integer. 
      The origin is indicated by an open dot in each figure. 
    }
    \label{contour0}
  \end{figure}
\end{center}

\section{Generalized Luttinger theorem}\label{sec:GLT}

Based on the finite-temperature formulation, we shall now show that 
(i) the Luttinger volume can be represented as the winding number of the determinant of the single-particle 
Green's function in the zero-temperature limit, 
(ii) the winding number of a ratio between the determinants of the interacting and noninteracting single-particle 
Green's functions provides the topological interpretation 
of the generalized Luttinger theorem, 
(iii) 
the topological interpretation can naturally separates two qualitatively different types (types I and II) 
of the condition for the validity of the generalized Luttinger theorem, 
(iv) a sufficient condition for the validity of the generalized Luttinger theorem of type I follows 
by directly applying the Rouche's theorem in complex analysis, 
and 
(v) the generalized Luttinger theorem is valid for generic interacting fermions 
as long as the particle-hole symmetry is preserved. 
Let us first define the Luttinger volume.

\subsection{Luttinger volume}\label{sec:lv}

From Dyson's equation for the single-particle Green's function, 
\begin{equation}
  \bs{G}(z)^{-1} = \bs{G}_0(z)^{-1} - \bs{\Sigma}(z),
\end{equation}
we can derive an identity 
\begin{equation}\label{eq:tr_G}
  \tr [\bs{G}(z)] = \frac{\partial \ln \det {\bs{G}(z)}^{-1}}{\partial z} 
  + \tr\left[ \bs{G}(z) \frac{\partial \bs{\Sigma} (z)}{\partial z} \right], 
\end{equation}
where $\bs{G}_{0}(z)=\left( z-H_0\right)^{-1}$ ($H_0$: the noninteracting part of Hamiltonian $H$) is 
the noninteracting single-particle Green's function and $\bs{\Sigma}(z)$ is the self-energy. 
Here we have used that 
\begin{equation}
  \frac{\partial\ln\det{\bs{G}(z)}^{-1}}{\partial z}=\tr\left[ \bs{G}(z)\frac{\partial  {\bs{G}(z)}^{-1}}{\partial z}\right].
\end{equation}
By substituting this identity in Eq.~(\ref{NLuttinger}), 
we can readily show that 
\begin{equation}
  \label{identity}
  N = V_{\rm L} + \oint_{\Gamma} \frac{\dd z}{2 \pi \imag} n_{\rm F} (z)
  \tr \left[ \bs{G}(z) \frac{\partial \bs{\Sigma} (z)}{\partial z} \right] 
\end{equation}
where we define the Luttinger volume $V_{\rm L}$ as 
\begin{equation}
  \label{eq.VL}
  V_{\rm L} = \oint_{\Gamma} 
  \frac{\dd z}{2 \pi \imag} n_{\rm F} (z) 
  \frac{\partial \ln  \det \bs{G}(z)^{-1}}{\partial z}. 
\end{equation}
Notice that $V_{\rm L}$ defined here is comparable to the particle number $N$ 
rather than the particle density. 
As shown in Appendix~\ref{sec:ad}, Eq.~(\ref{identity}) can also be derived directly from 
the derivative of the grand potential $\Omega$ with respect to the chemical potential $\mu$ 
(see also Ref.~\cite{Luttinger-Ward1960}).

There are three remarks in order. 
First, 
the Luttinger volume $V_{\rm L}$ in the zero-temperature limit is identical with 
the volume enclosed by the Fermi surface in metallic systems 
as originally proposed by Luttinger~\cite{Luttinger1960}, and  
the volume enclosed by the Luttinger surface in single-particle gapful systems, as generalized 
by Dzyaloshinskii~\cite{Dzyaloshinskii2003} 
(examples for this remark will be given in Sec.~\ref{sec:examples}). 
Second, the Luttinger volume $V_{\rm L}$ is an extensive quantity. 
For example, if the single-particle Green's function $\bs{G}(z)$ is diagonalized 
with respect to a single-particle index $\alpha$ (e.g., band index and momentum), i.e., 
$\bs{G}(z) = \oplus_\alpha {G}_{\alpha\alpha}(z)$, 
then the Luttinger volume is given as 
\begin{equation}\label{eq:vl_a}
  V_{\rm L} = \sum_\alpha V_{{\rm L},\alpha}, 
\end{equation}
where 
\begin{equation}
V_{{\rm L},\alpha} = \oint_\Gamma \frac{\dd z}{2\pi \imag} n_{\rm F} (z) \frac{\partial \ln {G}_{\alpha\alpha}^{-1}(z)}{\partial z}
\end{equation}
is the Luttinger volume labeled with $\alpha$. 
Therefore, the Luttinger volume $V_{\rm L}$ defined here is apparently an 
extensive quantity with respect to the single-particle index $\alpha$.  
Third, 
in the noninteracting limit, $V_{\rm L}=N$ simply because the self-energy $\bs{\Sigma}(z)=0$.

From Eq.~(\ref{detG}) and the Cauchy's integral theorem 
(or the argument principle)~\cite{Ahlfors}, we can now show that 
\begin{equation}
  \label{VLuttinger}
  V_{\rm L} = 
  \sum_{m=1}^{P_{\det}} n_{\rm F} (\tilde\w_{m}) - 
  \sum_{l=1}^{Z_{\det}} n_{\rm F} (\tilde{\zeta}_{l}), 
\end{equation}
where $\tilde{\zeta}_{l}$ and $\tilde\w_{m}$ are zeros and poles of the determinant of the single-particle 
Green's function given in 
Eqs.~(\ref{det_zeta}) and (\ref{det_w}), respectively. 
Note that $V_{\rm L}$ is a well-defined quantity and is unambiguously 
evaluated even for insulating states at zero temperature. 
This is simply because the chemical potential is always uniquely determined
in the zero-temperature limit even when it is located in a single-particle gap. 
It should also be noticed in Eq.~(\ref{VLuttinger}) that, in the zero-temperature limit, 
each pole (zero) exactly at the chemical potential contributes a factor of $1/2$ ($-1/2$)
to the Luttinger volume $V_{\rm L}$ since $n_{\rm F}(0)=1/2$. 
This implies that the Luttinger volume can be fractionalized when the zero-energy singularities 
exist in the determinant of the single-particle Green's function.

The generalized Luttinger theorem states that 
\begin{equation}
  \lim_{T \rightarrow 0}N = \lim_{T \rightarrow 0} V_{\rm L}, 
\end{equation}
or more explicitly, by equating Eqs.~(\ref{NLuttinger}) and (\ref{VLuttinger}), 
\begin{equation}
  \label{LSR}
  \sum_{\alpha=1}^{L_{\rm s}} \sum_{m=1}^{N_{\rm ex}} n_{\rm F}(\w_m) |Q_{\alpha m}|^2 =
  \sum_{m=1}^{P_{\det}} n_{\rm F} (\tilde\w_m) - 
  \sum_{l=1}^{Z_{\det}} n_{\rm F} (\tilde{\zeta}_l) ,
\end{equation}
and taking the zero-temperature limit. 
It is now obvious in Eq.~(\ref{LSR}) that the generalized Luttinger theorem is 
represented with the number of zeros and poles of the determinant of the single-particle Green's function.

We should note that an equation similar to Eq.~(\ref{LSR}) has been 
reported by Ortloff {\it et al.}~\cite{Ortloff2007} for single-band systems directly using the zero-temperature formulation  
where the Heaviside step function $\Theta(\w)$ appears, 
instead of the Fermi-Dirac distribution function $n_{\rm F}(\w)$. 
However, in their zero-temperature formulation, 
the value of the Heaviside step function at zero energy, $\Theta(0)$, 
is not specified~\cite{Ortloff2007, Eder2011}.  
Our finite-temperature formulation described here clarifies that the Heaviside step function at 
zero energy in the zero-temperature formulation should be regarded as $\Theta(0) = n_{\rm F}(0) = 1/2$. 
The ambiguity in treating poles and zeros of the single-particle Green's function 
(or the determinant of the single-particle Green's function) at the chemical potential is therefore 
clearly resolved in the finite-temperature formulation.

\subsection{Luttinger volume and winding number of $\det \bs{G}(z)$ }\label{sec:lv_wn}

Here we shall show that, in the zero-temperature limit, 
the Luttinger volume $V_{\rm L}$ defined in Eq.~(\ref{eq.VL}) is represented exactly as the winding number of the 
determinant of the single-particle Green's function. 
Since the Fermi-Dirac distribution function $n_{\rm F}(\w)$ in the zero-temperature limit 
takes three values depending on $\w$, i.e., 
\begin{eqnarray}
  \lim_{T \rightarrow 0} n_{\rm F} (\w) =
  \left\{
    \begin{array}{c}
      0           \ \ {\rm for} \,\,\w > 0 \\
      \frac{1}{2} \ \ {\rm for} \,\,\w = 0 \\
      1           \ \ {\rm for} \,\,\w < 0
    \end{array}
  \right.,
\end{eqnarray}
we divide contour $\Gamma$ into three pieces, $\Gamma_<$, $\Gamma_0$, and $\Gamma_>$, 
as shown in Fig.~\ref{contour0}(b), where contour $\Gamma_<$ ($\Gamma_>$) encloses the 
negative (positive) real axis and contour $\Gamma_0$ encloses the origin. 

Accordingly, the Luttinger volume can be divided into three parts,  
\begin{equation}
  V_{\rm L} =
  \left(\oint_{\Gamma_<} + \oint_{\Gamma_0} + \oint_{\Gamma_>} \right) 
  \frac{\dd z}{2 \pi \imag} n_{\rm F} (z) \frac{\partial \ln \det {\bs{G}(z)^{-1}}}{\partial z}. 
\end{equation}
In the zero-temperature limit, the integral along contour $\Gamma_>$ vanishes  
because $n_{\rm F}(\w)=0$ for $\w > 0$. Thus, we find that 
\begin{equation}\label{eq:vl_zero}
  \lim_{T \rightarrow 0} V_{\rm L}  = 
  n_{\det \bs{G}^{-1}}(\Gamma_<)  + \frac{1}{2} n_{\det \bs{G}^{-1}}(\Gamma_0)
\end{equation}
where 
\begin{eqnarray}\label{wn_detG}
  n_{\det \bs{G}^{-1}}(\mcal{C})
  &=& \oint_{\mcal{C}} \frac{\dd z}{2 \pi \imag} \frac{\partial \ln \det {\bs{G}(z)^{-1}}}{\partial z} \notag \\ 
  &=& \oint_{\det{\bs{G}^{-1}}(\mcal{C})} \frac{\dd (\det {\bs{G}^{-1}})}{2 \pi \imag} \frac{1}{\det {\bs{G}^{-1}}}.
\end{eqnarray}
Here $\det \bs{G}^{-1}(\mcal{C})$ represents the contour in complex $\det \bs{G}^{-1}$ plane 
which is parametrized by $z \in \mcal{C}\, (= \Gamma_<\, {\rm and}\,\Gamma_0)$. 
Therefore, $n_{\det \bs{G}^{-1}}(\mcal{C})$ is the winding number of $\det {\bs{G}^{-1}}$
around the origin of the complex $\det \bs{G}^{-1}$ plane (see Fig.~\ref{contour_detG}) and thus it is 
necessarily integer. Notice also that 
\begin{equation}
  n_{\det{\bs{G}^{-1}}}(\mcal{C}) = - n_{\det{\bs{G}}}(\mcal{C})
\end{equation} 
by definition.

\begin{center}
  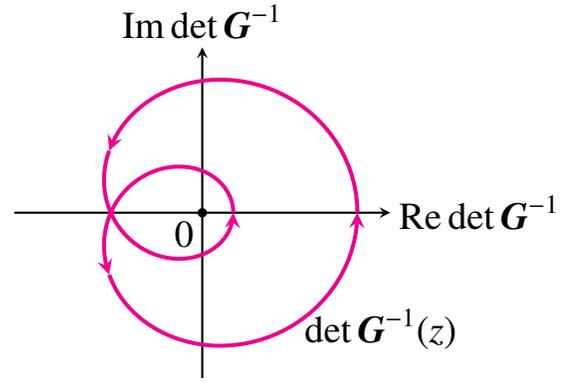
\begin{figure}
    \def\xmin{-2.5}
    \def\xmax{2.5}
    \def\ymin{-2.2}
    \def\ymax{2.2}
    \def\a{1.5}
    \def\b{1.4}
    \begin{tikzpicture}
      \draw [thick, -stealth](\xmin, 0.0)--(\xmax,0.0) node [anchor=west]  {\Large ${\rm Re}\det\bs{G}^{-1}$};
      \draw [thick, -stealth]( 0.0,\ymin)--(0.0,\ymax) node [anchor=south] {\Large ${\rm Im}\det\bs{G}^{-1}$};

      \draw[magenta,-stealth,domain=2.0*pi:1.0*pi,scale=1.5,line width = 1.5pt, samples=40] 
      plot ({0.55*(-\a*cos(\x r)+1*sin(\x*0.5 r))},{0.55*(\b*sin(\x r)+cos(\x*0.5 r))});
      \draw[magenta,-stealth,domain=1.0*pi:0*pi,scale=1.5,line width = 1.5pt, samples=40] 
      plot ({0.55*(-\a*cos(\x r)+1*sin(\x*0.5 r))},{0.55*(\b*sin(\x r)+cos(\x*0.5 r))});
      \draw[magenta,-stealth,domain=0*pi:-1.0*pi,scale=1.5,line width = 1.5pt, samples=40] 
      plot ({0.55*(-\a*cos(\x r)+1*sin(\x*0.5 r))},{0.55*(\b*sin(\x r)+cos(\x*0.5 r))});
      \draw[magenta,-stealth,domain=-1.0*pi:-2.0*pi,scale=1.5,line width = 1.5pt, samples=40] 
      plot ({0.55*(-\a*cos(\x r)+1*sin(\x*0.5 r))},{0.55*(\b*sin(\x r)+cos(\x*0.5 r))});

      \node [black, anchor=west]     at (\xmax * 0.5,\ymin * 0.7)        {\Large $\det\bs{G}^{-1}(z)$};
      \node [black, anchor=north east] at (0,0)  {\Large $0$};
      \filldraw[draw=black,fill=black] (0.0,0.0) circle (0.055);

    \end{tikzpicture}

    \caption{
      Schematic figure for the contour of 
      $\det \bs{G}^{-1}(\mcal{C})$ 
      in Eq.~(\ref{wn_detG}), parametrized by 
      $z \in \mcal{C}$, 
      on the complex $\det \bs{G}^{-1}$ plane. 
      The arrowheads indicate the direction of the trajectory in 
      $\det\bs{G}(z)^{-1}$ with $z \in \mcal{C}$ where 
      $\mcal{C}\, (= \Gamma_<,\, \Gamma_0,\,{\rm and}\, \Gamma_>)$ is shown in Fig.~\ref{contour0}(b). 
      The winding number of $\det \bs{G}^{-1}$ 
      around the origin corresponds to $n_{\det \bs{G}^{-1}}(\mcal{C})$ defined in Eq.~(\ref{wn_detG}). 
      The winding number in this figure is $n_{\det \bs{G}^{-1}}(\mcal{C}) = 2$. 
    }
    \label{contour_detG}
  \end{figure}
\end{center}

We should also emphasize here that 
$n_{\det \bs{G}^{-1}}(\Gamma_<)$, 
$n_{\det \bs{G}^{-1}}(\Gamma_0)$, and 
$n_{\det \bs{G}^{-1}}(\Gamma_>)$ are given simply by 
counting the number of poles and zeros of the determinant of the single-particle Green's function
below, exactly at, and above the chemical potential, i.e., 
\begin{equation}\label{eq:n_detG<}
  n_{\det \bs{G}^{-1}}(\Gamma_<) 
  = \sum_{m=1}^{P_{\det}} \Theta_{0} (-\tilde{\w}_{m})
  - \sum_{l=1}^{Z_{\det}} \Theta_{0} (-\tilde{\zeta}_{l}),
\end{equation}
\begin{equation}\label{eq:n_detG0}
  n_{\det \bs{G}^{-1}}(\Gamma_0)   
  = \sum_{m=1}^{P_{\det}} \delta_{\tilde{\w}_{m},0}
  - \sum_{l=1}^{Z_{\det}} \delta_{\tilde{\zeta}_{l}, 0}, 
\end{equation}
and 
\begin{equation}\label{eq:n_detG>}
  n_{\det \bs{G}^{-1}}(\Gamma_>) 
  = \sum_{m=1}^{P_{\det}} \Theta_{0} (\tilde{\w}_{m})
  - \sum_{l=1}^{Z_{\det}} \Theta_{0} (\tilde{\zeta}_{l})
\end{equation}
respectively.  Here $\Theta_c(\w)$ is the Heaviside step function defined as 
\begin{eqnarray}\label{eq:Theta_c}
  \Theta_c (\w) =
  \left\{
    \begin{array}{c}
      1           \ (\w > 0) \\
      c           \ (\w = 0) \\
      0           \ (\w < 0)
    \end{array}
  \right.,
\end{eqnarray}
and $\delta_{\alpha,\beta}$ is the Kronecker delta, which is 
1 only when $\alpha=\beta$ and zero otherwise.

\subsection{Topological interpretation 
of the generalized Luttinger theorem}\label{sec:ti}

We shall now examine the condition under which the generalized Luttinger theorem is valid.  
For this purpose, we analyze the deviation of the Luttinger volume from the noninteracting limit, which 
can be represented as the winding number of the ratio $D(z)$ between the determinants of the interacting 
and noninteracting single-particle Green's functions defined in Eq.~(\ref{fredholm}).

The Luttinger volume $V_{\rm L}^0$ for the noninteracting system is $N$. This can be shown directly 
by comparing Eq.~(\ref{NLuttinger}) and the definition of $V_{\rm L}$ given in Eq.~(\ref{eq.VL}),  
\begin{equation}\label{eq.VL0}
  V_{\rm L}^0 = \oint_{\Gamma} 
  \frac{\dd z}{2 \pi \imag} n_{\rm F} (z) 
  \frac{\partial \ln  \det \bs{G}_0(z)^{-1}}{\partial z}=N,  
\end{equation}
because  
\begin{equation}
\frac{\partial \ln\det{\bs{G}_0(z)}^{-1}}{\partial z}=\tr\left[ \bs{G}_0(z)\right]  
\end{equation}
when $\bs{\Sigma} (z)=0$ in Eq.~(\ref{eq:tr_G}). 
Therefore, the deviation $\Delta V_{\rm L}$ of the Luttinger volume from the noninteracting 
limit is the second term of the right-hand side in Eq.~(\ref{identity}), i.e., 
\begin{equation}
\Delta V_{\rm L}= V_{\rm L}-V_{\rm L}^0 = - \oint_{\Gamma} \frac{\dd z}{2 \pi \imag} n_{\rm F} (z)
  \tr \left[ \bs{G}(z) \frac{\partial \bs{\Sigma} (z)}{\partial z} \right].
\end{equation} 
By introducing the ratio between the determinants of the interacting and noninteracting single-particle Green's 
functions 
\begin{equation}
  \label{fredholm}
  D(z) 
  = \frac{\det \bs{G}_0(z)}{\det \bs{G}(z)} 
  = \det [\bs{I} - \bs{G}_0(z)  \bs{\Sigma}(z)]  
\end{equation}
directly in Eqs.~(\ref{eq.VL}) and (\ref{eq.VL0}), we can show that 
\begin{equation}
  \label{dlnD}
  \Delta V_{\rm L} = 
  \oint_{\Gamma} \frac{\dd z}{2 \pi \imag} n_{\rm F} (z) 
  \frac{\partial \ln D(z) }{\partial z},  
\end{equation}
where contour $\Gamma$ is defined in Fig.~\ref{contour0}(a).

We first notice that, in the zero-temperature limit, 
contour $\Gamma$ for the integral of Eq.~(\ref{dlnD}) in complex $z$ plane 
is reduced to contours $\Gamma_{<}$ and $\Gamma_{0}$
(see Fig.~\ref{contour0}) because $n_{\rm F}(\w)=0$ for $\w>0$, as discussed in Sec.~\ref{sec:lv_wn}.
Therefore, at zero temperature, the deviation of the Luttinger volume from the noninteracting one, 
$\Delta V_{\rm L}$, given in 
Eq.~(\ref{dlnD}) corresponds exactly to the winding number $n_D(\mcal{C})$ of $D(z)$ 
around the origin of complex $D$ plane, i.e., 
\begin{equation}\label{D_VL}
  \lim_{T \rightarrow 0}\Delta V_{\rm L}=  n_D(\Gamma_<) + \frac{1}{2} n_D(\Gamma_0)
\end{equation}
where 
\begin{equation} 
  \label{condition}
  n_D (\mcal{C})
  = \oint_{\mcal{C}} \frac{\dd z}{2 \pi \imag} \frac{\partial \ln D(z) }{\partial z} 
  = \oint_{D(\mcal{C})} \frac{\dd D}{2 \pi \imag} \frac{1}{D}
\end{equation}
and $D(\mcal{C})$ represents the contour in complex $D$ plane, which is parametrized 
by $z\in\mcal{C}\, (=\Gamma_<\,{\rm and}\,\Gamma_0)$ (see Fig.~\ref{contour}). 
Notice that $n_D (\mcal{C})=0$ in the noninteracting limit as $D(z)=1$. 
It should be emphasized that the quantity $n_D(\mcal{C})$ evaluated 
in Eq.~(\ref{condition}) must be integer 
as it is the winding number. 
Since the Fredholm-type determinant $D(z)$ can be defined for infinite dimensional matrices, 
Eq.~(\ref{condition}) is valid even in the thermodynamic limit. 
It should be also noticed that from the definition of $D(z)$ in Eq.~(\ref{fredholm})  
\begin{eqnarray}\label{eq:n_D}
  n_D(\mcal{C})
  &=& n_{\det {\bs{G}^{-1}}} (\mcal{C}) - n_{\det {\bs{G}^{-1}_0}} (\mcal{C}) \notag \\
  &=& n_{\det {\bs{G}_0}}    (\mcal{C}) - n_{\det {\bs{G}}} (\mcal{C}), 
\end{eqnarray}
where $n_{\det {\bs{G}^{-1}}} (\mcal{C})$ is defined in Eq.~(\ref{wn_detG}).

\begin{center}
  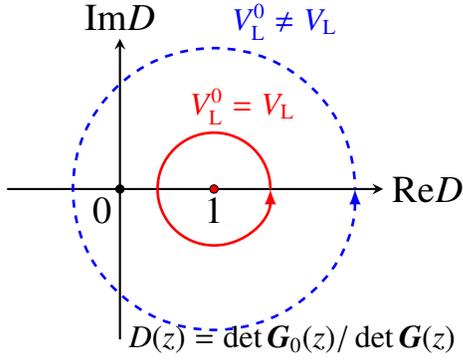
\begin{figure}
    \tikzstyle{contour1}=[red, line width=1.0pt]
    \tikzstyle{contour2}=[blue!100, dashed, line width=1.0pt]
    \def\xmin{-1.5}
    \def\xmax{3.5}
    \def\ymin{-2.0}
    \def\ymax{2.0}
    \def\angle{7.5}
    \def\dy{0.55}
    \def\unit{1.25}
    \def\r{\unit*0.6}
    \def\rr{\unit*1.5}
    \begin{tikzpicture}
      \draw [thick, -stealth](\xmin, 0.0)--(\xmax,0.0) node [anchor=west]  {\Large ${\rm Re}D$};
      \draw [thick, -stealth]( 0.0,\ymin)--(0.0,\ymax) node [anchor=south] {\Large ${\rm Im}D$};

      \filldraw[draw=black,fill=black] (0.0,0.0) circle (0.055);
      \filldraw[draw=black,fill=red]   (\unit,0.0) circle (0.055);
      
      \node [red, anchor=south]      at (\unit+\r*0.5,\r)   {\large $V_{\rm L}^0=V_{\rm L}$};
      \node [blue!100, anchor=south] at (\unit+\rr*0.5,\rr) {\large $V_{\rm L}^0\not=V_{\rm L}$};
      \node [black, anchor=west]     at (0,\ymin)           {\large $D(z)=\det\bs{G}_0(z)/\det\bs{G}(z)$};
      \node [black, anchor=north east] at (0,0)  {\Large $0$};
      \node [black, anchor=north] at (\unit,0)  {\Large $1$};

      \draw [contour1,latex-] (0:\unit+\r) arc (360:0:\r);
      \draw [contour2,latex-] (0:\unit+\rr) arc (360:0:\rr);

    \end{tikzpicture}
    
    \caption{
      Schematic figure to explain the relation between $D(z)=\det \bs{G}_0(z)/\det \bs{G}(z)$ 
      and the generalized Luttinger theorem. 
      The red solid and blue dashed lines represent 
      the integral contours $D(\mcal{C})$ in Eq.~(\ref{condition}) parametrized 
      by $z\in\mcal{C}\, (=\Gamma_<\,{\rm and}\,\Gamma_0)$. 
      The generalized Luttinger theorem of type I with Eq.~(\ref{Luttinger00}) 
      is valid (violated) when $D(\mcal{C})$ does not (does) 
      enclose the origin of complex $D$ plane, i.e., zero (non zero) winding number of $D(z)$ around the origin, 
      as indicated by red solid (blue dashed) line. 
      The red dot on the positive real axis at $D(z)=1$ represents 
      the noninteracting limit. 
    }
    \label{contour}
  \end{figure}
\end{center}

It is now apparent in Eq.~(\ref{D_VL}) that there exists two cases where the 
generalized Luttinger theorem is valid. The first case (type I) is when 
$n_D(\Gamma_<)$ and $n_D(\Gamma_0)$ are both zero, i.e., 
\begin{equation}
  \label{Luttinger00}
  n_D (\Gamma_<) =  n_D (\Gamma_0) = 0.
\end{equation}
Figure~\ref{contour} schematically shows contour $D(\mcal{C})$ in complex $D$ plane 
and explains the relation between the winding number $n_D(\mcal{C})$ and the generalized Luttinger theorem. 
Applying the weak version of Rouche's theorem~\cite{Ahlfors} to Eq.~(\ref{condition}), 
we find that 
\begin{equation}
|D(z) - 1| < 1
\label{sufficient} 
\end{equation}
for $z \in \Gamma_<$ and $\Gamma_0$ is a {\it sufficient} condition 
for $V_{\rm L} = V_{\rm L}^0$ (see Fig.~\ref{contour}).  
Considering the fact that $D(z)=1$ in the noninteracting limit, 
the inequality~(\ref{sufficient}) represents the robustness of the theorem 
against the perturbation of fermion interactions.  
In fact, the generic inequality~(\ref{sufficient}) can reproduce a particular 
condition which ensures the convergence of the perturbation expansion of the self-energy 
for a spin-density-wave state reported in Ref.~\cite{Altshuler1998}.

Another case (type II) which ensures the validity of the generalized Luttinger theorem 
is when neither $n_D(\Gamma_0)$ nor $n_D(\Gamma_{<})$ is zero 
but they cancel each other, i.e., 
\begin{equation}
  \label{genLuttinger}
  n_D (\Gamma_0) = - 2 n_D (\Gamma_<) \not = 0.
\end{equation}
The condition $n_D(\Gamma_0) \not = 0$ or, equivalently, 
\begin{equation} 
  n_{\det \bs{G}^{-1}}(\Gamma_0) \not = n_{\det \bs{G}^{-1}_0}(\Gamma_0)
\end{equation}
implies that the number of singularities of the 
determinant of the single-particle Green's function 
at the chemical potential is altered by introducing interactions. 
This happens, for example, if the whole Fermi surface (or a portion of the Fermi surface) 
is gapped out by introducing interactions. 
Nevertheless, as long as Eq.~(\ref{genLuttinger}) is satisfied, 
the generalized Luttinger theorem is guaranteed to be valid.

\subsection{Validity of the generalized Luttinger theorem for particle-hole symmetric systems}\label{sec:VGLT} 

Based on the analytical properties of the single-particle Green's function derived above, we shall now prove 
rigorously that the generalized Luttinger theorem is valid for generic interacting fermions as long as the 
particle-hole symmetry is preserved. 
For this purpose, it is important to recall that 
when the particle-hole symmetry is preserved, the Hamiltonian is 
invariant under a transformation, for example, $c_{\alpha}\to c_{\alpha}^\dag$. 
It then follows that $G_{\alpha\beta}(z) = - G_{\beta\alpha}(-z)$ 
and thus $\det\bs{G}(z)=(-1)^{L_s}\det\bs{G}(-z)$, where $L_{\rm s}$ is the dimension of $\bs{G}(z)$. 
Therefore, $\det\bs{G}(-z)=0$ when $\det\bs{G}(z)=0$. 
Similarly, it can be shown that $\left[\det\bs{G}(-z)\right]^{-1}=0$ when $\left[\det\bs{G}(z)\right]^{-1}=0$.

Thus, for particle-hole symmetric systems, 
(i) there exist a pair of states $m=(r,s)$ and ${\bar m}=({\bar r},{\bar s})$ with excitation energies 
$\tilde\w_m$ and $\tilde\w_{\bar m}$, 
respectively, distributed symmetrically with respect to zero energy, i.e., 
$\tilde\w_{\bar m} = - \tilde\w_{m}$, 
at which $\det\bs{G}(z)$ has poles, 
and similarly (ii) zeros of $\det \bs{G}(\w)$ 
appear symmetrically with respect to zero energy at $\tilde{\zeta}_l$ and $\tilde{\zeta}_{\bar l}$ 
where $\tilde{\zeta}_{\bar l} = -\tilde{\zeta}_{l}$. 
Note also that, 
for particle-hole symmetric systems, 
the chemical potential is exactly zero, independently of the temperature~\cite{Gebhard}. 
Using these properties as well as the identity for the Fermi-Dirac distribution function 
\begin{equation}
  \label{nF}
  n_{\rm F}(\w) + n_{\rm F} (-\w) = 1
\end{equation}  
in Eq.~(\ref{VLuttinger}), we can readily evaluate the Luttinger volume 
\begin{eqnarray}
  \label{Vph}
  V_{\rm L} 
  = \frac{1}{2} \left(P_{\det} - Z_{\det} \right) = \frac{L_{\rm s}}{2}. 
\end{eqnarray}
Here Eq.~(\ref{NpNzdet}) is used in the second equality~\cite{note5}. 
Apparently, in the noninteracting limit, 
$V_{\rm L}^0 = N$, as shown in Eq.~(\ref{eq.VL0}), and  $N= L_{\rm s}/2$ for the particle-hole 
symmetric case. 
Therefore, $\Delta V = 0 $, completing the proof that the generalized Luttinger theorem  
is valid for generic interacting fermions with the particle-hole symmetry. 
Notice that Eq.~(\ref{Vph}) is satisfied for all temperatures, including zero temperature. 

Finally, we should note that 
the the generalized Luttinger theorem is satisfied with the condition of either type I in Eq.~(\ref{Luttinger00}) 
or type II in Eq.~(\ref{genLuttinger}) for systems with the particle-hole symmetry. However, obviously, 
it is not necessarily the case that 
the particle-hole symmetry is preserved when either condition of type I or type II is satisfied.

\section{Examples for single-band interacting electrons with translational symmetry}\label{sec:examples}

In this section, we first summarize the 
analytical properties of the single-particle Green's function 
for a single-band, paramagnetic, and translationally symmetric system. 
We then explore the generalized Luttinger theorem of types I and II 
by examining a simple metal and a one-dimensional Mott insulator.  
As an example of multi-orbital systems with translational symmetry, 
the Hubbard model on the honeycomb lattice 
is examined within the Hubbard-I approximation in Appendix~\ref{sec:hubbard-I}.

\subsection{Summary of analytical properties}\label{sec:sum_sb}
When a system is paramagnetic and translationally symmetric, 
the single-particle Green's function $\bs{G}(z)$ is diagonal 
with its elements $G_{\mb{k}}(z)$ for each momentum $\mb{k}$ and therefore 
\begin{equation}\label{eq:detG_Gk}
  \det \bs{G}(z) = \left[\prod_{\mb{k}} G_{\mb{k}}(z) \right]^2,
\end{equation} 
where an exponent 2 is due to the spin degeneracy. 
Applying the argument in Sec.~\ref{diagG}, 
the single-particle Green's function $G_{\mb k}(z)$ for momentum $\mb k$ with spin $\sigma$ 
is generally given as
\begin{equation}\label{eq:gz_k_sb}
  G_{\mb k}(z) = \frac
  {\prod_{l=1}^{Z_{\mb k}} \left(z - \zeta_l^{(\mb k)}\right) }
  {\prod_{m=1}^{P_{\mb k}} \left(z - \w_m^{(\mb k)}\right)},  
\end{equation}
where real frequencies $\zeta_l^{(\mb k)}$ $(l=1,2,\cdots,Z_{\mb{k}})$ and 
$\w_m^{(\bm k)}$ $(m=1,2,\cdots,P_{\mb{k}})$ are poles and zeros of $G_{\mb{k}}(\w)$ 
for momentum ${\mb k}$, 
respectively, with 
\begin{equation}\label{eq:pz_k}
  \w_1^{(\mb k)} < \zeta_1^{(\mb k)} < \w_2^{(\mb k)} < \zeta_2^{(\mb k)} < \cdots < \zeta_{Z_{\mb k}}^{(\mb k)} < \w_{P_{\mb k}}^{(\mb k)} 
\end{equation}
and 
\begin{equation}\label{eq:pz3}
  P_{\mb k} - Z_{\mb k} = 1. 
\end{equation}
This is a simple example of Eqs.~(\ref{prod}) and (\ref{eq:pz}) for the single band system. 
The number $P_{\rm det}$ of poles and 
the number $Z_{\rm det}$ of zeros in $\det{\bs G}(z)$ is $P_{\rm det}=2\sum_{\mb k}P_{\mb k}$ and 
$Z_{\rm det}=2\sum_{\mb k}Z_{\mb k}$, 
respectively, and hence $P_{\rm det} - Z_{\rm det} = 2\sum_{\mb k}1$, which corresponds to Eq.~(\ref{NpNzdet}).

Therefore, for example, Eq.~(\ref{wn_detG}) is now simply given as 
\begin{eqnarray}\label{wn_detGk}
  n_{\det \bs{G}^{-1}}(\mcal{C})
  &=& 2\sum_{\mb{k}} n_{G^{-1}_{\mb{k}}}(\mcal{C}) 
\end{eqnarray} 
and the Luttinger volume in Eq.~(\ref{eq:vl_zero}) is 
\begin{equation}
  \label{eq:vl_k}
  \lim_{T \rightarrow 0} V_{\rm L} = 
  2 \sum_{\mb{k}} \left[
    n_{{G_{\mb{k}}^{-1}}} (\Gamma_<) + \frac{1}{2} 
    n_{{G_{\mb{k}}^{-1}}} (\Gamma_0) \right], 
\end{equation}
where the factor 2 is due to the spin degeneracy. 
We have also introduced that 
\begin{eqnarray}\label{eq:nGk}
  n_{G_{\mb{k}}^{-1}}(\mcal{C}) 
  = \oint_{\mcal{C}} \frac{\dd z}{2 \pi \imag} \frac{\partial \ln  {G}_{\mb{k}}^{-1}(z)}{\partial z} 
  = \oint_{G_{\mb{k}}^{-1}(\mcal{C})} \frac{\dd  {G}_{\mb{k}}^{-1}}{2 \pi \imag} \frac{1}{ {G}_{\mb{k}}^{-1}},  
\end{eqnarray}  
where $G_{\mb{k}}^{-1}(\mcal{C})$ represents the contour in complex $G_{\mb{k}}^{-1}$ plane 
parametrized by $z \in \mcal{C}$. Thus, $n_{G_{\mb{k}}^{-1}}(\mcal{C})$ is the winding number of 
${G}_{\mb{k}}^{-1}$ around the origin of the complex ${G}_{\mb{k}}^{-1}$ plane 
(for example, see Fig.~\ref{contour_detG}) and it must be integer. 
Notice also that $n_{G_{\mb{k}}^{-1}}(\mcal{C}) = - n_{G_{\mb{k}}}(\mcal{C}) $ 
by definition. 
As shown in Eqs.~(\ref{eq:n_detG<})--(\ref{eq:n_detG>}), $n_{G_{\mb{k}}^{-1}}(\mcal{C})$ can also be given 
by counting the number of poles and zero of the single-particle Green's function $G_{\mb{k}}(z)$, i.e., 
\begin{equation}\label{eq:n_G<}
  n_{G^{-1}}(\Gamma_<) 
  = \sum_{m=1}^{P_{\mb k}} \Theta_{0} \left(-\w_{m}^{(\mb k)}\right)
  - \sum_{l=1}^{Z_{\mb k}} \Theta_{0} \left(-\zeta_{l}^{(\mb k)}\right),
\end{equation}
\begin{equation}\label{eq:n_G0}
  n_{G^{-1}}(\Gamma_0)   
  = \sum_{m=1}^{P_{\mb k}} \delta_{\w_{m}^{(\mb k)},0}
  - \sum_{l=1}^{Z_{\mb k}} \delta_{\zeta_{l}^{(\mb k)}, 0}, 
\end{equation}
and 
\begin{equation}\label{eq:n_G>}
  n_{G^{-1}}(\Gamma_>) 
  = \sum_{m=1}^{P_{\mb k}} \Theta_{0} \left(\w_{m}^{(\mb k)}\right)
  - \sum_{l=1}^{Z_{\mb k}} \Theta_{0} \left(\zeta_{l}^{(\mb k)}\right). 
\end{equation}

Defining the ratio between the noninteracting and interacting single-particle Green's functions for momentum 
${\mb k}$
\begin{equation}\label{eq:Dk}
D_{\mb k} (z)= \frac{G_{0{\mb k}}(z)}{G_{\mb k}(z)} = 1-G_{0{\mb k}}(z)\Sigma_{\mb k}(z),
\end{equation}
where $G_{0{\mb k}}(z)$ is the single-particle Green's function in the noninteracting limit 
and $\Sigma_{\mb k}(z)$ is the 
self-energy, the Fredholm-type determinant of the single-particle Green's function 
in Eq.~(\ref{fredholm}) is now 
simply 
\begin{equation}
  D(z) = \left[\prod_{\mb k}D_{\mb k}(z) \right]^2, 
\end{equation}
including the spin degree of freedom. 
Therefore, the deviation of the Luttinger volume from the noninteracting one in the zero-temperature limit is 
\begin{equation}
  \label{eq:vl_k2}
\lim_{T\to0}\Delta V_{\rm L} = 2\sum_{\mb k}\left[ n_{D_{\mb k}}(\Gamma_<) + \frac{1}{2} n_{D_{\mb k}}(\Gamma_0)   \right], 
\end{equation}
where
\begin{equation}\label{eq:nD_k}
n_{D_{\mb k}}({\mcal C}) 
  = \oint_{\mcal{C}} \frac{\dd z}{2 \pi \imag} \frac{\partial \ln D_{\mb k}(z) }{\partial z} 
  = \oint_{D_{\mb k}(\mcal{C})} \frac{\dd D_{\mb k}}{2 \pi \imag} \frac{1}{D_{\mb k}}
\end{equation}
and $D_{\mb k}(\mcal{C})$ represents the contour of $D_{\mb k}(z)$, parametrized 
by $z\in\mcal{C} (=\Gamma_<\,{\rm and}\,\Gamma_0)$, 
in the complex $D_{\mb k}$ plane. 
Thus, $n_D(\mcal C)$ in Eq.~(\ref{condition}) is now simply 
\begin{equation}\label{eq:n_Dk}
n_D(\mcal C) = 2\sum_{\mb k} n_{D_{\mb k}}(\mcal C). 
\end{equation}
Notice also that by comparing Eqs.~(\ref{eq:nGk}) and (\ref{eq:nD_k}), 
\begin{equation}\label{eq:n_Dk_d}
n_{D_{\mb k}}(\mcal C) = n_{G_{\mb k}^{-1}}(\mcal C) - n_{G_{0{\mb k}}^{-1}}(\mcal C). 
\end{equation}

\subsection{type I: Simple metal}\label{sec:typeI}

Let us first consider the noninteracting limit. The single-particle Green's function $G_{0{\mb{k}}}(z)$ 
in the noninteracting limit is given as 
$G_{0 \mb{k}}(z)=1/(z-\w^{(\mb{k})}$), where $\w^{(\mb{k})}=\varepsilon_{\mb{k}}$ and $\varepsilon_{\mb{k}}$ is the single-particle 
energy dispersion in the noninteracting limit. 
Therefore, we find that 
$n_{{{G}_{0{\mb{k}}}^{-1}}}(\Gamma_<)=1$ and 
$n_{{{G}_{0{\mb{k}}}^{-1}}}(\Gamma_0)=n_{{{G}_{0{\mb{k}}}^{-1}}}(\Gamma_>)=0$ for $\mb{k}$ inside the Fermi surface, 
$n_{{{G}_{0{\mb{k}}}^{-1}}}(\Gamma_0)=1$ and 
$n_{{{G}_{0{\mb{k}}}^{-1}}}(\Gamma_<)=n_{{{G}_{0{\mb{k}}}^{-1}}}(\Gamma_>)=0$ for $\mb{k}$ on the Fermi surface, and 
$n_{{{G}_{0{\mb{k}}}^{-1}}}(\Gamma_>)=1$ and 
$n_{{{G}_{0{\mb{k}}}^{-1}}}(\Gamma_0)=n_{{{G}_{0{\mb{k}}}^{-1}}}(\Gamma_<)=0$ for $\mb{k}$ outside the Fermi surface.
Thus, 
$n_{G_{0{\mb k}}^{-1}}(\Gamma_<)$ [$n_{G_{0{\mb k}}^{-1}}(\Gamma_>)$] 
gives the number of occupied (unoccupied) 
single-particle states inside (outside) the Fermi surface, and 
a set of momenta where $n_{G_{0{\mb k}}^{-1}} (\Gamma_0)=1$  
forms the Fermi surface and 
the number of these $\mb{k}$ points corresponds to the area of the Fermi surface.  
The analytical properties of $G_{0 \mb{k}}(\w)$, including the sign of  $G_{0\mb{k}}(0)$, are summarized in Table~\ref{tableG0k}.

\begin{table}
  \caption{
    \label{tableG0k}    
    Analytical properties of the single-particle Green's function $G_{0 \mb{k}}(\w) = 1/(\w - \w^{(\mb{k})})$ in 
    the noninteracting limit, 
    where $\w^{(\mb{k})}=\varepsilon_{\mb{k}}$ denotes the noninteracting single-particle energy dispersion. 
    FS stands for Fermi surface. $n_{\mb{k}}^{(0)}$ is defined in Eq.~(\ref{QPTzero}).
  }
  \begin{tabular}{c|ccc}
    \hline \hline
    location of ${\mb{k}}$& inside FS & on FS & outside FS \\
    \hline
    position of a singularity  &$ \w^{(\mb{k})} < 0$ & $\w^{(\mb{k})} =0 $ & $\w^{(\mb{k})} > 0 $ \\
    \hline
    sign of $G_{\mb{k}}(0)$ & $G_{\mb{k}}(0) > 0$ & $G_{\mb{k}}^{-1}(0)=0$ & $G_{\mb{k}}(0) < 0$ \\
    \hline
    $n_{G_{0\mb{k}}^{-1}}(\Gamma_<)$ &  1 & 0 & 0 \\
    $n_{G_{0\mb{k}}^{-1}}(\Gamma_0)$ &  0 & 1 & 0 \\
    $n_{G_{0\mb{k}}^{-1}}(\Gamma_>)$ &  0 & 0 & 1 \\
    \hline
    $n_{\mb{k}}^{(0)}$ & 1 & 1/2 & 0 \\
    \hline \hline
  \end{tabular}
\end{table}

\begin{center}
  \begin{figure}
    \includegraphics[width=7.5cm]{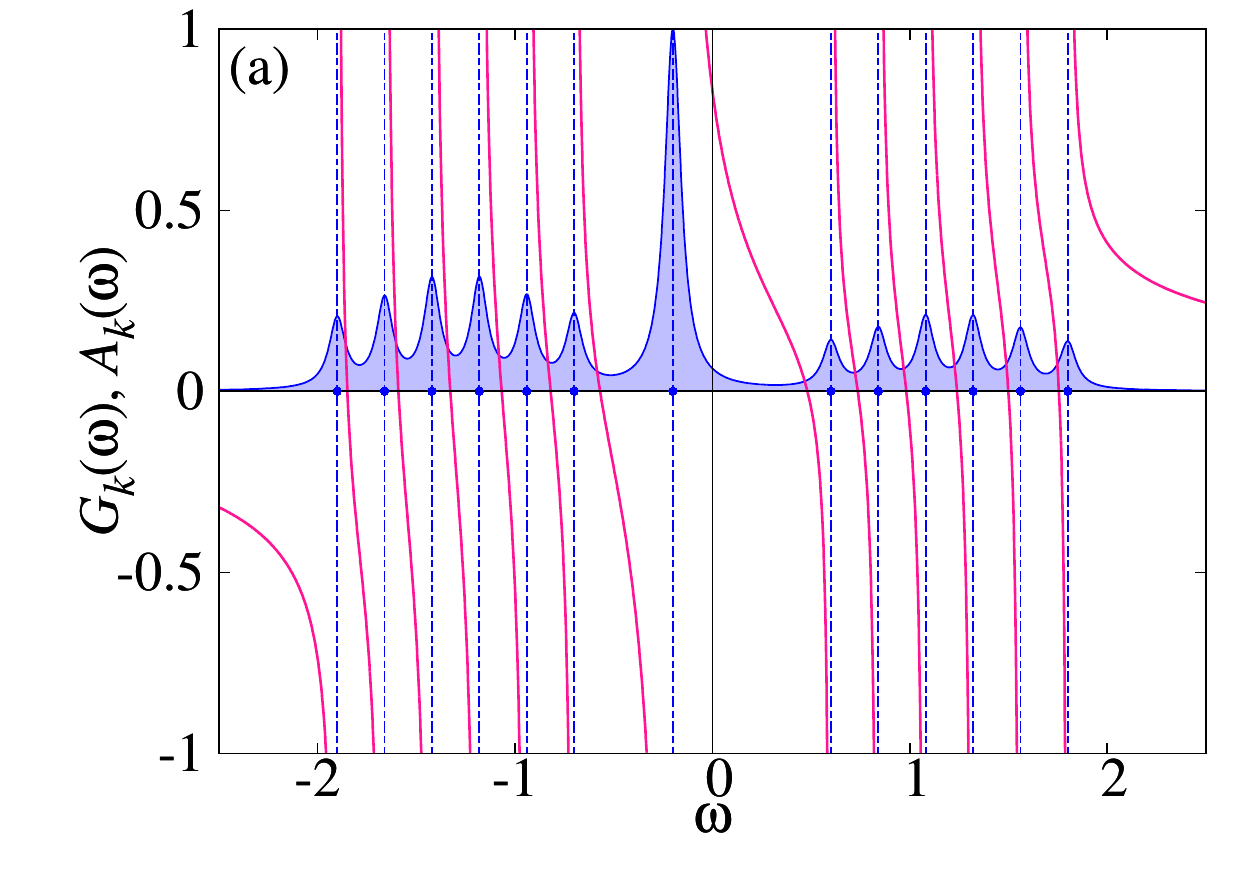}\\
    \includegraphics[width=7.5cm]{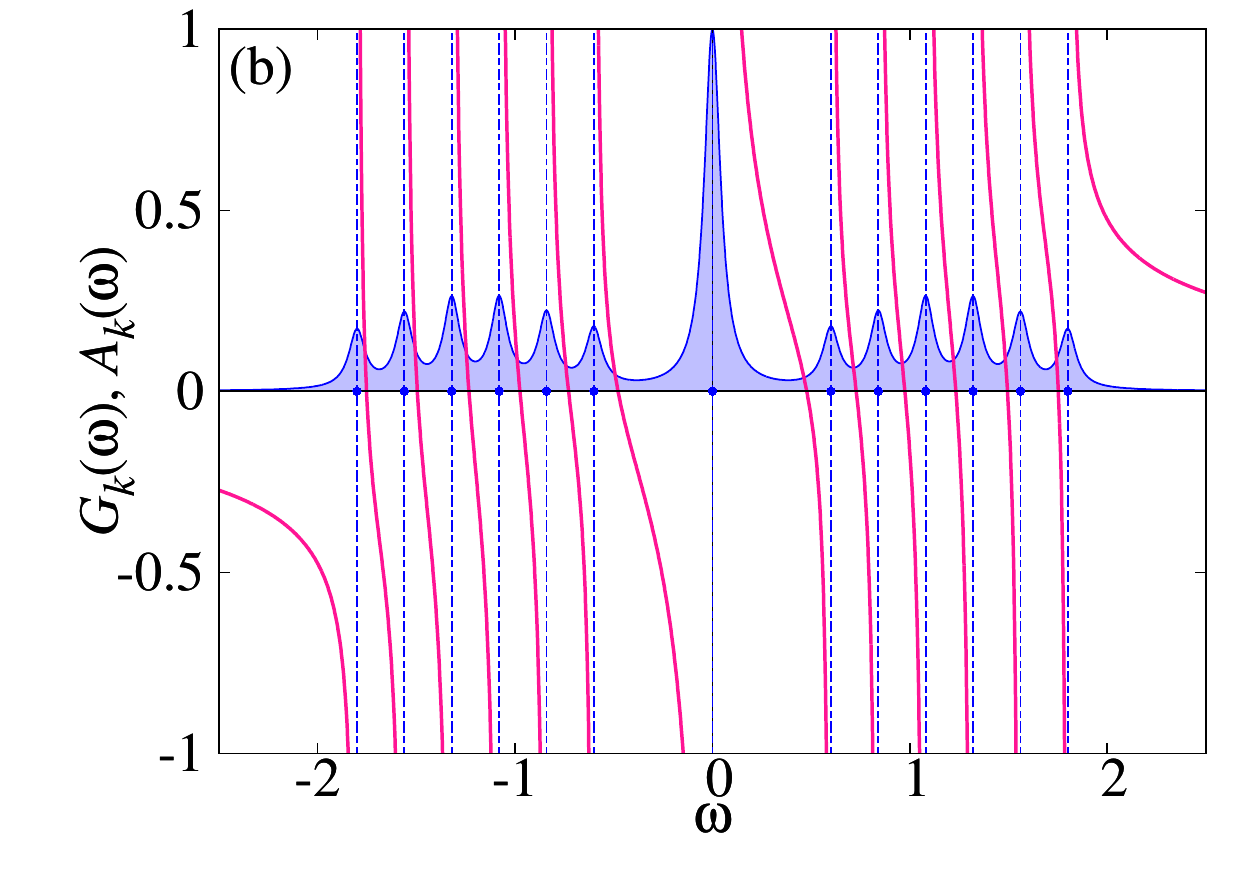}\\
    \includegraphics[width=7.5cm]{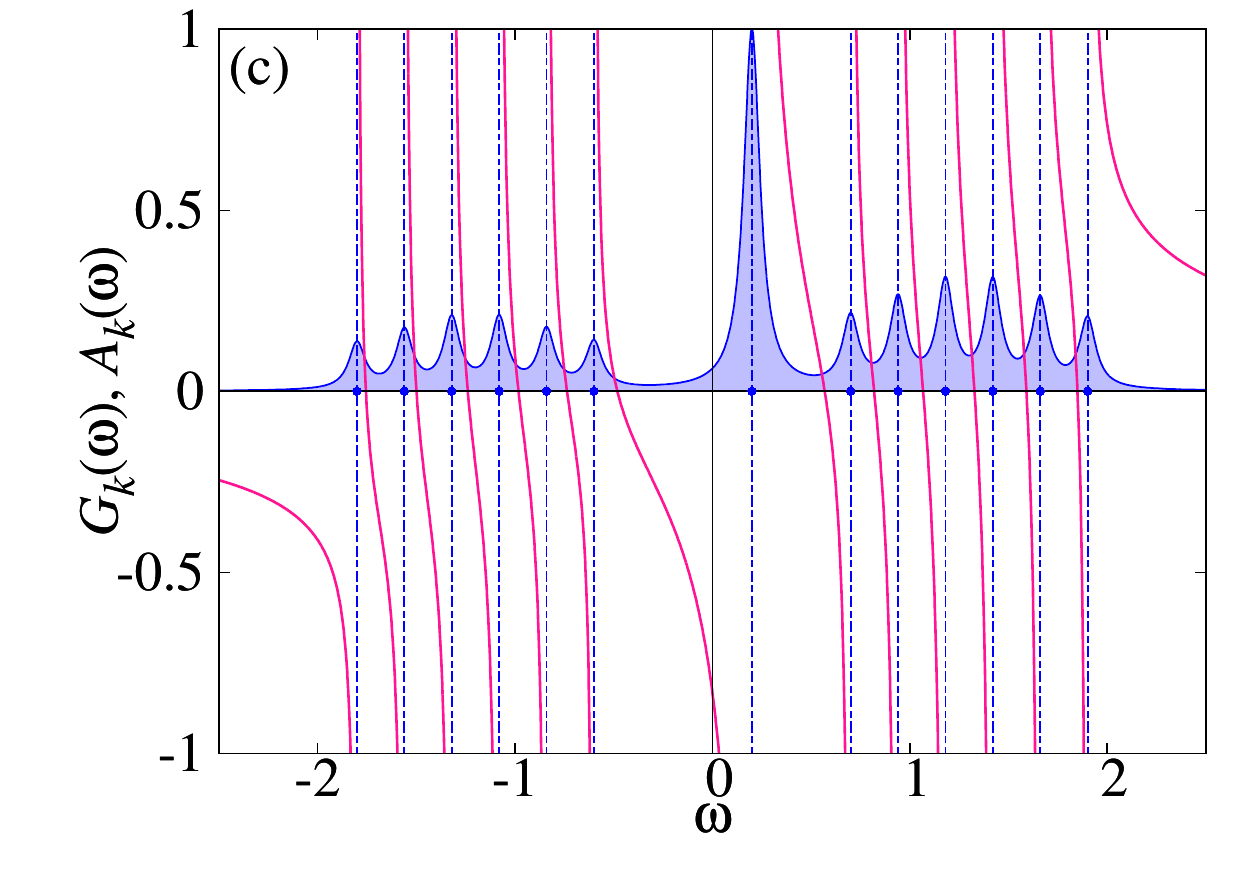}
    \caption{
      Schematic figures of 
      $A_{\mb{k}} (\w) = -{\rm Im} G_{\mb{k}} (\w+i\delta^+)/\pi$ (blue shaded region) and 
      $G_{\mb{k}}(\w)$ (red solid lines) at momentum $\mb{k}$ 
      (a) inside, (b) on, and (c) outside  
      the Fermi surface for a simple metal. Here $\omega=0$ corresponds to the Fermi energy. 
      The poles of $G_{\mb{k}}(\w)$ are indicated by dots along with dashed vertical lines. 
    }
    \label{fig:Akw}
  \end{figure}
\end{center}

Once the interactions are considered, $G_{\mb{k}}(\w)$ can have many poles as well as many zeros for 
each momentum $\mb{k}$. 
However, according to Eq.~(\ref{eq:pz3}), the number of poles is larger than the number of zeros exactly by one 
and thus  
\begin{equation}
n_{{{G}_{\mb{k}}^{-1}}}(\Gamma_<) + n_{{{G}_{\mb{k}}^{-1}}}(\Gamma_0) + n_{{{G}_{\mb{k}}^{-1}}}(\Gamma_>) = 1. 
\end{equation}
Typical behaviors of the single-particle spectral function
\begin{equation}
  A_{\mb{k}}(\w) = -\frac{1}{\pi}{\rm Im}{ G}_{\mb{k}}(\w+i\delta^+) 
  \label{eq:akw}
\end{equation}
and $G_{\mb{k}}(\w)$ are schematically shown in Fig.~\ref{fig:Akw}, 
where $\delta^+$ is a positively small real number.  

Here, following the Luttinger's argument 
on the {\it interior} of the Fermi surface for Fermi liquids~\cite{Luttinger1960}, 
we define that momentum $\mb{k}$ is inside the Fermi surface 
when the sign of the zero-energy Green's function is positive, i.e., $G_{\mb{k}}(0) > 0$, 
and similarly momentum $\mb{k}$ is outside the Fermi surface when $G_{\mb{k}}(0) < 0$. 
This implies that $G_{\mb{k}}(0)$ changes the sign only when momentum $\mb{k}$ crosses the Fermi surface.

Apparently, for momentum $\mb{k}$ on the Fermi surface, there exists a pole exactly at the chemical potential, i.e., 
\begin{equation}
  \w_1^{(\mb{k})} < \zeta_1^{(\mb{k})} < \cdots 
  < \zeta_{m_{\rm FS}-1}^{(\mb{k})} < \w_{m_{\rm FS}}^{(\mb{k})} = 0 < \zeta_{m_{\rm FS}}^{(\mb{k})} < \cdots,
  \label{wfs}
\end{equation}
with 
$n_{G^{-1}_{\mb{k}}}(\Gamma_<)=0$ and 
$n_{G^{-1}_{\mb{k}}}(\Gamma_0)=1$. Therefore, according to Eq.~(\ref{eq:vl_k}), 
this momentum contributes to the Luttinger volume $V_{\rm L}$  by one, including the spin degrees of freedom. 
Since $G_{\mb{k}}(0)$ exhibits a pole, its sign is not defined. 
A typical behavior of $A_{\mb{k}}(\w)$ and $G_{\mb{k}}(\w)$ 
is schematically shown in Fig.~\ref{fig:Akw} (b).

For momentum $\mb{k}$ inside the Fermi surface, 
the topmost singularity of the 
Green's function below the chemical potential must be a pole, i.e., 
\begin{equation}
  \w_1^{(\mb{k})} < \zeta_1^{(\mb{k})} < \cdots 
  < \zeta_{m_{\rm top}-1}^{(\mb{k})} < \w_{m_{\rm top}}^{(\mb{k})} < 0 < \zeta_{m_{\rm top}}^{(\mb{k})} < \cdots  
  \label{wtop}
\end{equation}
because $G_{\mb{k}}(0)>0$.
Since the number of poles below the chemical potential is larger 
than the number of zeros below the chemical potential exactly by one, 
it is shown that $n_{{{G}_{\mb{k}}^{-1}}}(\Gamma_<)=1$ and 
$n_{{{G}_{\mb{k}}^{-1}}}(\Gamma_0)=0$, and thus 
this momentum contributes to the Luttinger volume $V_{\rm L}$  
by two, including the spin degrees of freedom [see Eq.~(\ref{eq:vl_k})]. 
Recall here that in a simple metal such as Fermi liquid~\cite{Landau1956,Nozieres1962,Luttinger1962,Pines} 
the topmost pole at $\w_{m_{\rm top}}^{(\mb{k})}$ 
below and in the vicinity of  
the chemical potential corresponds to the quasiparticle, 
and the other poles form the incoherent part of the single-particle excitation, 
as depicted in Fig.~\ref{fig:Akw} (a).

For momentum $\mb{k}$ outside the Fermi surface, the number of poles below the chemical 
potential is exactly the same as the number of zeros below the chemical potential because 
\begin{equation}
  \w_1^{(\mb{k})} < \zeta_1^{(\mb{k})} < \cdots 
  < \w_{m_{\rm bot}-1}^{(\mb{k})} < \zeta_{m_{\rm bot}-1}^{(\mb{k})} < 0 < \w_{m_{\rm bot}}^{(\mb{k})} < \cdots,
  \label{wbot}
\end{equation}
satisfying that $G_{\mb{k}}(0) < 0$, as shown in Fig.~\ref{fig:Akw} (c). 
Therefore, $n_{G_{\mb{k}}^{-1}}(\Gamma_<)=n_{G_{\mb{k}}^{-1}}(\Gamma_0)=0$ 
and hence this momentum does not contribute to $V_{\rm L}$. 
The analytical properties of the single-particle Green's function $G_{\mb{k}}(\omega)$ are summarized in Table~\ref{tableGk}.   
Since $G_{\mb{k}}^{-1}(0)=0$ for $\mb{k}$ on the Fermi surface and 
$\Theta_c(G_{\mb{k}}(0)) = \Theta_c(G_{\mb{k}}^{-1} (0))$ for $|G_{\mb{k}}(0)| < \infty$, 
the Luttinger volume $V_{\rm L}$ in Eq.~(\ref{eq:vl_k}) can be given as 
\begin{equation}
  \label{normalLSR}
  \lim_{T \rightarrow 0 } V_{\rm L} = 2 \sum_{\mb{k}} \Theta_{\frac{1}{2}} \left(G_{\mb{k}}^{-1}(0) \right)    
\end{equation}
for a paramagnetic single-band system, where $\Theta_c(\w)$ is defined in Eq.~(\ref{eq:Theta_c}). 
Equation.~(\ref{normalLSR}) clearly shows that 
the Luttinger volume, defined as the winding number of the single-particle Green's function in 
Eq.~(\ref{eq:vl_k}), indeed corresponds to the momentum volume surrounded by the Fermi surface, as expected for 
a simple metal.

\begin{table}
  \caption{
    \label{tableGk}
    Analytical properties of the single-particle Green's function $G_{\mb{k}}(\w)$ for a simple metal. 
    FS stands for Fermi surface. $n_{\mb{k}}^{(0)}$ is defined in Eq.~(\ref{QPTzero}).
  }
  \begin{tabular}{c|ccc}
    \hline \hline
    location of ${\mb{k}}$ & inside FS & on FS & outside FS \\
    \hline 
    position of singularities  & Eq.~(\ref{wtop}) & Eq.~(\ref{wfs})& Eq.~(\ref{wbot}) \\
    \hline   
    sign of $G_{\mb{k}}(0)$ & $G_{\mb{k}}(0) > 0$ & $G_{\mb{k}}^{-1}(0)=0$ & $G_{\mb{k}}(0) < 0$ \\
    \hline
    $n_{G_{\mb{k}}^{-1}}(\Gamma_<)$ &  1 & 0 & 0 \\
    $n_{G_{\mb{k}}^{-1}}(\Gamma_0)$ &  0 & 1 & 0 \\
    $n_{G_{\mb{k}}^{-1}}(\Gamma_>)$ &  0 & 0 & 1 \\
    \hline
    $n_{\mb{k}}^{(0)}$ & 1 & 1/2 & 0 \\    
    \hline \hline
  \end{tabular}
\end{table}

If we assume that the shape of the Fermi surface does not change with and without 
introducing electron interactions, then obviously the analysis given above shows that 
$n_D(\Gamma_<)=0$ and $n_{D}(\Gamma_0)=0$, thus satisfying the condition of 
type I for the Luttinger theorem in Eq.~(\ref{Luttinger00}). 
However, in general, the condition of type I can be satisfied even if electron interactions alter 
the shape of the Fermi surface. Thus, we now simply 
assume that the type-I condition in Eq.~(\ref{Luttinger00}) 
is satisfied. Then, it follows immediately that in the zero-temperature limit 
\begin{equation}
  \label{original}
  N = 2 \sum_{\mb{k}} \Theta_{\frac{1}{2}} \left(G_{\mb{k}}^{-1}(0) \right). 
\end{equation}
This is the well known expression of the Luttinger theorem~\cite{Luttinger1960,AGD}, 
originally proved by the many-body perturbation theory in which 
the second-term of the right-hand side in Eq.~(\ref{identity}) 
vanishes under the assumption that the self-energy is regular 
at the chemical potential and thus the perturbation expansion is converged~\cite{Luttinger-Ward1960} 
(see also Ref.~\cite{Karlsson2016}).

Let us now discuss how the condition in Eq.~(\ref{Luttinger00}), 
obtained independently of the many-body perturbation theory, 
is related to Luttinger's original statement~\cite{Luttinger1960}. 
The first condition $n_D(\Gamma_<)=0$ in Eq.~(\ref{Luttinger00}) implies that the number of
$\mb{k}$ points inside the Fermi surface remains the same with and without introducing electron interactions.
This is exactly the original statement of the Luttinger theorem,
{\it ``The interaction may deform the FS (Fermi surface), but it cannot change its volume''}~\cite{Luttinger1960}.
The second condition $n_{D}(\Gamma_0)=0$ in Eq.~(\ref{Luttinger00}) implies that 
the number of zero-energy quasiparticle excitation, 
i.e., the number of ${\mb{k}}$ points on the Fermi surface, 
is unchanged by introducing electron interactions. 

The implication of the second condition is seemingly 
stronger than the original statement of the Luttinger theorem. 
However, the essential point of the second condition is to prohibit 
the appearance of the zeros of the single-particle Green's function, 
or equivalently the emergence of the poles of the self-energy, at the chemical potential 
by introducing electron interactions 
in order to ensure the convergence of the many-body perturbation theory.  
Therefore, the Luttinger theorem with 
the type-I condition in Eq.~(\ref{Luttinger00}) falls into 
the original statement of the theorem by Luttinger~\cite{Luttinger1960}.

We also note that, from Table~\ref{tableGk},  
it is plausible to regard the quantity 
in the parentheses of Eq.~(\ref{eq:vl_k}), i.e., 
\begin{equation}
  \label{QPTzero}
  n^{(0)}_{\mb{k}} = n_{{G_{\mb{k}}^{-1}}} (\Gamma_<) + \frac{1}{2} 
  n_{{G_{\mb{k}}^{-1}}} (\Gamma_0), 
\end{equation}  
as the distribution function of quasiparticles labeled by momentum $\mb{k}$ 
in the Fermi-liquid theory at zero temperature (see for example Eq.~(1.1) of Ref.~\cite{Luttinger1962}). 
Thus, the winding number $n_{{G_{\mb{k}}^{-1}}} ({\cal C})$ 
of the interacting single-particle Green's function $G_{\mb{k}}(z)$ embodies 
the concept of the quasiparticle distribution function (not the bare particle one). 
Therefore, the Luttinger volume in the zero temperature limit, 
  $\lim_{T\to0}V_{\rm L}=2\sum_{\mb k} n^{(0)}_{\mb{k}}$, 
represents nothing but the number of quasiparticles. 
Recall now that in the Landau's Fermi-liquid theory the number $N$ of particles is 
{\it a priori} assumed to be equal to the number of quasiparticles at zero temperature~\cite{Landau1956,Luttinger1962}.  
Hence, the argument here guarantees this fundamental assumption of the Landau's Fermi-liquid theory 
if the Luttinger theorem is valid since the theorem equates $N$ with the Luttinger volume.
In Appendix~\ref{sec:QP}, we generalize $n_{\mb{k}}^{(0)}$ for 
finite (but still low) temperatures and discuss the physical meaning.

\subsection{type II: Mott insulator}\label{sec:typeII}

\begin{center}
  \begin{figure}
    \includegraphics[width=7.5cm]{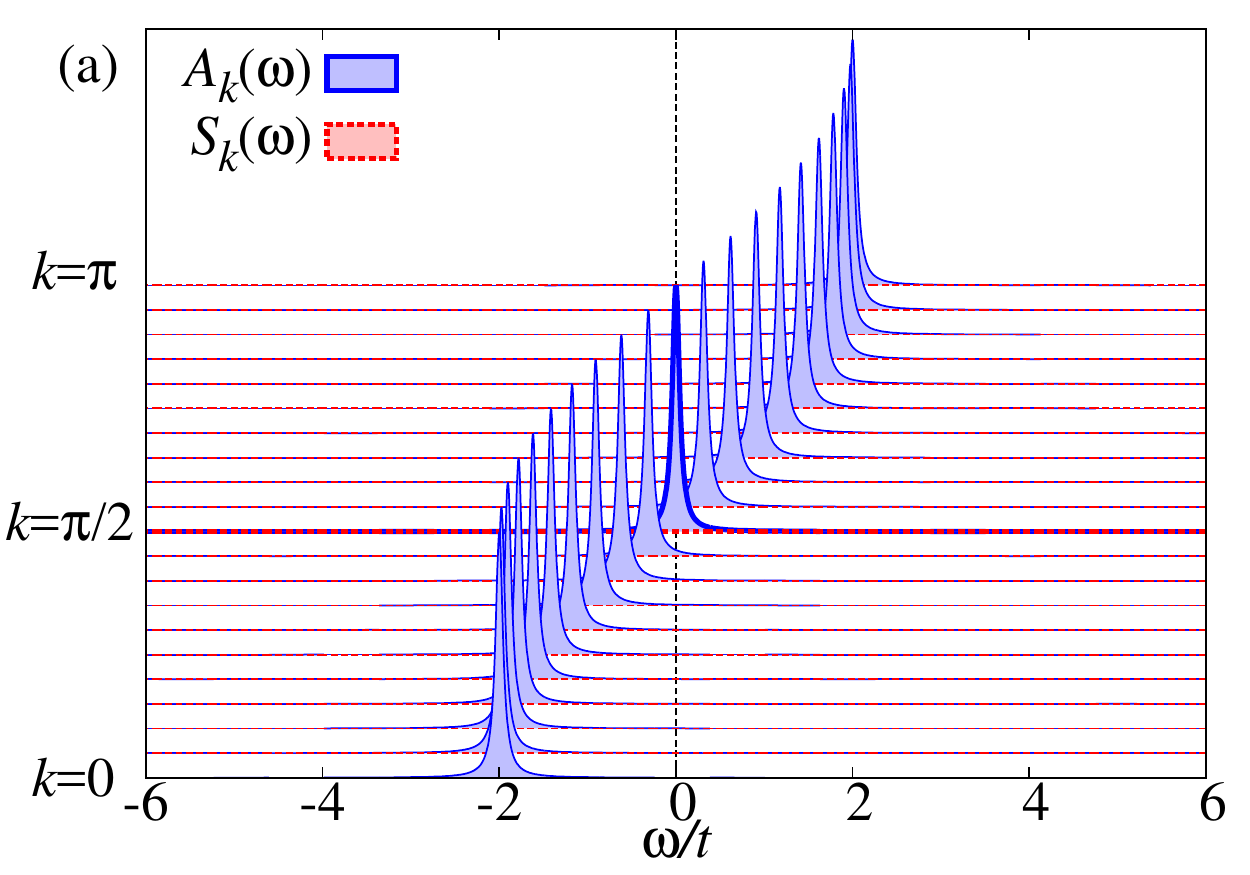}\\
    \includegraphics[width=7.5cm]{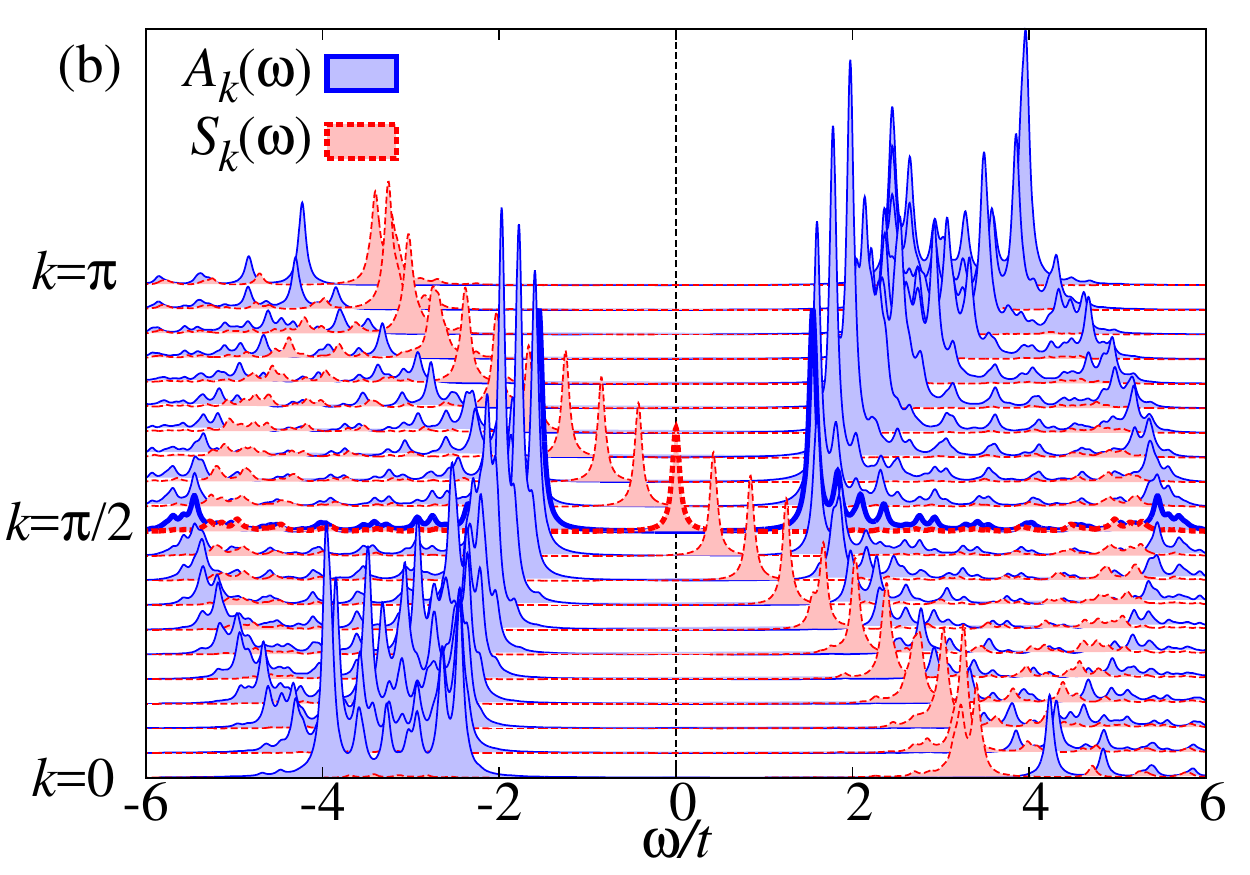}\\
    \includegraphics[width=7.5cm]{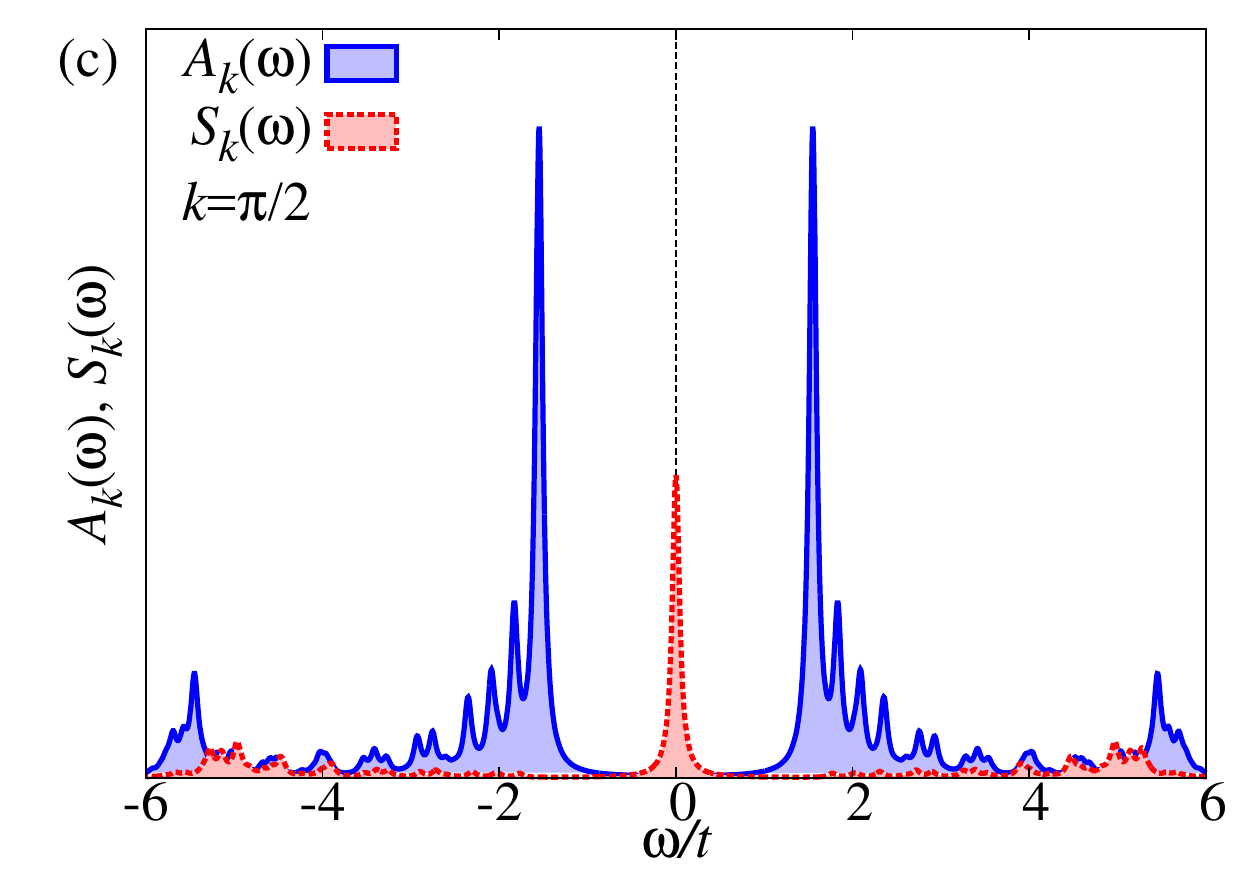}
    \caption{
      $A_{\mb{k}}(\w) = -{\rm Im} G_{\mb{k}}      (\w + \imag \delta^{+})/\pi$ (blue solid lines) and
      $S_{\mb{k}}(\w) = -{\rm Im} \Sigma_{\mb{k}} (\w + \imag \delta^{+})/\pi$ (red dashed lines) 
      of the one-dimensional single-band Hubbard model at half-filling for 
      (a) $U=\mu=0$, 
      (b) $U=2\mu=6t$, and 
      (c) same as (b) but only for $k=\pi/2$. 
      $\delta^+/t = 0.05$ is set for all calculations. 
      Note that different figures use different intensity scales. 
      For visibility, $S_{\mb{k}}(\w)$ is divided by $(U/t)^2$ in (b) and (c). 
      The results for $k=\pi/2$ in  
      (a) and (b) are indicated by thick lines.
    }
    \label{1dCPT}
  \end{figure}
\end{center}

As an example of type II for the generalized Luttinger theorem with $n_D(\Gamma_0)\ne0$　
in Eq.~(\ref{genLuttinger}), let us consider 
a system where a metal-insulator transition is induced by introducing fermion interactions. 
In the noninteracting limit, there should exist zero-energy poles in $\det \bs{G}_0$ at the Fermi energy since the 
system is metallic. 
However, once the interactions are introduced and the metal-insulator transition occurs, 
these zero-energy poles are moved away from the chemical potential 
and replaced with the zeros of ${\det \bs{G}}(0)$ 
due to the appearance of poles in the self energy (for example, see Fig.~\ref{1dCPT}). 
This immediately implies that $n_D(\Gamma_0) \not = 0$ because 
$n_{{\det \bs{G}^{-1}_0}}(\Gamma_0) > n_{{\det \bs{G}^{-1}}} (\Gamma_0)$, 
and thus the case where the metal-insulator transition is induced by introducing 
fermion interactions should in general 
corresponds to type II for the generalized Luttinger theorem when the theorem is valid.

To demonstrate this, here we calculate the single-particle Green's function 
of the one-dimensional single-band Hubbard model at half-filling by using the CPT~\cite{Senechal2000, Senechal2012}. 
The Hamiltonian is described as 
\begin{eqnarray}
  \label{Hubbard}
  H = -t \sum_{\bra i,j \ket, \s}
  \left( 
  c_{i \s}^\dag c_{j \s} + {\rm H.c.} 
  \right)  \\
  + U \sum_{i} n_{i \up} n_{i \dn}
  - \mu \sum_{i, \s} n_{i \s},
\end{eqnarray}
where 
the sum in the first term of the right-hand side, indicated by $\langle i, j\rangle$, runs over 
all pairs of nearest neighbor sites $i$ and $j$ with the hopping integral $-t$. 
The on-site interaction interaction (chemical potential) is represented by 
$U$ ($\mu$) and $n_{i \s}= c_{i \s}^\dag c_{i \s}$. 
We set $\mu = U/2$ at half-filling for which the particle-hole symmetry is preserved.

The CPT allows to approximately evaluate the single-particle Green's function $G_{\mb{k}}(z)$ 
at any momentum ${\mb k}$ 
with arbitrary fine resolution from the numerically exact single-particle Green's function of a small 
cluster~\cite{Senechal2000,Senechal2012}. 
In the CPT, the infinitely large cluster on which $H$ is defined is divided into a set of identical clusters, 
each of which is described by the cluster Hamiltonian $H_{\rm c}$, the same Hamiltonian $H$ 
in Eq.~(\ref{Hubbard}) but with open-boundary conditions, and  
the single-particle Green's function for $H$ is approximated as 
\begin{equation}
  G_{\mb{k}}(z) = \frac{1}{L_{\rm c}} \sum_{i,j} 
  \e^{-\imag \mb{k} \cdot {(\mb{r}_i - \mb{r}_j)}} 
  \left[{\bs{G}'}^{-1}_{\s}(z)-\bs{V}_{\mb{k}} \right]^{-1}_{ij},
\end{equation}
where $L_{\rm c}$ is the number of sites in the cluster 
and $\mb{r}_i$ denotes the spatial location of site $i\, (=1,2,\cdots, L_{\rm c})$ 
in the cluster. ${\bs{G}}'_\s(z)$ 
is the exact single-particle Green's function of the cluster, i.e., 
\begin{eqnarray}
  \left[{\bs{G}}'_\s(z)\right]_{ij} 
  &=& \langle 0| c_{i\s}\left[z-H_{\rm c}+E_0^{\rm c}\right]^{-1}c_{j\s}^\dag|0\rangle  \nonumber \\
  &+& \langle 0| c^\dag_{j\s}\left[z+H_{\rm c}-E_0^{\rm c}\right]^{-1}c_{i\s}|0\rangle. 
\end{eqnarray}
where 
$|0\rangle$ is the ground state of $H_{\rm c}$ with the eigenvalue $E_0^{\rm c}$. 
$\left[\bs{V}_{\mb{k}}\right]_{ij}$ is the matrix element between sites $i$ and $j$ for the inter-cluster hopping term 
represented in momentum space.

We evaluate ${\bs{G}'}_\sigma(z)$ for a one-dimensional 12-site cluster, i.e., $L_{\rm c}=12$, 
with the Lanczos exact-diagonalization method~\cite{Dagotto1994,Weisse}. 
Note that the single-particle Green's function $G_{\mb{k}}(z)$ obtained by the CPT 
can be represented in the Lehmann representation~\cite{Aichhorn2006,Senechal2012} and 
satisfies the spectral-weight sum rule in Eq.~(\ref{sumrule}) with positive-definite spectral 
weight. Therefore, the CPT is an appropriate method to demonstrate 
the formalism derived in Sec.~\ref{sec:GLT}.

To examine the analytical properties of the single-particle Green's function $G_{\mb{k}}(z)$,  
here we calculate the single-particle spectral function $A_{\mb{k}}(\w)$ defined in Eq.~(\ref{eq:akw}) and 
the imaginary part of the self-energy 
\begin{equation}
  S_{\mb{k}}(\w) = -\frac{1}{\pi} {\rm Im} \Sigma_{\mb{k}}(\w+\imag \delta^+) 
\end{equation}
where $\Sigma_{\mb{k}}(z) = z - 2t \cos{k} - {G^{-1}_{\mb{k}}}(z)$. 
Since the singularities (i.e., poles and zeros) of 
the single-particle Green's function $G_{\mb{k}}(z)$ in complex $z$ plane occur only in the real frequency  
$\omega$ axis, 
these two quantities $A_{\mb{k}}(\w)$ and $S_{\mb{k}}(\w)$ can capture the structure of poles and zeros of 
$G_{\mb{k}}(z)$: a divergence of $A_{\mb{k}}(\w)$ [$S_{\mb{k}}(\w)$] corresponds to a pole (zero) of 
$G_{\mb{k}}(\w+\delta^+)$ in the limit of $\delta^+\to0$. 
The CPT has been employed to study intensively the single-particle excitation spectra $A_{\mb{k}}(\w)$ of 
the one-dimensional Hubbard model~\cite{Senechal2000}, and therefore 
we shall focus only on the analytical properties of $G_{\mb{k}}(\w)$ in the following.

The results of $A_{\mb{k}}(\w)$ and $S_{\mb{k}}(\w)$ for $U=0$ and $U/t=6$ 
at zero temperature are shown in Figs.~\ref{1dCPT}(a) and \ref{1dCPT}(b), respectively. 
Here, we set that $\delta^+/t=0.05$ and thus the diverging behavior of $A_{\mb{k}}(\w)$ and $S_{\mb{k}}(\w)$ is 
replaced by sharp peak structures. The value of $U/t=6$ is 
chosen merely for better visibility of the spectra, although the one-dimensional Hubbard model at half-filling 
is insulating for any $U\, (>0)$~\cite{Lieb1968}, which can be correctly reproduced 
by the CPT~\cite{Senechal2012}.

\begin{center}
  \begin{figure}
    \includegraphics[width=7.5cm]{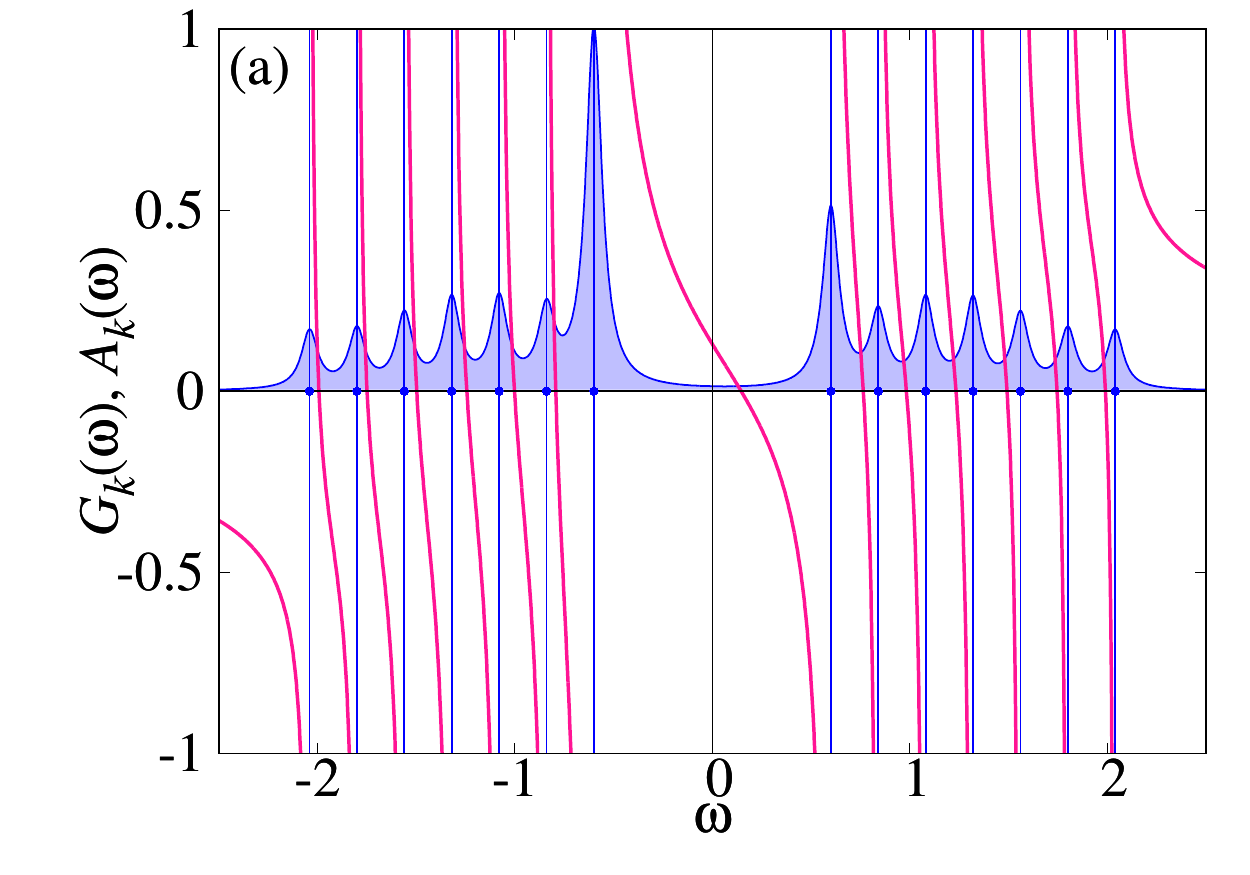}
    \includegraphics[width=7.5cm]{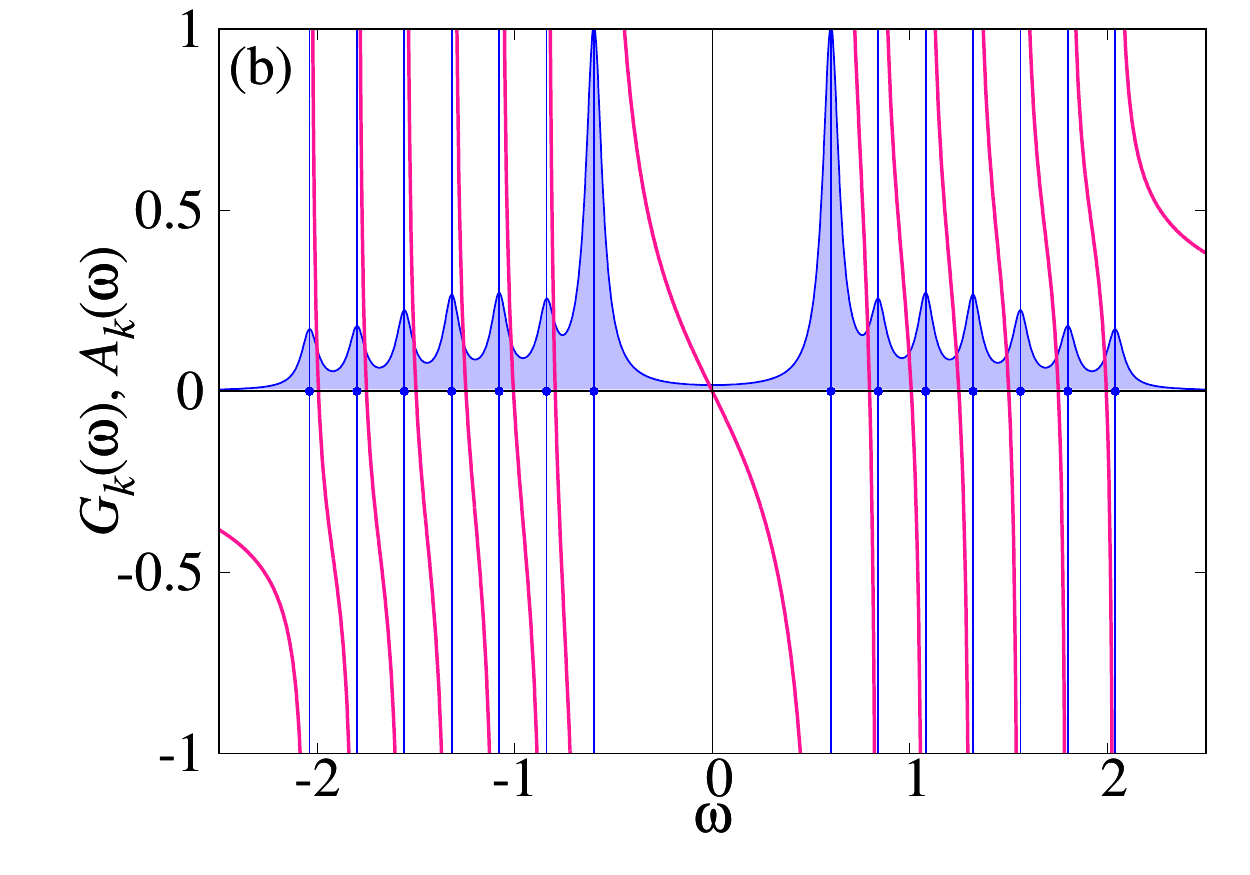}
    \includegraphics[width=7.5cm]{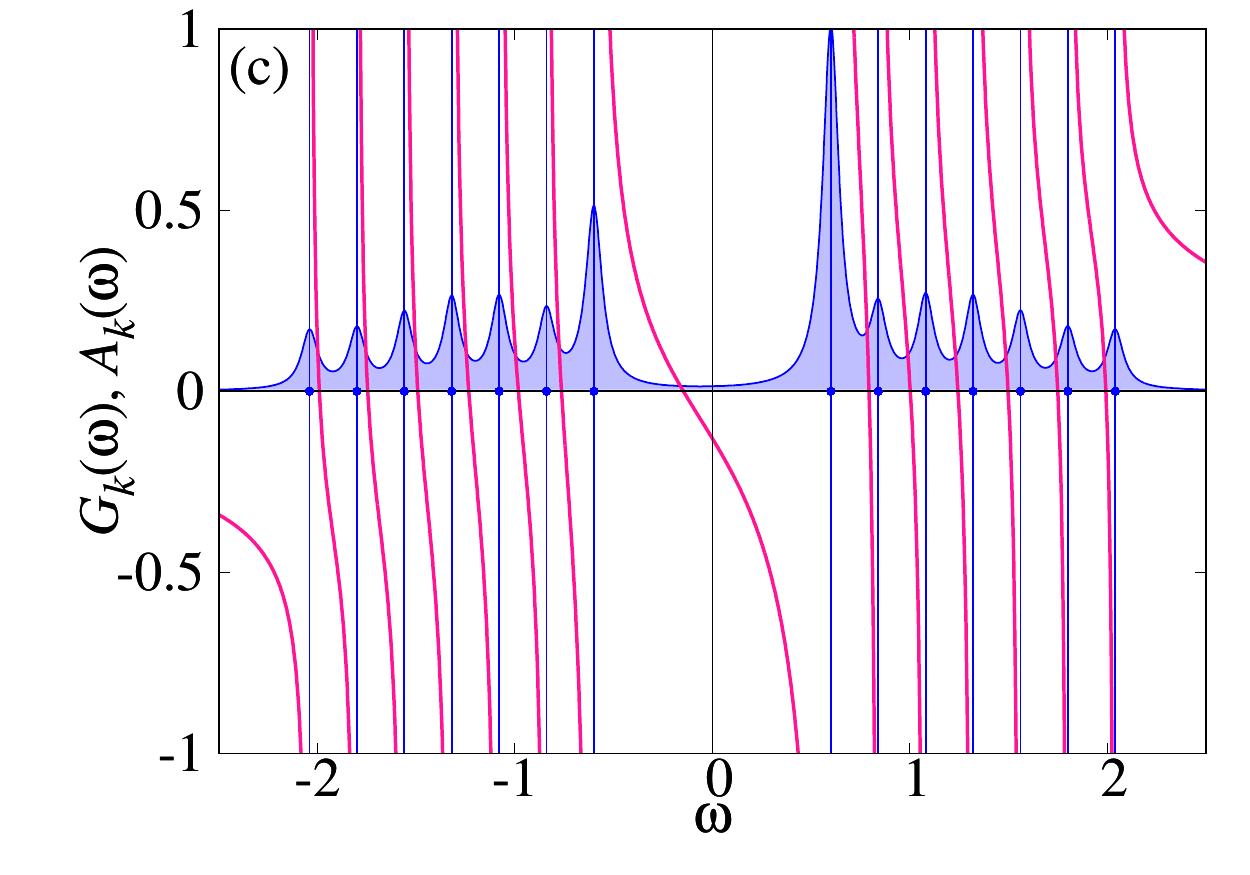}
    \caption{ 
      Schematic figures of 
      $A_{\mb{k}} (\w) = -{\rm Im} G_{\mb{k}} (\w+i\delta^+)/\pi$ (blue shaded region) and 
      $G_{\mb{k}}(\w)$ (red solid lines) with momentum $\mb{k}$ 
      (a) inside, (b) on, and (c) outside 
      the Luttinger surface for a Mott insulator. 
      Here $\omega=0$ corresponds to the chemical potential in the zero-temperature limit. 
      The poles of $G_{\mb{k}}(\w)$ are indicated by dots along with dashed vertical lines. 
    }
    \label{fig:AkwLS}
  \end{figure}
\end{center}

For the noninteracting case, the Fermi points locate at $k = \pm\pi/2$ 
and the self-energy is zero by definition [see Fig.~\ref{1dCPT}(a)]. 
On the other hand, for $U/t=6$, $A_{\mb{k}}(\w)$ exhibits the single-particle excitation gap, 
as shown in Fig.~\ref{1dCPT}(b). 
More interestingly, we find in Figs.~\ref{1dCPT}(b) and \ref{1dCPT}(c) that a peak of $S_{\mb{k}}(\w)$ intersects the zero energy, i.e., 
$\w=0$, exactly at $k = \pm\pi/2$. 
Since the peak of $S_{\mb{k}}(\w) $ corresponds to the zero of $G_{\mb{k}}(\w)$, 
the result indicates that $G_{k=\pm\pi/2}(0) = 0$. 
The momenta $k=\pi/2$ and $-\pi/2$ thereby form the Luttinger surface~\cite{Dzyaloshinskii2003}, 
which is defined as a set of momenta $\mb{k}$ such that $G_{\mb{k}}(0)=0$. 
A typical behavior of $A_{\mb k}(\w)$ and $G_{\mb k}(\w)$ on the Luttinger surface 
is schematically shown in Fig.~\ref{fig:AkwLS}(b).

Because of $G_{\mb{k}}(0) = 0$ on the Luttinger surface by definition, 
the order of singularities in $G_{\mb{k}}(\w)$ for momentum ${\mb k}$ 
on the Luttinger surface must be 
\begin{equation}
  \w_1^{(\mb{k})} < \zeta_1^{(\mb{k})} < \cdots 
  < \w_{m_{\rm LS}}^{(\mb{k})} < \zeta_{m_{\rm LS}}^{(\mb{k})} = 0 
  < \w_{m_{\rm LS}+1}^{(\mb{k})} < \cdots, 
  \label{wls}
\end{equation}
where $\zeta_{m_{\rm LS}}^{(\mb{k})}$ is the $m_{\rm LS}$-th zero of $G_{\mb{k}}(\w)$, i.e., 
$G_{\mb{k}}(\w=\zeta_{m_{\rm LS}}^{(\mb{k})})=0$, and exactly zero~\cite{note7}. Therefore, 
the number of poles $\{\w^{(\mb k)}_1, \w^{(\mb k)}_2,\dots\}$ below (above) the chemical potential is larger 
than the number of zeros $\{\zeta^{(\mb k)}_1, \zeta^{(\mb k)}_2,\dots\}$ below (above) the chemical potential by one. 
From Eqs.~(\ref{eq:n_G<})--(\ref{eq:n_G>}), we can now easily find that 
$n_{G^{-1}_{\mb{k}}}(\Gamma_<) =  n_{G^{-1}_{\mb{k}}}(\Gamma_>) =  1 $ and 
$n_{G^{-1}_{\mb{k}}}(\Gamma_0) = -1 $, i.e., $n_{D_{\mb k}}(\Gamma_<)=1$ and 
$n_{D_{\mb k}}(\Gamma_0)=-2$, where $n_{D_{\mb k}}(\cal C)$ is given in Eq.~(\ref{eq:n_Dk_d}). 
Thus, this momentum contributes to the Luttinger volume $V_{\rm L}$ by one, 
including the spin degree of freedom [see Eq.~(\ref{eq:vl_k})].

We also find in Fig.~\ref{1dCPT}(b) that  
the topmost (bottommost) singularity below (above) the chemical potential for $|k| < \pi/2$ is a pole of 
$A_{\mb{k}}(\w)$ [$S_{\mb{k}}(\w)$], implying that 
\begin{equation}
  \w_1^{(\mb{k})} < \zeta_1^{(\mb{k})} < \cdots 
  < \w_{m_{\rm LS}'}^{(\mb{k})} < 0 < \zeta_{m_{\rm LS}'}^{(\mb{k})} 
  < \w_{m_{\rm LS}'+1}^{(\mb{k})} < \cdots  
  \label{wls<}
  \end{equation}
and thus $G_{\mb k}(0)>0$ [see also Fig.~\ref{fig:AkwLS}(a)]. 
From Eqs.~(\ref{eq:n_G<})--(\ref{eq:n_G>}), we find that $n_{G^{-1}_{\mb{k}}}(\Gamma_<) = 1 $ 
and $n_{G^{-1}_{\mb{k}}}(\Gamma_0)=  n_{G^{-1}_{\mb{k}}}(\Gamma_>)  = 0 $, 
which contributes two to the Luttinger volume $V_{\rm L}$ in the 
zero-temperature limit, including the spin degree of freedom. 
It is also apparent that $n_{D_{\mb k}}(\Gamma_<)= n_{D_{\mb k}}(\Gamma_0)=0$ 
for momentum below the Luttinger surface. 

On the other hand, as shown in Fig.~\ref{1dCPT}(b), the topmost (bottommost) singularity 
below (above) the chemical potential for $|k| > \pi/2$ is a pole of 
$S_{\mb{k}}(\w)$ [$A_{\mb{k}}(\w)$]. This implies that 
\begin{equation}
  \w_1^{(\mb{k})} < \zeta_1^{(\mb{k})} < \cdots 
  < \w_{m_{\rm LS}''}^{(\mb{k})} < \zeta_{m_{\rm LS}''}^{(\mb{k})} <0
  < \w_{m_{\rm LS}''+1}^{(\mb{k})} < \cdots  
  \label{wls>}
\end{equation}
and thus $G_{\mb k}(0)<0$ [see also Fig.~\ref{fig:AkwLS}(c)]. 
From Eqs.~(\ref{eq:n_G<})--(\ref{eq:n_G>}), we find that 
$n_{G^{-1}_{\mb{k}}}(\Gamma_<) = n_{G^{-1}_{\mb{k}}}(\Gamma_0) = 0 $ and $ n_{G^{-1}_{\mb{k}}}(\Gamma_>) = 1$, 
which contributes zero to the Luttinger volume $V_{\rm L}$ in the zero-temperature limit. 
These analytical properties of $G_{\mb k}(z)$ are summarized in Table~\ref{tableLS}.

\begin{table}
  \caption{
    \label{tableLS}
    Analytical properties of the single-particle Green's function $G_{\mb{k}}(\w)$ for a Mott insulator. 
    LS stands for Luttinger surface. $n_{\mb{k}}^{(0)}$ is defined in Eq.~(\ref{QPTzero}).   
  }
  \begin{tabular}{c|ccc}
    \hline \hline
    location of ${\mb{k}}$ & inside LS & on LS & outside LS \\
    \hline 
    position of singularities  & Eq.~(\ref{wls<}) & Eq.~(\ref{wls})& Eq.~(\ref{wls>}) \\
    \hline 
    sign of $G_{\mb{k}}(0)$ & $G_{\mb{k}}(0) > 0$ & $G_{\mb{k}}(0)=0$ & $G_{\mb{k}}(0) < 0$ \\
    \hline
    $n_{G_{\mb{k}}^{-1}}(\Gamma_<)$ &  1 & 1 & 0 \\
    $n_{G_{\mb{k}}^{-1}}(\Gamma_0)$ &  0 & -1 & 0 \\
    $n_{G_{\mb{k}}^{-1}}(\Gamma_>)$ &  0 & 1 & 1 \\
    \hline
    $n_{\mb{k}}^{(0)}$ & 1 & 1/2 & 0 \\
    \hline \hline
  \end{tabular}
\end{table}

Counting the momentum volume surrounded by the Luttinger surface in Fig.~\ref{1dCPT}, 
we can find that the Luttinger volume $V_{\rm L}$ is exactly $N$, the number of total electrons, therefore 
satisfying the generalized Luttinger theorem. 
Indeed, as discussed above, we also find from zeros and poles of the single-particle Green's function 
that the type II condition is fulfilled, i.e., 
$n_D(\Gamma_0)=-2n_D(\Gamma_<)\ne0$, where $n_D({\cal C})$ is given in Eq.~(\ref{eq:n_Dk}). 
Since the system studied here is particle-hole symmetric, these results simply demonstrate 
the general statement in Sec.~\ref{sec:VGLT}.

Finally, it is instructive to directly count how many times $G_{\mb k}^{-1}(z)$ and $D_{\mb k}(z)$ wind around 
the origin in the complex $G_{\mb k}^{-1}$ and $D_{\mb k}$ planes, respectively, when $z$ moves along contour 
$\mcal C$ shown in Fig.~\ref{contour0} (b). 
Recall that $n_{G_{\mb k}^{-1}}(\mcal C)$ and $n_{D_{\mb k}}(\mcal C)$ can be evaluated either by 
counting the number of zeros and poles of $G_{\mb k}(z)$, as shown above, 
or by directly counting the winding 
numbers of $G_{\mb k}^{-1}(z)$ and $D_{\mb k}(z)$ [see Eq.~(\ref{eq:nGk}) and (\ref{eq:nD_k})]. 
For this purpose, it should be noted that as long as the poles and zeros are properly included, contour $\mcal C$ 
($=\Gamma_0$, $\Gamma_<$, and $\Gamma_>$) 
can be chosen rather freely. Therefore, as shown in Fig.~\ref{contour2}, 
we consider the following contours 
\begin{eqnarray}
  z= \left\{
  \begin{array}{ll}
    r_0 \e^{\imag \phi}  & {\rm for}\,\, \Gamma_0 \\
    r_< \e^{\imag \phi} -a & {\rm for}\,\, \Gamma_< \\
    r_> \e^{\imag \phi} +a & {\rm for}\,\, \Gamma_>, 
  \end{array}
  \right. 
  \label{eq:contour}
\end{eqnarray}
parametrized by angle $\phi$ ($-\pi < \phi \leq \pi$). 
Here, we set $r_0 = t$, $r_< = r_> = 2t$, and $a = 3t$.

\begin{center}
  \begin{figure}
    \tikzstyle{branchpoint}=[black, line width=1.0pt]
    \def\RC{0.9} 
    \def\RL{1.2}
    \def\AA{2.35}
    
    \def\angle{55} 
    \begin{tikzpicture}
      \draw [thick, -stealth](-3.75, 0.0)--(3.75,0.0) node [anchor=west]{\Large ${\rm Re}z$};      
      \draw [thick, -stealth]( 0.0,-3.0)--(0.0,3.0) node [anchor=south] {\Large ${\rm Im}z$};
      \foreach \x in {-1,0,1}{
        \filldraw[draw=black,fill=black] (\AA*\x, 0.0) circle (0.055);
      }

      \draw [magenta, thick, -latex] (-\AA,0) + (  0:\RL) arc (  0:135:\RL);
      \draw [magenta, thick]         (-\AA,0) + (134:\RL) arc (134:360:\RL);
      \node [magenta, anchor=south east] at (-\AA-0.5*\RL,0.866*\RL) {\Large $\Gamma_<$};
      \draw[black] (-\AA,0) -- +(\angle:\RL);
      \draw[black] (-\AA,0) -- +(0.3*\RL,0.0) arc (0:\angle:0.3*\RL) -- cycle;
      \node [anchor=west] at (-\AA+0.3*\RL*0.866,0.3*\RL*0.866){\Large $\phi$};
      \node [anchor=west] at (-\AA-0.1*\RL,0.75*\RL){\Large $r_<$};
      \node [anchor=north]      at (-\AA,0){\Large $-a$};

      \draw [magenta, thick, -latex] (\AA,0) + (  0:\RL) arc (  0:135:\RL);
      \draw [magenta, thick]         (\AA,0) + (134:\RL) arc (134:360:\RL);
      \node [magenta, anchor=south east] at (\AA-0.5*\RL,0.866*\RL) {\Large $\Gamma_>$};
      \draw[black] (\AA,0) -- +(\angle:\RL);
      \draw[black] (\AA,0) -- +(0.3*\RL,0.0) arc (0:\angle:0.3*\RL) -- cycle;
      \node [anchor=west] at (\AA+0.3*\RL*0.866,0.3*\RL*0.866){\Large $\phi$};
      \node [anchor=west] at (\AA-0.1*\RL,0.75*\RL){\Large $r_>$};
      \node [anchor=north]      at (\AA,0){\Large $a$};

      \draw [magenta, thick, -latex] (0,0) + (  0:\RC) arc (  0:135:\RC);
      \draw [magenta, thick]         (0,0) + (134:\RC) arc (134:360:\RC);
      \node [magenta, anchor=south east] at (0-0.5*\RC,0.866*\RC) {\Large $\Gamma_0$};
      \draw[black] (0,0) -- +(\angle:\RC);
      \draw[black] (0,0) -- +(0.3*\RC,0.0) arc (0:\angle:0.3*\RC) -- cycle;
      \node [anchor=west] at (0+0.3*\RC*0.866,0.3*\RC*0.866){\Large $\phi$};
      \node [anchor=west] at (-0.1*\RC,0.75*\RC){\Large $r_0$};
      \node [anchor=north east] at (0,0){\Large $0$};
    \end{tikzpicture}

    \caption{
      Contours $\Gamma_0$, $\Gamma_<$, and $\Gamma_>$ in the complex $z$ plane, 
      parametrized by $\phi$ ($-\pi<\phi\leq\pi$), 
      which correspond to those in Fig.~\ref{contour0}(b). 
    }
    \label{contour2}
  \end{figure}
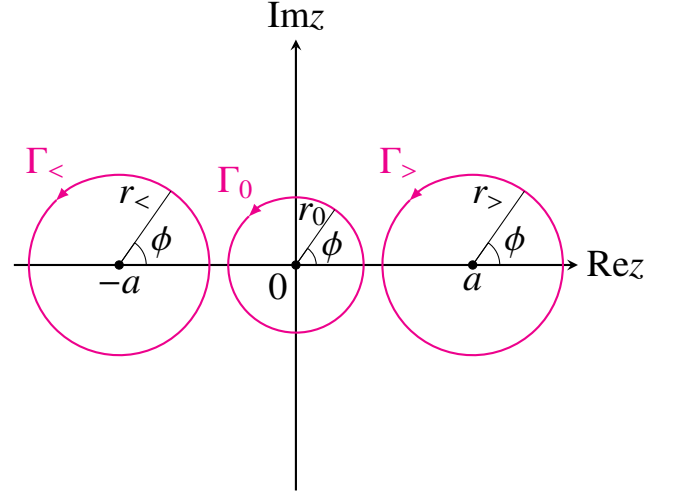
\end{center}

\begin{center}
  \begin{figure*}
    \includegraphics[width=12.5cm]{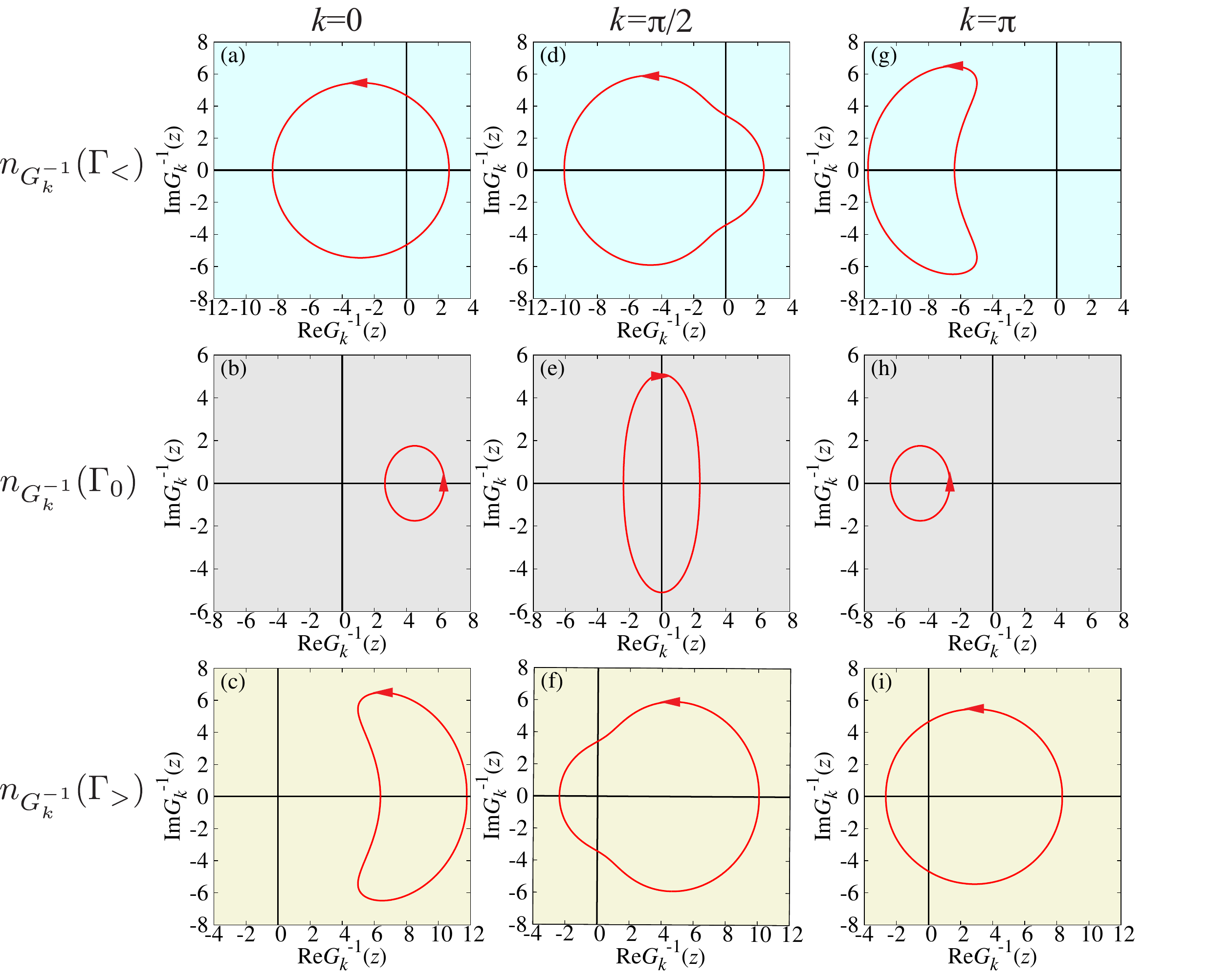}
    \caption{
    Trajectory of $G_{\mb k}^{-1}(z)$ in the complex $G_{\mb k}^{-1}$ plane 
    when $z$ moves along contour $\mcal C$ ($=\Gamma_0$, $\Gamma_<$, and $\Gamma_>$) 
    given in Eq.~(\ref{eq:contour}) and also in Fig.~\ref{contour2} for three representative momenta, i.e., 
    $k=0$ (inside the Luttinger  surface), 
    $k=\pi/2$ (on the Luttinger surface), and 
    $k=\pi$ (outside the Luttinger surface). 
    The CPT is employed for the one-dimensional one-band 
    Hubbard model defined in Eq.~(\ref{Hubbard}) with $U/t=6$ at half-filling. 
    By directly counting how many times and which direction (indicated by arrows) 
    the trajectory winds around the origin in the complex $G_{\mb k}^{-1}$ plane, 
    we find that 
    (a) $n_{G_{k=0}^{-1}}(\Gamma_<) = 1$, 
    (b) $n_{G_{k=0}^{-1}}(\Gamma_0) = 0$,
    (c) $n_{G_{k=0}^{-1}}(\Gamma_>) = 0$,
    (d) $n_{G_{k=\pi/2}^{-1}}(\Gamma_<) = 1$,
    (e) $n_{G_{k=\pi/2}^{-1}}(\Gamma_0) = -1$,
    (f) $n_{G_{k=\pi/2}^{-1}}(\Gamma_>) = 1$,
    (g) $n_{G_{k=\pi}^{-1}}(\Gamma_<) = 0$,
    (h) $n_{G_{k=\pi}^{-1}}(\Gamma_0) = 0$, and
    (i) $n_{G_{k=\pi}^{-1}}(\Gamma_>) = 1$.
    These are exactly the same as those obtained by counting the number of poles and zeros of the 
    single-particle Green's function $G_{\mb k}(z)$ given in Eqs.~(\ref{wls})--(\ref{wls>}) 
    (see also Table~\ref{tableLS}). 
    }
    \label{fig:nq}
  \end{figure*}
\end{center}

Figure~\ref{fig:nq} summarizes the results for the trajectory of $G_{\mb k}^{-1}(z)$ in the 
complex ${G_{\mb k}^{-1}}$ plane at three representative momenta, i.e., $\mb k$ below, on, and above 
the Luttinger surface. 
Directly counting how many times and which direction the trajectory winds around the origin in the complex 
${G_{\mb k}^{-1}}$ plane, we find in Fig.~\ref{fig:nq} 
that i) $n_{G_{\mb k}^{-1}}(\Gamma_<)=1$ and $G_{\mb k}^{-1}(\Gamma_0)=G_{\mb k}^{-1}(\Gamma_>)=0$ for 
$\mb k$ below the Luttinger surface, 
ii) $n_{G_{\mb k}^{-1}}(\Gamma_<)=G_{\mb k}^{-1}(\Gamma_>)=1$ and $G_{\mb k}^{-1}(\Gamma_0)=-1$ for 
$\mb k$ on the Luttinger surface, and 
iii) $n_{G_{\mb k}^{-1}}(\Gamma_<)=G_{\mb k}^{-1}(\Gamma_0)=0$ and $G_{\mb k}^{-1}(\Gamma_>)=1$ for 
$\mb k$ above the Luttinger surface. 
These results are indeed exactly the same as those obtained above in Table~\ref{tableLS} by counting the 
number of zeros and poles of $G_{\mb k}(z)$ given in Eqs.~(\ref{wls})--(\ref{wls>}).

\begin{center}
  \begin{figure*}
    \includegraphics[width=12.5cm]{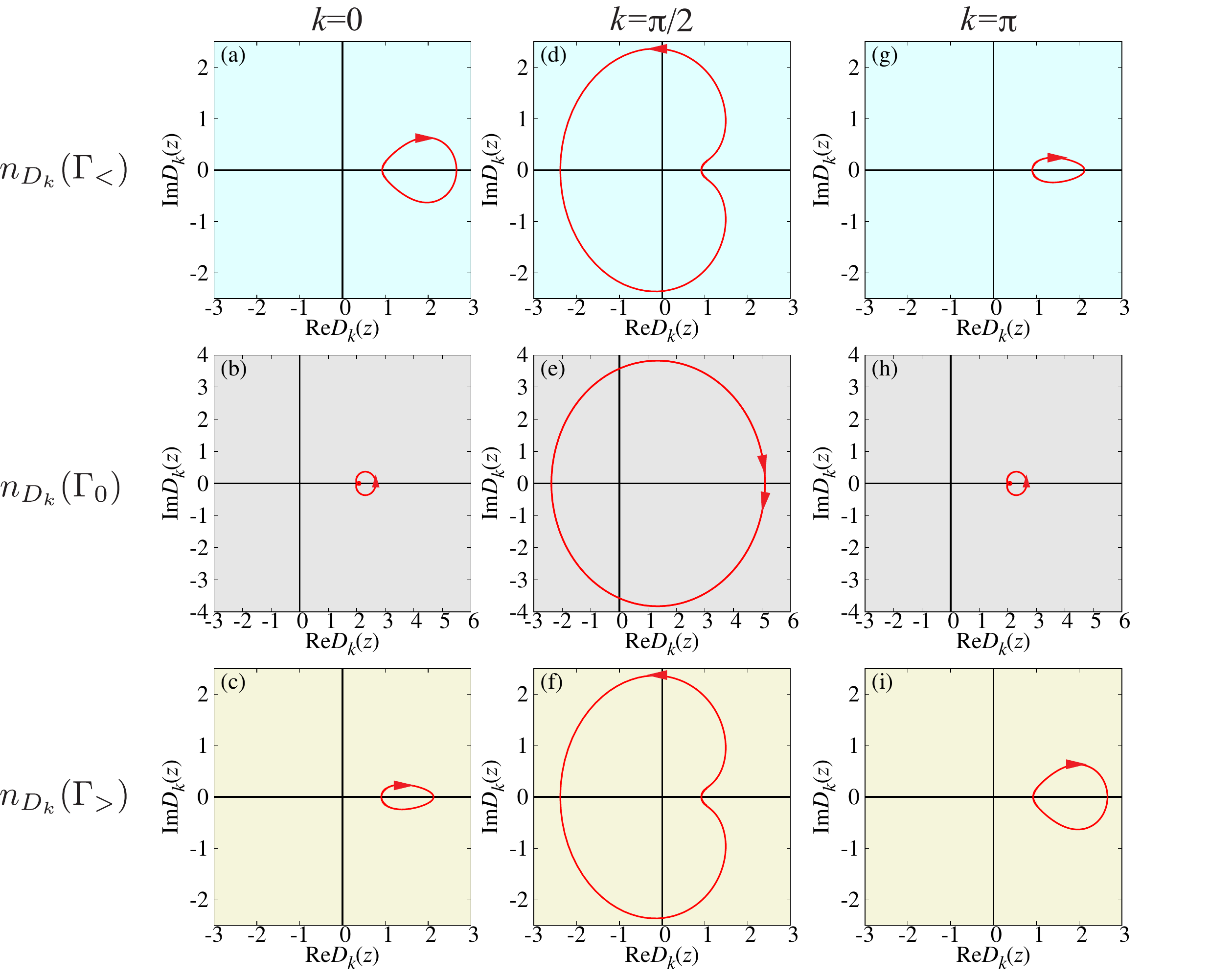}
    \caption{
    Trajectory of $D_{\mb k}(z)=G_{0{\mb k}}(z)/G_{\mb k}(z)$ in the complex $D_{\mb k}$ plane when $z$ 
    moves along contour $\mcal C$ ($=\Gamma_0$, $\Gamma_<$, and $\Gamma_>$) 
    given in Eq.~(\ref{eq:contour}) and also in Fig.~\ref{contour2} for three representative momenta, i.e., 
    $k=0$ (inside the Luttinger surface), 
    $k=\pi/2$ (on the Luttinger surface), and 
    $k=\pi$ (outside the Luttinger surface). 
    The CPT is employed for the one-dimensional one-band 
    Hubbard model defined in Eq.~(\ref{Hubbard}) with $U/t=6$ at half-filling. 
    By directly counting how many times and which direction (indicated by arrows) the trajectory winds around 
    the origin in the complex $D_{\mb k}$ plane, we find that 
    (a) $n_{D_{k=0}}(\Gamma_<) = 0$, 
    (b) $n_{D_{k=0}}(\Gamma_0) = 0$,
    (c) $n_{D_{k=0}}(\Gamma_>) = 0$,
    (d) $n_{D_{k=\pi/2}}(\Gamma_<) = 1$,
    (e) $n_{D_{k=\pi/2}}(\Gamma_0) = -2$,
    (f) $n_{D_{k=\pi/2}}(\Gamma_>) = 1$,
    (g) $n_{D_{k=\pi}}(\Gamma_<) = 0$,
    (h) $n_{D_{k=\pi}}(\Gamma_0) = 0$, and
    (i) $n_{D_{k=\pi}}(\Gamma_>) = 0$. 
    Notice that the trajectory winds twice around the origin in (e). 
    Therefore, we find that $n_D (\Gamma_0) = - 2 n_D (\Gamma_<) = -4$, satisfying the condition 
    of type II for the generalized Luttinger theorem in Eq.~(\ref{genLuttinger}).   
    }
    \label{fig:nD}
  \end{figure*}
\end{center}

Figure~\ref{fig:nD} shows the results for the trajectory of $D_{\mb k}(z)$ in the complex $D_{\mb k}$ plane 
at the three representative momenta. 
Counting how many times and which direction the trajectory winds around the origin 
in the complex $D_{\mb k}$ plane, we find in Fig.~\ref{fig:nD} 
that i) $n_{D_{\mb k}}(\Gamma_<)=n_{D_{\mb k}}(\Gamma_0)=n_{D_{\mb k}}(\Gamma_>)=0$ for $\mb k$ 
below and above the Luttinger surface, and 
ii) $n_{D_{\mb k}}(\Gamma_<)=n_{D_{\mb k}}(\Gamma_>)=1$ and $n_{D_{\mb k}}(\Gamma_0)=-2$ for $\mb k$ 
on the Luttinger surface. 
Therefore, we can show that $n_{D_{\mb k}}(\Gamma_<)+\frac{1}{2}n_{D_{\mb k}}(\Gamma_0)=0$ for 
each momentum ${\mb k}$ and hence the generalized Luttinger theorem is valid since 
$\lim_{T\to0}\Delta V_{\rm L}=0$ [see Eq.~(\ref{eq:vl_k2})], the same conclusion reached 
above by counting the number of poles and zeros of $G_{\mb k}(z)$.

\section{Remarks}\label{sec:remarks}
First, 
the sign of $G_{\mb{k}}(0)$ has been originally utilized 
to quantify interior and exterior of the Fermi or Luttinger surface~\cite{Luttinger1960,Dzyaloshinskii2003}. 
However, as summarized in Tables~\ref{tableG0k}, \ref{tableGk}, and \ref{tableLS},  
$n^{(0)}_{\mb{k}}$ defined in Eq.~(\ref{QPTzero}) can also quantify the location of momentum 
$\mb{k}$ which may be inside, outside, or on the Fermi or Luttinger surface. 
Indeed, as already discussed in Sec.~\ref{sec:typeI} (also see Appendix~\ref{sec:QP}), 
$n^{(0)}_{\mb{k}}$ can be interpreted as 
the quasiparticle distribution function in the Fermi-liquid theory. 
It should be emphasized that $n^{(0)}_{\mb{k}}$ itself has the topological nature since it is 
associated with the winding number of the single-particle Green's function $G_{\mb k}(z)$.

Second, it is interesting to notice that the phase shift discussed in impurity scattering problems 
can be described with the similar form of Eqs.~(\ref{fredholm}) and (\ref{condition}), 
where the scattering potential or $T$-matrix replaces 
the many body self-energy $\bs{\Sigma}(z)$. In the impurity scattering problems, 
the integer winding number $n_D({\mathcal C})$ corresponds to the number of bound states 
(Levinson's theorem)~\cite{Levinson1949,Dewitt1956,Schiff,Umezawa1986} 
or the number of external charges (Friedel sum rule)~\cite{Friedel1952, Langer1961, Doniach}.  
Furthermore, according to the Levinson's theorem, a fractional factor of $1/2$ should be added in the 
phase shift when a bound state exists at zero energy~\cite{Schiff,Umezawa1986}, which is again analogous to the fractional contribution 
to $\Delta V_{\rm L}$ in Eq.~(\ref{D_VL}) when 
the determinant of the single-particle Green's function exhibits 
singularities 
at the chemical potential.

Third, 
the topological aspect of the Luttinger theorem found here clearly differs from 
the topological approach to the Luttinger theorem reported in Ref.~\cite{Oshikawa2000}. 
The main difference is 
twofold: 
(i) the topological nature examined and 
(ii) the resulting topological quantity. 
We have derived here the topological nature of the single-particle Green's function for general systems, 
whereas in Ref.~\cite{Oshikawa2000} the topological nature of the ground-state wave function is studied 
for Fermi liquids. 
We have shown that the winding number $n_D({\mathcal C})$ is the topological quantity which quantifies whether 
the generalized Luttinger theorem is valid. 
In Ref.~\cite{Oshikawa2000}, the difference of the Fermi surface volume and the filling factor of a partially filled band is 
the topological quantity $n$, which is nothing but the number of completely filled bands. Therefore, $n=0$ corresponds to the 
case where no filling band exists and Luttinger theorem is always satisfied regardless of values of $n$. 
Note also that the topological approach in Ref.~\cite{Oshikawa2000} is formulated for 
periodic systems with a particular set of system sizes, while our approach can be applied to general systems.

\section{Summary and discussion}\label{sec:summary} 

Based solely on analytical properties of the single-particle Green's function of fermions at finite temperatures, 
we have shown that the Luttinger volume is represented as the winding number of 
(the determinant of) the single-particle Green's functions. 
Therefore, this inherently introduces the topological interpretation of the generalized Luttinger theorem, 
and naturally leads to two types of conditions (types I and II) for the validity of the generalized Luttinger theorem. 
Type I falls into the Fermi-liquid case originally discussed by Luttinger in the 1960's, 
where the Fermi surface is well defined at momenta where $G_{\mb{k}}(\w=0)$ 
exhibits a pole. Type II includes the non-metallic case such as the Mott insulator to which Dzyaloshinskii 
has extended the Luttinger's argument in the 2000's by introducing the new concept of the Luttinger surface 
defined as a set of momenta where the sign of $G_{\mb{k}}(\w=0)$ changes. 
We have also derived the sufficient condition for the validity of the Luttinger theorem of type I, 
representing the robustness of the theorem against the perturbation. 
We have also shown rigorously that the generalized Luttinger theorem of both types 
is valid for generic interacting fermions as long as the particle-hole symmetry is preserved. 
Moreover, we have shown that the winding number of the single-particle Green's function can be considered as 
the distribution function of quasiparticles.

We should emphasize that 
these general statements can be made by noticing that 
the generalized Luttinger volume is expressed as the winding number of the single-particle Green's 
function at finite temperatures, for which the complex analysis can be exploited readily and successfully 
without any ambiguity. This allows us to explore the intrinsic features of interacting fermions, independently 
of details of a microscopic Hamiltonian.

To be more specific in terms of these general analysis of interacting fermions, 
first we have examined the single-band simple metallic system with 
translational symmetry and discussed how the original statement of the theorem by Luttinger 
is understood with respect to our present analysis. 
We have also demonstrated our general analysis for a Mott insulator 
by examining the one-dimensional single-band Hubbard model at half-filling. 
Furthermore, using the Hubbard-I approximation, we have analyzed the half-filled 
Hubbard model on the honeycomb lattice where no apparent Fermi surface 
exists in the noninteracting limit.

It should be emphasized that the fermionic anti-commutation relation 
$\{c_\alpha^\dag, c_\beta\} = \delta_{\alpha \beta}$ plays a central role 
to determine the analytical properties of the single-particle Green's function, including 
the asymptotic behavior for large $|z|$ and 
the number of zeros and poles of the single-particle Green's function. 
Therefore, our analysis can also be extend to any fermionic systems with spin larger than 1/2. 
However, it is not straightforward to extend the present formalism 
to the $t$-$J$ model, which is a prototypical model of the strongly correlated electron systems studied 
extensively for cuprates~\cite{Dagotto1994}. 
In the $t$-$J$ model, an electron moves between sites via the correlated hopping, represented  
in terms of the projected electron creation and annihilation operators to exclude the double occupancy. 
For example, the correlated hopping of an electron with spin $\s$ from site $j$ to site $i$ 
with hopping amplitude $t$ is expressed as 
\begin{eqnarray}
  t {\bar{c}}_{i \s}^\dag {\bar{c}}_{j \s} 
  &=& t (1-{n}_{i\bar{\s}}) {c}_{i\s}^\dag {c}_{j\s} (1-{n}_{j \bar{\s}}), 
\end{eqnarray}
where ${\bar{c}}^\dag_{i \s} = (1-{n}_{i\bar{\s}}) {c}_{i\s}^\dag$ and 
${\bar{c}}_{j \s} = {c}_{j\s} (1-{n}_{j\bar{\s}})$ 
exclude the double occupancy on each site. Here, $\bar\sigma$ represents the opposite 
spin of $\sigma$. Since the projected electron creation and annihilation operators do not 
satisfy the anti-commutation relation, $\{\bar{c}_\alpha^\dag,  \bar{c}_\beta \}\ne \delta_{\alpha \beta}$, 
we can no longer directly apply the same analytical argument of the single-particle Green's function 
in Sec.~\ref{sec:formulation} and Sec.~\ref{sec:GLT}. 
Nonetheless, it is interesting to note that the violation of the Luttinger theorem in the two-dimensional 
$t$-$J$ model has been reported, based on the high-temperature expansion analysis of the momentum distribution 
function~\cite{Putikka1998} and  
the exact-diagonalization analysis of the single-particle Green's function for finite-size clusters~\cite{Kokalj2007}, 
although the opposite had been concluded in the earlier study~\cite{Stephan1991}.   

\acknowledgments
The authors gratefully acknowledge enlightening discussions with R. Eder  
and constructive comments from T. Shirakawa. 
The numerical computations have been performed with 
the RIKEN supercomputer system (HOKUSAI GreatWave). 
This work has been supported in part 
by RIKEN iTHES Project and Molecular Systems.

\appendix
\section{Another derivation of Eq.~(\ref{identity})}\label{sec:ad}

Here, we derive Eq.~(\ref{identity}) 
from the derivative of the grand potential $\Omega$ with respect to 
the chemical potential $\mu$. As shown in the following, this alternative analysis reveals 
how the Legendre transform of the Luttinger-Ward functional, $\mcal{F}$, 
is related to the deviation $\Delta V_{\rm L}$ of the Luttinger volume from the noninteracting 
one.

In the main text, we set the chemical potential $\mu$ as the origin of $z$ in the single-particle 
Green's function $G_{\alpha\beta}(z)$ [Eq.~(\ref{lehmann})] by including $\mu$ in Hamiltonian 
$H$ (see Sec.~\ref{sec:notation}), and thus $z=0$ in $G_{\alpha\beta}(z)$ corresponding to the chemical potential. 
However, it is more useful to express $\mu$ explicitly in the formulae for the present purpose, and this can be done  
simply by replacing the Matsubara frequency $\imag\w_\nu$ with 
$\mu+\imag\w_\nu$: 
\begin{equation}
  \imag \w_{\nu} \rightarrow z_\nu  =  \mu + \imag \w_\nu. 
\end{equation}
The formulae return to those given in the main text  
by setting $\mu = 0$ after the derivative with respect to $\mu$ is taken.

According to the self-energy-functional theory~\cite{Potthoff2003,Potthoff2006,Potthoff2012}, 
the grand potential $\Omega$ is given as 
\begin{equation}
  \label{SFTgrand}
  \Omega = T \sum_{\nu = -\infty}^{\infty} \ln \det \bs{G}(z_\nu) + \mcal{F},
\end{equation}
where $\mcal{F}[\Sigma]$ is the Legendre transform of Luttinger-Ward functional $\Phi[\bs{G}]$ defined as 
\begin{equation}
  \mcal{F}[\bs{\Sigma}] = \Phi[\bs{G}] - 
  T \sum_{\nu=-\infty}^{\infty} \e^{\imag \w_\nu 0^+}\tr [\bs{G}(\imag \w_\nu) \bs{\Sigma} (\imag \w_\nu)], 
\end{equation}
and all the quantities in Eq.~(\ref{SFTgrand}) are given as 
$\Omega = \Omega[\bs{\Sigma}^*]$, 
$\bs{G}=\bs{G}[\bs{\Sigma}^*]$, and 
$\mcal{F}= \mcal{F}[\bs{\Sigma}^*]$ with the self-energy $\bs{\Sigma}^*$ which satisfies 
the stationary condition $\delta \Omega/\delta \bs{\Sigma}|_{\bs{\Sigma}=\bs{\Sigma}^*} = \bs{0}$.
Note that, because $\Phi [\bs{G}]$ is the generating function of $\bs{\Sigma}$~\cite{Luttinger1960,Luttinger-Ward1960}, 
$\cal{F}[\bs{\Sigma}]$ is the generating function of $\bs{G}$, 
\begin{equation}
  \frac{\delta \mcal{F}[\bs{\Sigma}]}{\delta {\Sigma}_{\alpha \beta}(z)} 
  = - T G_{\beta \alpha}(z).
\end{equation}
Thus we can show that 
\begin{eqnarray}
  \frac{\delta  \mcal{F} [\bs{\Sigma}]  }{\delta z}
  = \sum_{\alpha, \, \beta=1}^{L_{\rm s}} 
  \frac{\delta \mcal{F} [\bs{\Sigma}]  }{\delta \Sigma_{\alpha \beta} }
  \frac{\delta \Sigma_{\alpha \beta}     }{\delta z} 
  = - T \tr \left[\bs{G}(z) \frac{\partial  \bs{\Sigma}(z)}{\partial z } \right].
  \label{fw}
\end{eqnarray}

Applying $-\partial/\partial \mu$ in both sides of Eq.~(\ref{SFTgrand}), we immediately find 
that the left-hand side gives the average particle number $N$, i.e., 
\begin{equation}
  -\frac{\partial \Omega}{\partial \mu} = N, 
\end{equation}
whereas the first term of the right-hand side in Eq.~(\ref{SFTgrand}) reads  
\begin{eqnarray}
  &-& T \frac{\partial}{\partial \mu} \sum_{\nu = -\infty}^{\infty} \ln \det \bs{G}(z_\nu) \notag \\
  &=& T \sum_{\nu = -\infty}^{\infty} \frac{\partial z_\nu}{\partial \mu} \frac{\partial}{\partial z_\nu} \ln \det \bs{G}(z_\nu)^{-1} \notag \\
  &=& \oint_{\Gamma} \frac{\dd z}{2 \pi \imag} n_{\rm F}(z - \mu) \frac{\partial \ln \det \bs{G}(z)^{-1}}{\partial z} = V_{\rm L}. 
\end{eqnarray}
Here, contour $\Gamma$ is indicated in Fig.~\ref{contour0}(a) 
with trivial modification due to non zero $\mu$
and Eq.~(\ref{eq.VL}) is used in the last equality.

Because the Luttinger volume of the noninteracting system $V_{\rm L}^0$ is $N$, 
we find form Eq.~(\ref{SFTgrand}) that the deviation of the Luttinger volume from the noninteracting one, 
$\Delta V_{\rm L}$,  
is given as the derivative of the Legendre transform of the Luttinger-Ward functional, $\mcal{F}$,
with respect to $\mu$, i.e., 
\begin{equation}\label{f_mu}
  \frac{\partial{\mcal{F}}}{\partial \mu} = V_{\rm L} - N = \Delta V_{\rm L}. 
\end{equation}
Finally, $\partial \mcal{F}/\partial \mu$ can also be directly evaluated as 
\begin{eqnarray}
  \frac{\partial{\mcal{F}}}{\partial \mu} 
  &=& \sum_{\nu = -\infty}^{\infty} \frac{\delta \mcal{F}}{\delta z_\nu} 
  = -T \sum_{\nu = -\infty}^{\infty} {\rm tr} 
  \left[\bs{G}(z) \frac{\partial \bs{\Sigma} (z)}{\partial z} \right]_{z=z_\nu}  \nonumber \\
  &=& -\oint_{\Gamma} \frac{\dd z}{2 \pi \imag} n_{\rm F}(z - \mu) \, {\rm tr} \left[\bs{G}(z) 
    \frac{\partial \bs{\Sigma} (z)}{\partial z} \right]     
\end{eqnarray}
where Eq.~(\ref{fw}) is used in the second equality. 
Therefore, the first equality of Eq.~(\ref{f_mu}) is nothing but Eq.~(\ref{identity}) and 
thus we have proved that Eq.~(\ref{identity}) 
can also be derived by the derivative of $\Omega$ with respect to $\mu$.

\section{Hubbard model on the honeycomb lattice: Hubbard-I approximation}\label{sec:hubbard-I}

The half-filled Hubbard model on the honeycomb lattice~\cite{Sorella1992, Meng2010, Sorella2012, Otsuka2016}
is a very instructive and yet non-trivial system to apply the analytical results in 
Sec.~\ref{sec:GLT} 
because only the Fermi points exist in the two-dimensional Brillouin zone 
and hence the concept of ``Fermi surface volume'' is 
absent in the noninteracting limit.

\subsection{Hubbard model on the honeycomb lattice}\label{sub:HBI} 

The Hubbard model on the honeycomb lattice is described by the following Hamiltonian: 
\begin{equation}\label{eq:hb_h}
H_h = H_h^0 + U\sum_i\sum_{\xi=A,B} n_{i\xi\up}n_{i\xi\dn},
\end{equation}
where $H_h^0$ is the noninteracting tight-banding Hamiltonian on the honeycomb lattice, 
\begin{equation}
  H_h^0 = \sum_{\mb{k} \s} 
  \left(c_{\mb{k} A\s}^\dag, c_{\mb{k} B\s}^\dag \right) 
  \left(
    \begin{array}{cc}
      -\mu & \gamma_{\mb{k}} \\
      \gamma_{\mb{k}}^* & -\mu 
    \end{array}
  \right)
  \left(
    \begin{array}{c}
      c_{\mb{k} A\s} \\ 
      c_{\mb{k} B\s}
    \end{array}
  \right).
  \label{Hhk}
\end{equation}
Here, $c_{\mb{k} \xi \s }^\dag = \frac{1}{\sqrt{L}} \sum_{i} c_{i \xi \s }^\dag \e^{-{\imag}  \mb{k} \cdot \mb{r}_i}$ 
is the Fourier transform of an electron creation operator
$c_{i \xi \s }^\dag$ at the $i$-th unit cell, the location being denoted as ${\mb r}_i$ in real space, 
on sublattice $\xi\,(=A,B)$ with spin $\s\,(=\up,\dn)$, and  
$\gamma_{\mb{k}} = -t ( 1+\e^{\imag \mb{k} \cdot \mb{a}_1} + \e^{\imag \mb{k} \cdot \mb{a}_2})$, 
where the hopping between the nearest neighbor sites is denoted as $-t$ and the primitive 
translational vectors are given as $\mb{a}_1 = (1/2, \sqrt{3}/2)$ 
and $\mb{a}_2 = (-1/2, \sqrt{3}/2)$, assuming that the lattice constant between 
the nearest neighbor sites is $1/\sqrt{3}$. 
The number of unit cells is $L$ and the 
chemical potential $\mu$ is explicitly included in $H_h^0$. 
$U$ is the on-site interaction and $n_{i\xi\s}=c^\dag_{i\xi\s}c_{i\xi\s}$. 
Note that the particle-hole symmetry is preserved when $\mu = U/2$ at half-filling. 
The number $L_{\rm s}$ of the single-particle states labeled by $\alpha = (\mb{k}, \s, \xi)$ 
(see Sec.~\ref{sec:notation}) is  
\begin{equation}
  L_{\rm s} = \sum_{\mb{k}} \sum_{\s= \up, \dn} \sum_{\xi = A,B} = 4L.
\end{equation}
In the following of this Appendix, we only consider zero temperature.

\subsection{Noninteracting limit}

Let us first consider the noninteracting limit with $U=0$. 
As shown in Eq.~(\ref{Hhk}), the noninteracting Hamiltonian $H_h^0$ is already diagonal with respect to 
momentum $\mb{k}$ and spin $\s$. Accordingly, the single-particle Green's function ${\bs G}_0(z)$ is block 
diagonalized with respect to $\mb k$ and $\s$, and each element is denoted here as ${\bs G}_{0{\mb k}\s}(z)$. 
Since 
\begin{equation}
  \bs{G}_{0\mb{k}\s}(z) = 
  \left(
    \begin{array}{cc}
      z+\mu                & -\gamma_{\mb{k}}       \\
      -\gamma^*_{\mb{k}} &      z+\mu         
    \end{array}
  \right)^{-1}, 
\end{equation}  
we can readily show that  
\begin{equation}
  \det \bs{G}_{0\mb{k}\s}(z) = \left(\frac{1}{z+\mu - |\gamma_{\mb{k}}|} \right)  \left( \frac{1}{z+\mu + |\gamma_{\mb{k}}|} \right)
  \label{eq:det_G0}
\end{equation}
for given $\mb k$ and $\s$. 
Notice that, because the single-particle energy dispersion $\pm|\gamma_{\mb{k}}|$ is $0$ at the $K$ and $K'$ points 
(Dirac points), i.e., $\mb{k} = \frac{4\pi}{3}( \frac{1}{2}, \frac{\sqrt{3}}{2})$ and 
$\frac{4\pi}{3}(-\frac{1}{2}, \frac{\sqrt{3}}{2})$, respectively, 
$\det \bs{G}_{0\mb{k}\s}(z)$ has zero-energy poles at these momenta when $\mu=0$ at half-filling.

The determinant of the single-particle Green's function is now evaluated as 
\begin{eqnarray}\label{eq:detG0}
  &&\det \bs{G}_{0}(z) 
  = \prod_{\bs k} \prod_{\s} \det \bs{G}_{0\mb{k}\s} (z)  \notag \\
  &=& \left(\frac{1}{z+\mu} \right)^8 \prod_{\mb{k} (\not = K,K')}  
    \left( \frac{1}{z+\mu - |\gamma_{\mb{k}}|} \right)^2  \left( \frac{1}{z+\mu + |\gamma_{\mb{k}}|} \right)^2.
\end{eqnarray}  
By counting the singularities as many times as its order, we find that at half-filling ($\mu=0$) 
the number of poles in $\det \bs{G}_{0}(z)$ at the chemical potential, corresponding to $z=0$, is 8
and the number of poles below the chemical potential is 
$2L-4$.
Since $\det \bs{G}_0(z)$ has no zeros, i.e., $\det \bs{G}_0(z) \not = 0$ for any $z$, 
  \begin{eqnarray}
    \left\{
      \begin{array}{l}
        n_{\det{\bs G}_0^{-1}}(\Gamma_0)= 8   \\ 
        n_{\det{\bs G}_0^{-1}}(\Gamma_<)= 2L-4 
      \end{array}
    \right. 
    \label{eq:wind_detG0}
  \end{eqnarray}
at half-filling [see Eqs.~(\ref{eq:n_detG<}) and (\ref{eq:n_detG0})]. 
Therefore, using Eq.~(\ref{eq:vl_zero}), we find that $\lim_{T\to0}V^0_{\rm L}=2L$.   
This is indeed expected for the noninteracting particle-hole symmetric systems since the number of 
electrons is $L_{\rm s}/2 = 2L$. 
However, we should note that this result is perhaps less obvious 
when we consider the Fermi surface volume 
because the Fermi surface here is composed of the Dirac points in the noninteracting limit at half-filling.

\subsection{Hubbard-I approximation}

Let us employ the Hubbard-I approximation~\cite{Hubbard1963} 
to treat the on-site interaction at half-filling. 
In this approximation, the on-site interaction is approximated in the atomic limit with 
the self-energy ${\bs\Sigma}_{{\mb k}\s}(z)$ given as 
\begin{eqnarray}
  &{}& 
  {\bs\Sigma}_{{\mb k}\s}(z) \nonumber \\
  &=&
  \left(
    \begin{array}{cc}
      Un_{A{\bar\sigma}} + \frac{U^2n_{A{\bar\sigma}}(1-n_{A{\bar\sigma}})}{z+\mu-U(1-n_{A{\bar\sigma}})}    & 0     \\
      0 &Un_{B{\bar\sigma}} + \frac{U^2n_{B{\bar\sigma}}(1-n_{B{\bar\sigma}})}{z+\mu-U(1-n_{B{\bar\sigma}})}         
    \end{array}
  \right), 
\end{eqnarray}  
where $n_{\xi\sigma}$ is the average electron density on sublattice $\xi$ with spin $\sigma$, and 
$\bar\sigma$ indicates the opposite spin of $\sigma$. 
Noticing that $n_{A\sigma}=n_{B\sigma}=1/2$ at half-filling with the chemical potential $\mu=U/2$, 
the interacting single-particle Green's function $\bs{G}_{\mb{k}\s}(z)$ for given $\mb{k}$ and $\sigma$ 
is now simply evaluated as 
\begin{eqnarray}
  \bs{G}_{\mb{k}\s}(z) &=&  
  \left[ \bs{G}_{0\mb{k}\s}^{-1}(z)  - \bs{\Sigma}_{\mb{k}\s} (z) \right]^{-1}\notag \\
  &=&
  \left(
    \begin{array}{cc}
      z - \frac{U^2}{4z}                & -\gamma_{\mb{k}}       \\
      -\gamma^*_{\mb{k}}  &      z - \frac{U^2}{4z}         
    \end{array}
  \right)^{-1} 
\end{eqnarray}
and hence 
\begin{equation}
  \det \bs{G}_{\mb{k}\s} (z) = \frac{z^2}
  {(z - \w_{\mb{k}}^{+})  (z - \w_{\mb{k}}^{-})  (z + \w_{\mb{k}}^{+})  (z + \w_{\mb{k}}^{-})} 
\end{equation}
with 
\begin{equation}
  \w_{\mb{k}}^{\pm} = \frac{1}{2} \left(|\gamma_{\mb{k}}| \pm \sqrt{|\gamma_{\mb{k}}|^2 + U^2}\right ). 
\end{equation}

Therefore, the determinant of the interacting single-particle Green's function $\bs{G}(z)$ is 
\begin{eqnarray}\label{eq:detG_hb1}
  \det \bs{G}(z) 
  &=& \prod_{\mb k} \prod_{\s} \det \bs{G}_{\mb{k}\s} (z) \notag \\
  &=& z^{4L}
  \prod_{\mb{k}}  
  \left( \frac{1}{z - \w_{\mb{k}}^+} \right)^2
  \left( \frac{1}{z - \w_{\mb{k}}^-} \right)^2  \notag \\
  &\times&
  \left( \frac{1}{z + \w_{\mb{k}}^+} \right)^2
  \left( \frac{1}{z + \w_{\mb{k}}^-} \right)^2.  
\end{eqnarray}
Since $\w_{\bs k}^+>0$ and $\w_{\bs k}^-<0$ for non-zero $U$, 
we find that the number of zeros of $ \det \bs{G}(z)$ at (below) the chemical potential, corresponding to $z=0$, 
is $4L$ (0) and the number of poles of $ \det \bs{G}(z)$ at (below) the chemical potential is 0 ($4L$). 
Thus, from Eqs.~(\ref{eq:n_detG<}) and (\ref{eq:n_detG0}), we find that 
\begin{eqnarray}
  \left\{
    \begin{array}{l}
      n_{\det{\bs G}^{-1}}(\Gamma_0)=-4L   \\
      n_{\det{\bs G}^{-1}}(\Gamma_<)= 4L
    \end{array}
  \right. ,
  \label{eq:wind_detG}
\end{eqnarray}
at half-filling.
Using Eq.~(\ref{eq:vl_zero}), we obtain that $\lim_{T\to0}V_{\rm L}=2L$, thus satisfying the generalized 
Luttinger theorem.

Knowing the number of zeros and poles of $\det \bs{G}(z)$, 
the winding number $n_D({\mcal C})$ [see Eq.~(\ref{eq:n_D})] of the Fredholm determinant $D(z)$ 
defined in Eq.~(\ref{fredholm}) is now evaluated as 
\begin{eqnarray}
  \left\{
    \begin{array}{l}
      n_D(\Gamma_0) =  - 4L-8  \\
      n_D(\Gamma_<)  = 2L + 4
    \end{array}
  \right. ,
  \label{eq:nD_hb1}
\end{eqnarray}
which indeed fulfills the condition of type II for the validity of the generalized Luttinger theorem 
in Eq.~(\ref{genLuttinger})~\cite{note8}. 
Note that in the Hubbard-I approximation the metal-insulator transition occurs 
as soon as a finite $U\, (>0) $ is introduced. Therefore, 
this example studied here also suggests that the (portion of) Fermi surface 
should disappear with the introduction of electron interactions when the condition of type II is satisfied.

\subsection{Direct counting of winding numbers}

\begin{center}
  \begin{figure*}[!htb]
    \includegraphics[width=5.5cm]{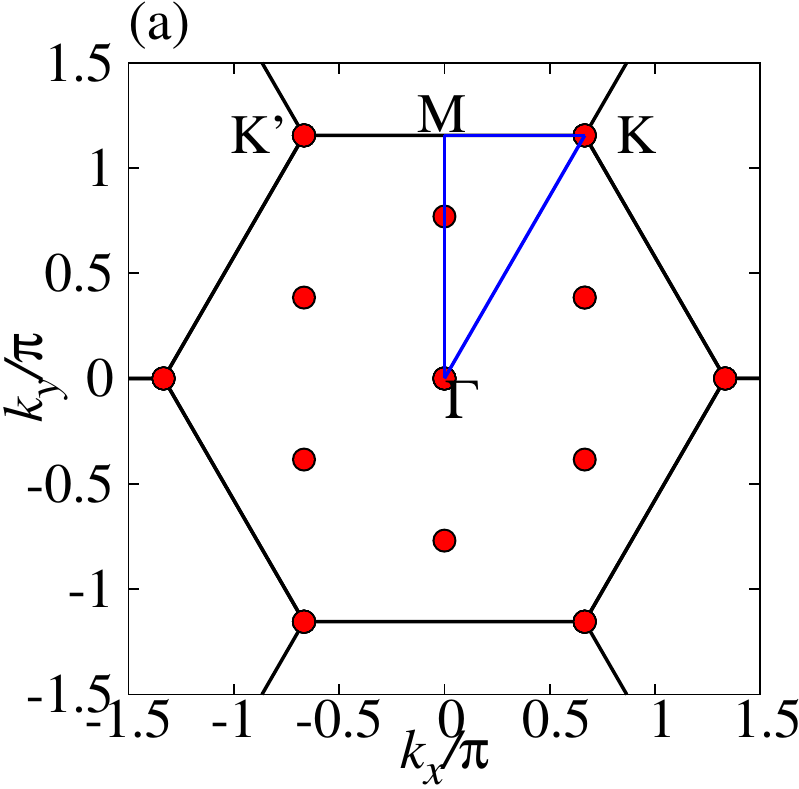}
    \includegraphics[width=5.3cm]{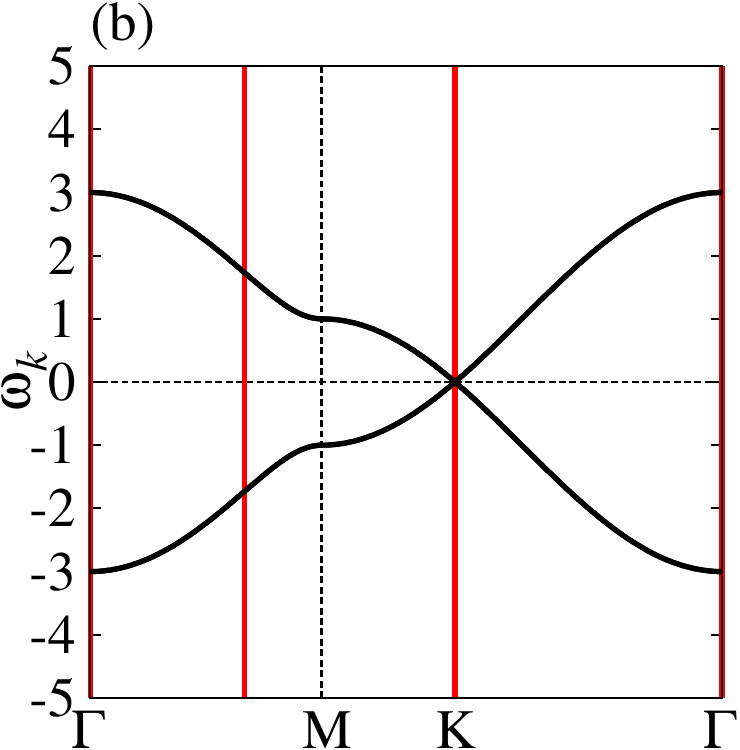}
    \includegraphics[width=5.3cm]{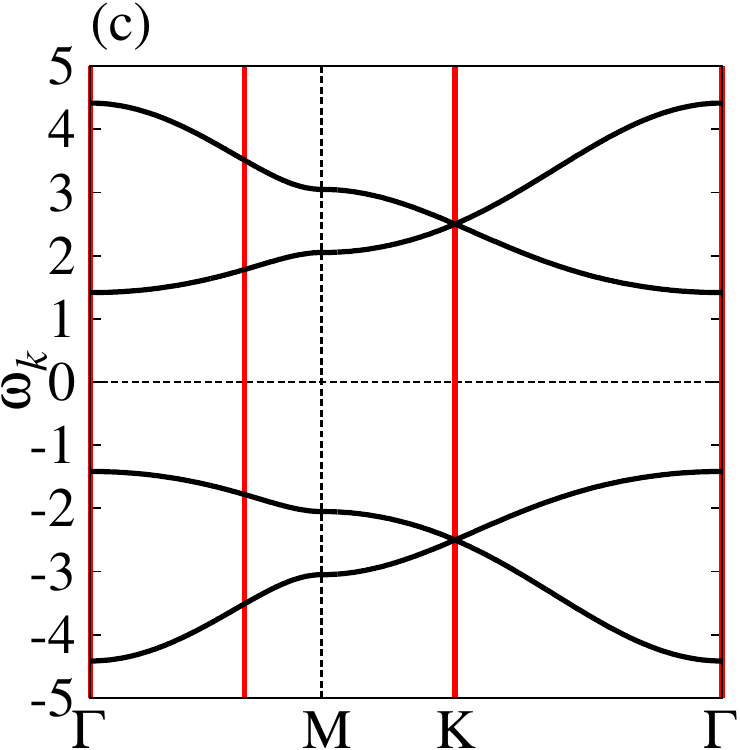}
    \caption{
      (a) The available $\mb{k}$ points (red solid circles) for the $3 \times 3$ unit cell cluster of the 
      honeycomb lattice with periodic boundary conditions. The black solid lines represent the boundaries 
      of the Brillouin zone and the blue solid lines indicate the momentum path along which the single-particle 
      energy-dispersion relations are shown in (b) and (c). 
      Four high symmetric momenta are denoted as $\Gamma$: $(0,0)$, $K$: $(2\pi/3,2\pi/\sqrt{3})$, 
      $M$: $(0,2\pi/\sqrt{3})$, and $K'$: $(-2\pi/3,2\pi/\sqrt{3})$.  
      (b) The single-particle energy-dispersion relation for $U=0$. 
      (c) Same as (b) but for $U=5t$ obtained within the Hubbard-I approximation at half-filling. 
      The red vertical lines in (b) and (c) 
      represent the available $\mb{k}$ points for the $3 \times 3$ unit cell cluster shown in (a). 
    }
    \label{HC18}
  \end{figure*}
\end{center}

As indicated in Fig.~\ref{contour_detG}, 
the winding numbers, $n_{\det{\bs G}_0^{-1}}({\mcal C})$ and $n_{\det{\bs G}^{-1}}({\mcal C})$, 
can be evaluated directly by counting how many times and which direction 
$\det{\bs G}_0^{-1}(z)$ and $\det{\bs G}^{-1}(z)$ 
wind around the origin in the complex $\det{\bs G}_0^{-1}$ and $\det{\bs G}^{-1}$ planes, respectively, 
when $z$ moves along contour $\mcal C$ shown in Fig.~\ref{contour0}(b). 
Since it is instructive, here we shall evaluate directly the winding numbers within the Hubbard-I approximation 
by considering $3 \times 3$ unit cells (i.e., $L=9$) with periodic boundary conditions, the smallest system size 
which contains both $K$ and $K'$ points in the Brillouin zone, as shown in Fig.~\ref{HC18}(a). 
The single-particle energy-dispersion relations for the noninteracting limit and for $U=5t$ are 
shown in Figs.~\ref{HC18}(b) and \ref{HC18}(c), respectively. 
Since contour $\mcal C$ can be chosen rather freely, as long as the poles and zeros are properly included, 
here we consider the contours given in Eq.~(\ref{eq:contour}) and in Fig.~\ref{contour2}.

\begin{center}
  \begin{figure*}
    \includegraphics[width=7.5cm]{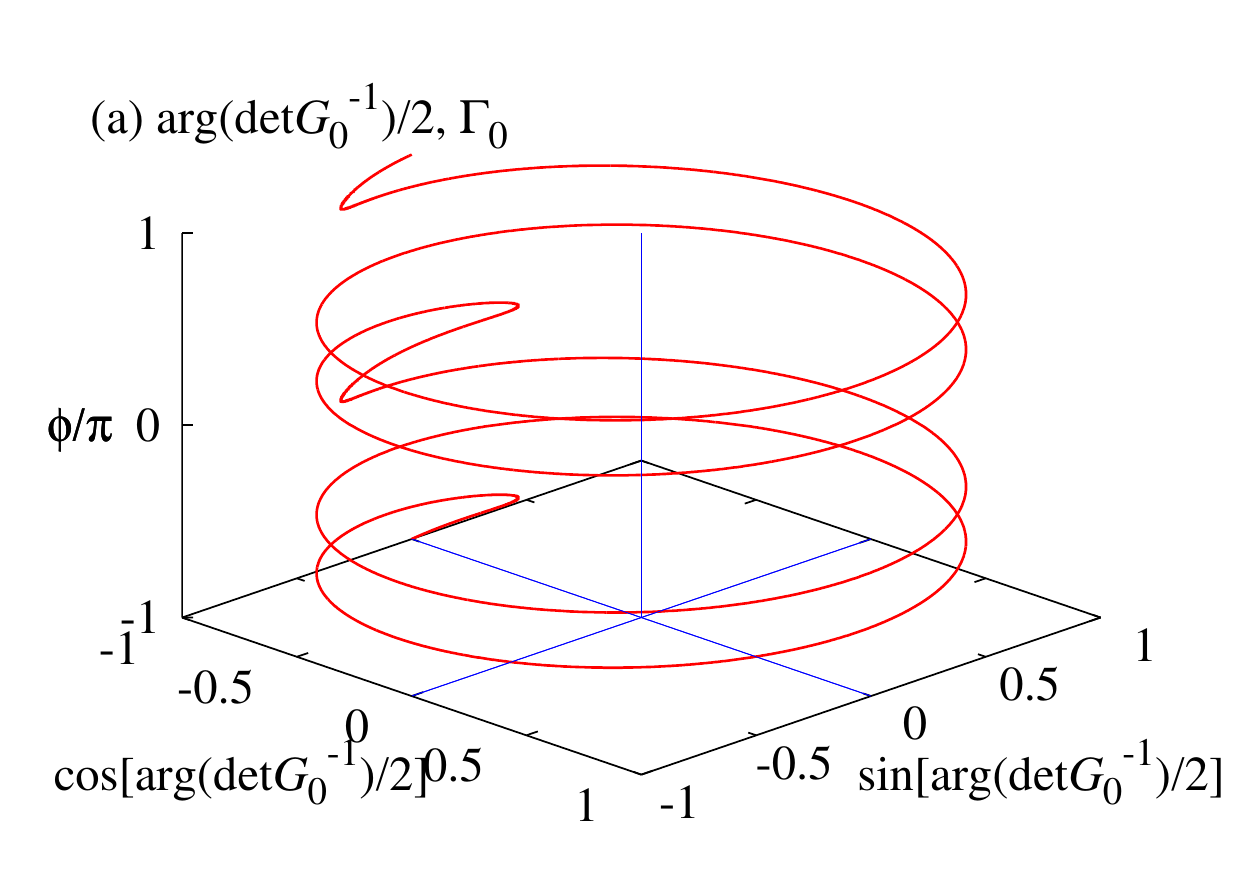}
    \includegraphics[width=7.5cm]{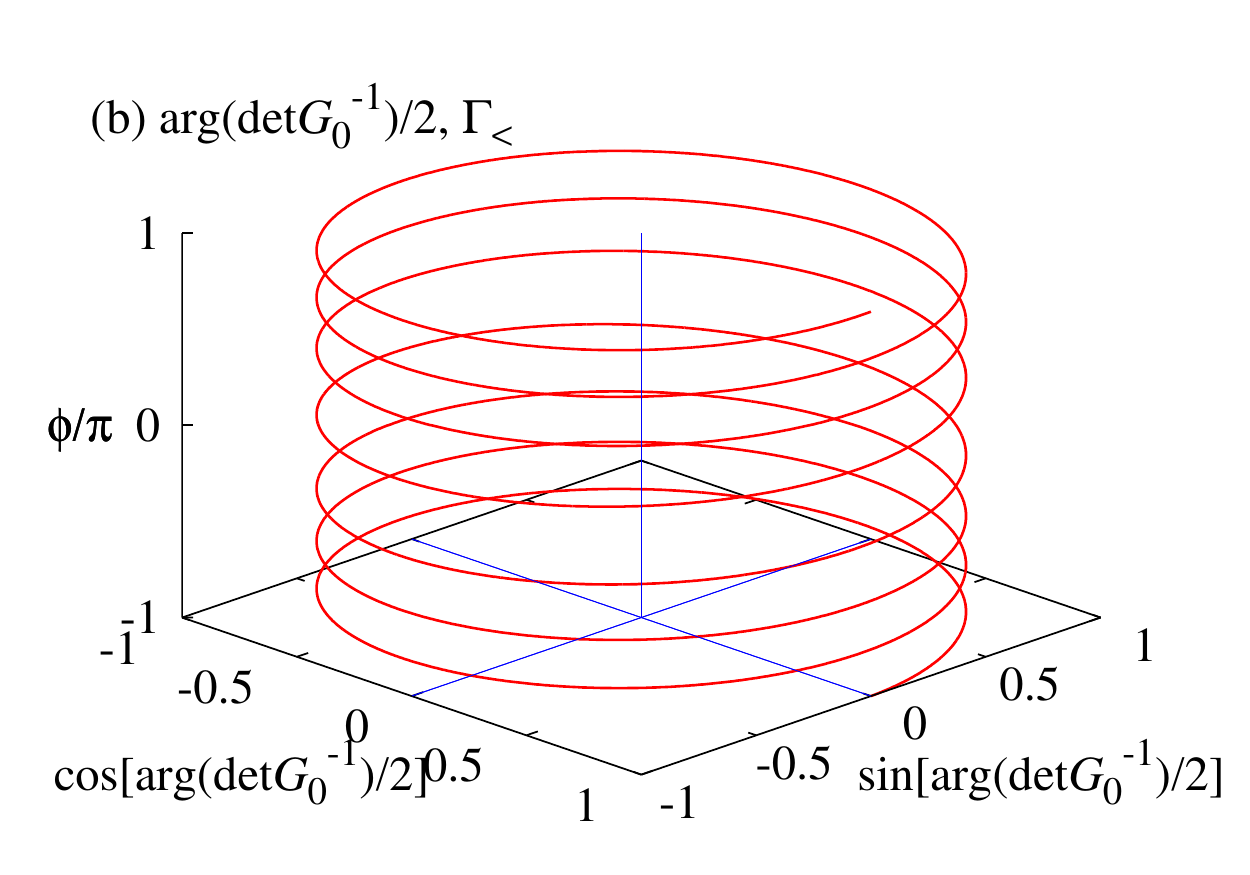}\\
    \includegraphics[width=7.5cm]{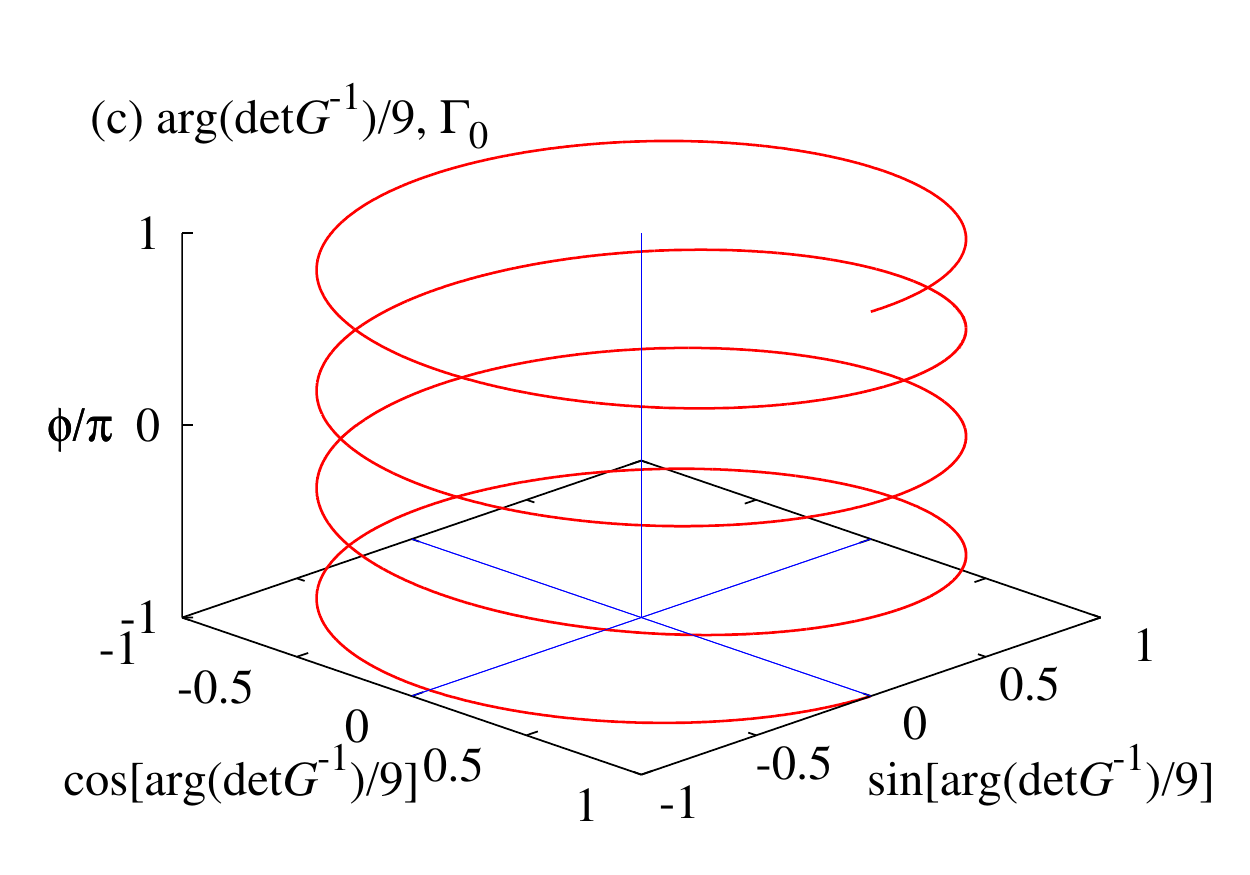}
    \includegraphics[width=7.5cm]{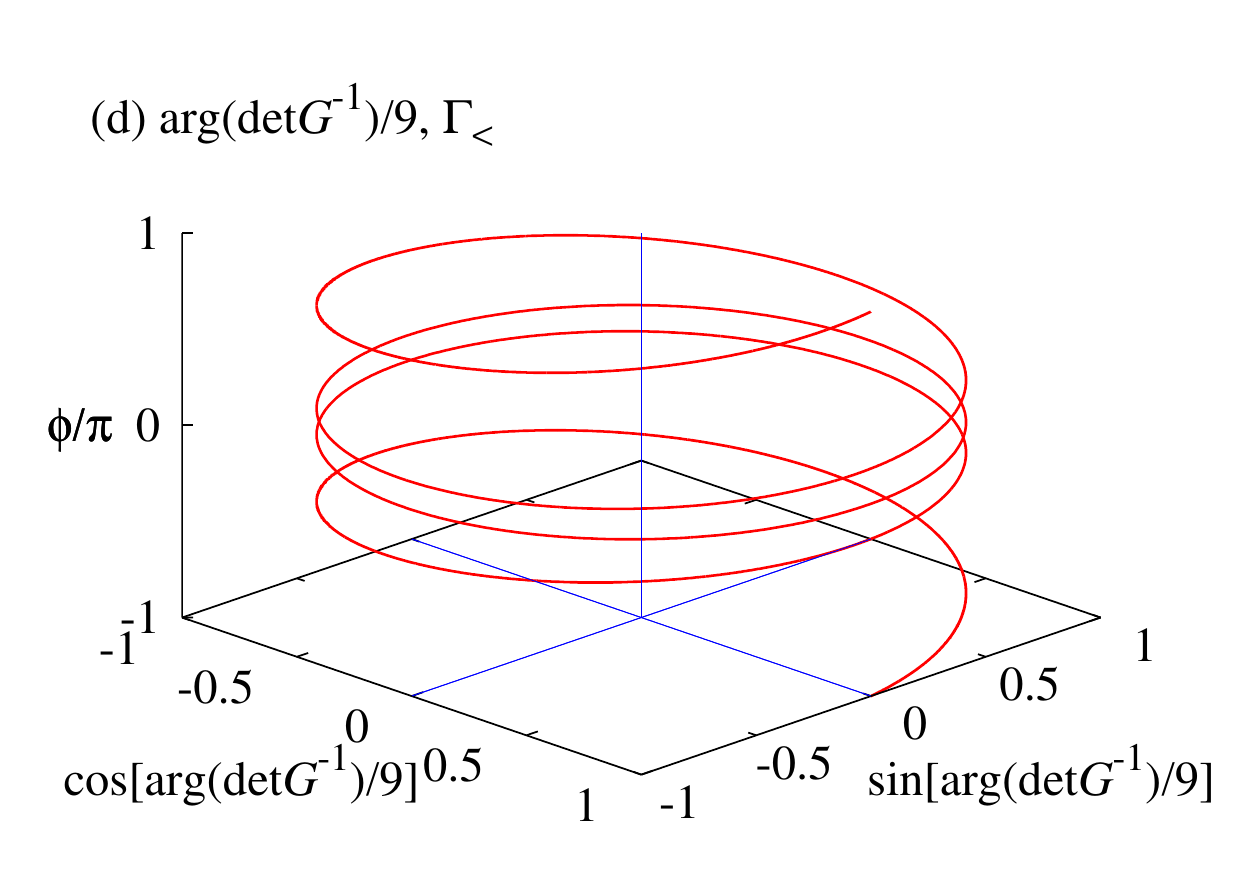}\\
    \includegraphics[width=7.5cm]{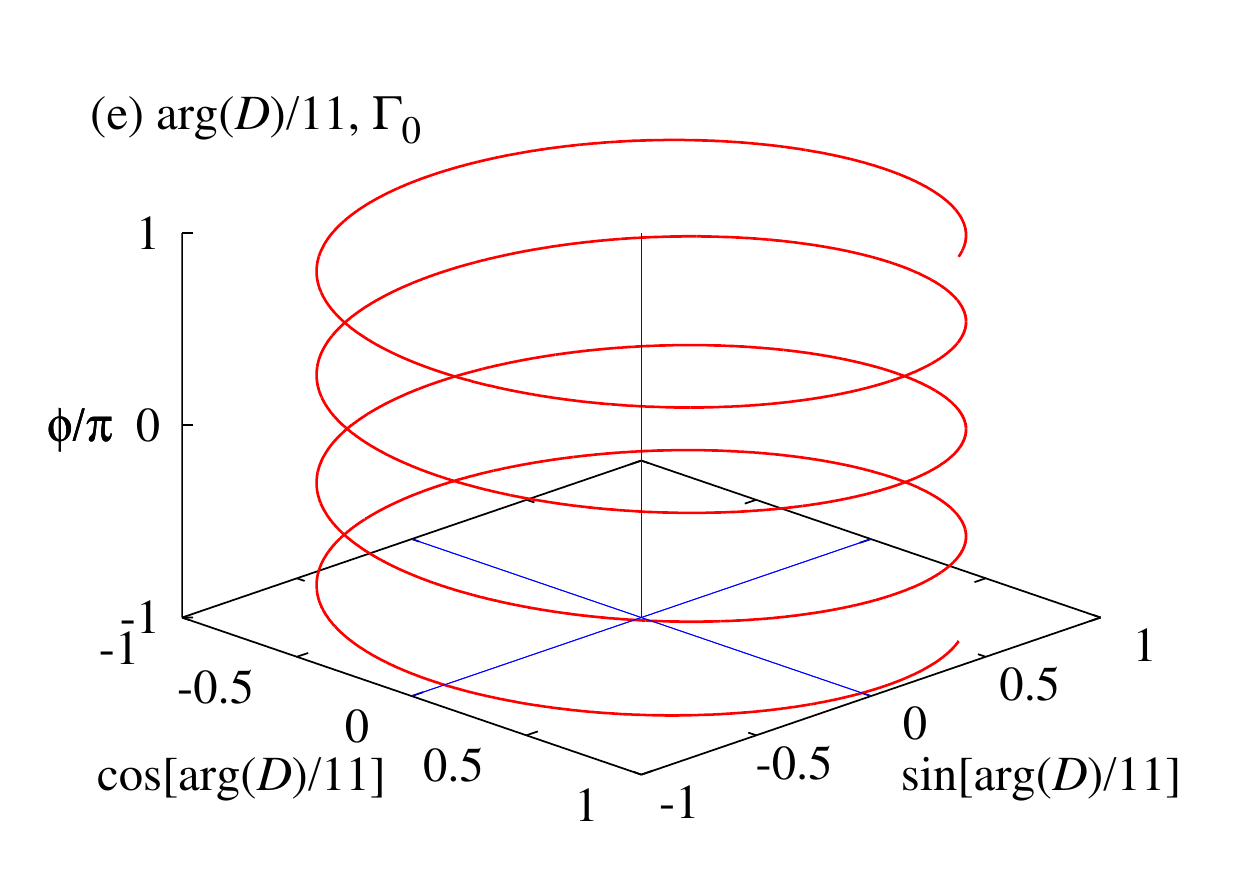}
    \includegraphics[width=7.5cm]{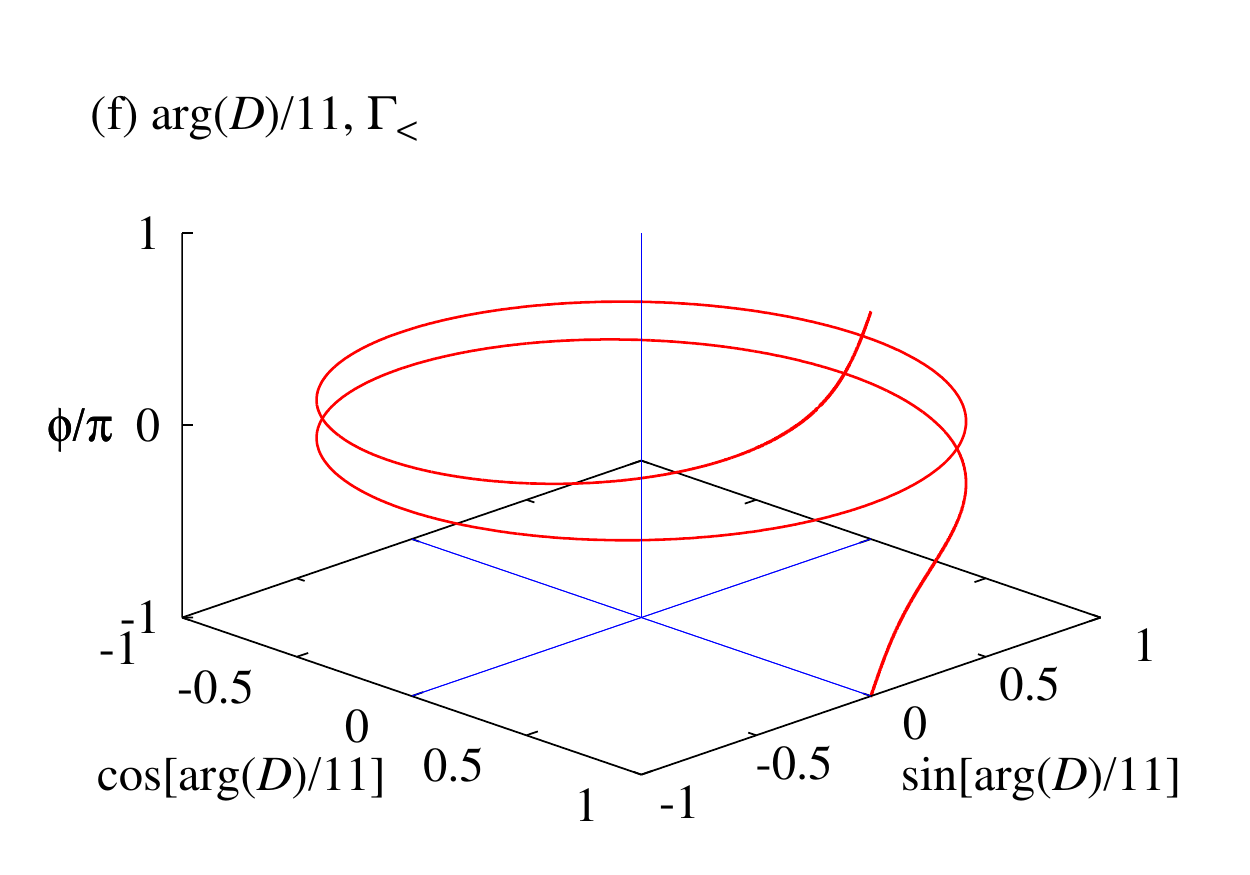}
    \caption{ 
      Arguments of $\det\bs{G}^{-1}(z)$ and $D(z)$ when $z$ moves along 
      contour ${\mcal C}\, (=\Gamma_0\,{\rm and}\, \Gamma_<\, {\rm in}\, {\rm Fig.~\ref{contour2}})$ 
      for the Hubbard model on the honeycomb lattice with $3\times 3$ unit cells at half-filling. 
      (a)--(d): $\arg[\det\bs{G}_0^{-1}(z)]$ for $z$ along contour $\Gamma_0$ (a) and $\Gamma_<$ (b) with $U=0$, and 
      $\arg[\det\bs{G}^{-1}(z)]$ for $z$ along contour $\Gamma_0$ (c) and $\Gamma_<$ (d) with $U=5t$. 
      (e) and (f): $\arg[D(z)]$ for $z$ along contour $\Gamma_0$ (e) and $\Gamma_<$ (f) with $U=5t$. 
      Notice that 
      the arguments are divided by 2, 9, or 11 (indicated in the figures), for clarity. 
      The determinants of the single-particle Green's functions, $\det \bs{G}_0(z)$ and $\det \bs{G}(z)$, 
      are analytically given in Eqs.~(\ref{eq:detG0}) and (\ref{eq:detG_hb1}), respectively, and 
      $D(z)=\det \bs{G}_0(z) / \det \bs{G}(z)$. 
      By directly counting how many times and which directions these quantities wind around the origin 
      while $\phi$ varies from $-\pi$ to $\pi$, 
      we find that 
      (a) $n_{\det \bs{G}_0^{-1}}(\Gamma_0)/2 = 4$, 
      (b) $n_{\det \bs{G}_0^{-1}}(\Gamma_<)/2 = 7$, 
      (c) $n_{\det \bs{G}^{-1}}(\Gamma_0)/9 = -4$,   
      (d) $n_{\det \bs{G}^{-1}}(\Gamma_<)/9 = 4$, 
      (e) $n_D(\Gamma_0)/11=-4$, and 
      (f) $n_D(\Gamma_<)/11=2$. 
    }
    \label{HC18arg}
  \end{figure*}
\end{center}

Figures~\ref{HC18arg}(a)--\ref{HC18arg}(d) show the results of 
$\arg[\det \bs{G}_0^{-1}(z)]$ in the noninteracting limit and 
$\arg[\det \bs{G}^{-1}(z)]$ with $U=5t$ 
for $z$ along contours $\Gamma_0$ and $\Gamma_<$. 
Here, we have used $\det \bs{G}_0(z)$ and $\det \bs{G}(z)$ obtained analytically 
in Eqs.~(\ref{eq:detG0}) and (\ref{eq:detG_hb1}), respectively. 
Notice in these figures that 
the arguments of $\det \bs{G}_0^{-1}(z)$ and $\det \bs{G}^{-1}(z)$ are divided by $2$ 
[Figs.~\ref{HC18arg}(a) and \ref{HC18arg}(b)] and 9 [Figs.~\ref{HC18arg}(c) and \ref{HC18arg}(d)], 
respectively, for clarity. 
By directly counting how many times and which direction these quantities wind around the origin 
we find that 
$n_{\det \bs{G}_0^{-1}} (\Gamma_0)/2 = 4$,
$n_{\det \bs{G}_0^{-1}} (\Gamma_<)/2 = 7$,
$n_{\det \bs{G}^{-1}} (\Gamma_0)/9 = -4$, and 
$n_{\det \bs{G}^{-1}} (\Gamma_<)/9 = 4$. 
These results are indeed the same as those obtained above in Eqs.~(\ref{eq:wind_detG0}) and (\ref{eq:wind_detG}) 
with $L=9$ 
by counting the number of zeros and poles of 
the determinant of the single-particle Green's functions.

Similarly, $n_D({\mcal C})$ in Eq.~(\ref{condition}) 
can also be evaluated directly by counting 
how many times and which direction $D(z)$ winds around the origin in the complex $D$ plane, 
when $z$ moves along contour $\mcal C$ shown in Fig.~\ref{contour2}.
The results of $\arg[D(z)]$ for $z$ along contours $\Gamma_0$ and $\Gamma_<$ with 
$U=5t$ are shown in Figs.~\ref{HC18arg}(e) and \ref{HC18arg}(f), respectively. 
It is clearly observed in these figures that the winding numbers are $n_D(\Gamma_0)/11=-4$ 
and $n_D(\Gamma_<)/11=2$. 
These results are again comparable with those evaluated above in Eq.~(\ref{eq:nD_hb1}) 
with $L=9$
by counting the number of zeros and poles of 
the determinant of the single-particle Green's functions.  
Indeed, we again find that $n_{D}(\Gamma_<)+{\frac{1}{2}}n_D(\Gamma_0)=0$, 
confirming the validity of the generalized Luttinger theorem with the condition of type II.

The analysis in this Appendix have clearly demonstrated that the formalism developed in 
Sec.~\ref{sec:GLT} can apply without any ambiguity even to systems with point-like Fermi surfaces, 
where the concept of Fermi surface volume is obscure. 
We also note that the formalism developed in Sec.~\ref{sec:GLT} can apply equally to, 
for example, particle-hole symmetric flat-band systems~\cite{Lieb1989, Seki2016} 
where the entire Brillouin zone is covered with zero-energy poles of the single-particle Green's function 
and thus the well-defined Fermi surface volume is absent in the noninteracting limit, 
and  where the ground state might be ferrimagnetic when electron interactions are introduced.

\section{Quasiparticle distribution function at low temperatures}\label{sec:QP}

In this Appendix, we shall generalize the quasiparticle distribution function $n_{\mb k}^{(0)}$ 
[Eq.~(\ref{QPTzero})] introduced in 
Sec.~\ref{sec:typeI} to finite temperatures. 
For a paramagnetic single-band metallic system with translational symmetry, 
the Luttinger volume $V_{\rm L}$ at finite temperatures defined in Eq.~(\ref{eq.VL}) 
is given as 
\begin{equation}
  V_{\rm L} = 2 \sum_{\mb{k}} n_{\mb{k}}, 
  \label{eq:vl_ft}
\end{equation}
where  
\begin{equation}
  \label{QPT}
  n_{\mb{k}} = \sum_{m=1}^{P_{\mb{k}}} n_{\rm F}(\w_m^{(\mb{k})}) - \sum_{l=1}^{Z_{\mb{k}}} n_{\rm F}(\zeta_l^{(\mb{k})}) 
\end{equation}
and the factor 2 in Eq.~(\ref{eq:vl_ft}) is due to the spin degrees of freedom. Here, we have used 
Eqs.~(\ref{eq:vl_a}) and (\ref{VLuttinger}), and 
$P_{\mb{k}}$ ($Z_{\mb{k}} = P_{\mb{k}} -1$) is the number of poles (zeros) of the single-particle 
Green's function $G_{\mb{k}}(\w)$ for momentum $\mb k$ with spin $\sigma$ at finite temperatures 
[see Eqs.~(\ref{eq:gz_k_sb}) and (\ref{eq:pz3})]. 
In the zero-temperature limit, 
$n_{\mb{k}}$ reduces to the winding number $n_{\mb{k}}^{(0)}$ given in Eq.~(\ref{QPTzero}). 
Here, we argue that $n_{\mb{k}}$ defined in Eq.~(\ref{QPT}) 
can be considered as the quasiparticle distribution function 
in the Fermi-liquid theory at temperatures. 
In order to well define quasiparticles, 
the temperature has to be sufficiently low as compared with the 
quasiparticle excitation energy $\w_{m_{\rm QP}}^{(\mb{k})}$ 
where $\w_{m_{\rm QP}}^{(\mb{k})}$ is either $\w_{m_{\rm top}}^{(\mb{k})}$, 
$\w_{m_{\rm FS}}^{(\mb{k})}$, or $\w_{m_{\rm bot}}^{(\mb{k})}$ in 
Eqs.~(\ref{wfs})--(\ref{wbot}), depending on momentum $\bs k$ 
(see Sec.~\ref{sec:typeI} and Fig.~\ref{fig:Akw}).  
Since $\w_{m_{\rm QP}}^{(\mb{k})}$ is bounded by 
$\zeta_{m_{\rm QP}-1}^{(\mb{k})} (< 0)$ and $\zeta_{m_{\rm QP}}(>0)$ 
from the lower and upper sides, respectively, 
we will assume that temperature $T$ satisfies 
$T \ll \zeta_{m_{\rm QP}}^{(\mb{k})} - \zeta_{m_{\rm QP}-1}^{(\mb{k})}$, 
implying that 
\begin{equation}
  T \ll - \zeta_{m_{\rm QP}-1}^{(\mb{k})},\,\,\, \zeta_{m_{\rm QP}}^{(\mb{k})}. 
  \label{eq:low-T}
\end{equation}

Let us consider momentum $\mb{k}$ at which the singularities of $G_{\mb{k}}(\w)$ 
are given as Eq.~(\ref{wtop}), i.e., $\mb{k}$ below the Fermi surface 
in the zero-temperature limit. Then we can write that 
\begin{eqnarray}
  \label{nk}
  n_{\mb{k}} =&& n_{\rm F}(\w_{m_{\rm top}}^{(\mb{k})}) 
  -   \sum_{m=1}^{m_{\rm top}-1}          \left[n_{\rm F}(\zeta_m^{(\mb{k})})     -n_{\rm F}(\w_m^{(\mb{k})}) \right] \notag \\
  &+& \sum_{m=m_{\rm top}}^{P_{\mb{k}}-1} \left[n_{\rm F}(\w_{m+1}^{(\mb{k})}) -n_{\rm F}(\zeta_m^{(\mb{k})}) \right],
\end{eqnarray}
where the first term represents the contribution from the quasiparticle excitation, 
i.e., the topmost excitation below the chemical potential for which $G_{\mb{k}}(\w)$ exhibits a pole, 
and the second (third) term from the incoherent part below (above) the chemical potential. 
In the following, we shall show that the contributions to $n_{\mb{k}}$ from the incoherent parts are 
exponentially small at low temperatures.

The second term on the right-hand side of Eq.~(\ref{nk}) can be approximated as 
\begin{eqnarray}
  &&\sum_{m=1}^{m_{\rm top}-1}   \left(\zeta_m^{(\mb{k})} - \w_m^{(\mb{k})} \right) 
  \frac{n_{\rm F}(\zeta_m^{(\mb{k})})     -n_{\rm F}(\w_m^{(\mb{k})})}{\zeta_m^{(\mb{k})} - \w_m^{(\mb{k})} }  \notag \\
  &\simeq& \int_{\zeta_1^{(\mb{k})}}^{\zeta_{m_{\rm top}-1}^{(\mb{k})}} \dd \w \frac{\dd n_{\rm F} (\w)}{\dd \w} 
  = n_{\rm F} (\zeta^{(\mb{k})}_{m_{\rm top} -1}) - n_{\rm F} (\zeta_{1}^{(\mb{k})}). 
  \label{eq:int<}
\end{eqnarray}
Here, in the second line, we have assumed that each energy interval 
between the successive pole and zero, $\zeta_m^{(\mb{k})} - \w_m^{(\mb{k})}$, in the incoherent part 
is small enough as compared with the whole energy width of the incoherent part itself, i.e., 
$\zeta^{(\mb{k})}_{m_{\rm top}-1} - \w_{1}^{(\mb{k})}$.    
Similarly, the third term on the right-hand side of Eq.~(\ref{nk}) can be approximated as
\begin{eqnarray}
  \sum_{m=m_{\rm top}}^{P_{\mb{k}}-1} \left[n_{\rm F}(\w_{m+1}^{(\mb{k})}) -n_{\rm F}(\zeta_m^{(\mb{k})}) \right] 
  \simeq n_{\rm F} (\zeta_{P_{\mb{k}}-1}^{(\mb{k})}) - n_{\rm F} (\zeta^{(\mb{k})}_{m_{\rm top}}).   
  \label{eq:int>}
\end{eqnarray}

Since we assume that $T \ll \zeta_{m_{\rm top}}^{(\mb{k})} - \zeta_{m_{\rm top}-1}^{(\mb{k})}$, 
we can set that $n_{\rm F}(\zeta_1^{(\mb{k})}) = 1$ and $n_{\rm F} (\zeta_{P_{\mb{k}}-1}^{(\mb{k})}) = 0$ 
in Eqs.~(\ref{eq:int<}) and (\ref{eq:int>}), respectively. 
Using $1-n_{\rm F}(z) = n_{\rm F}(-z)$, we thus find that 
\begin{eqnarray}
  \label{ntop}
  n_{\mb{k}} 
  &\simeq& n_{\rm F} (\w^{(\mb{k})}_{m_{\rm top}}) + n_{\rm{F}} (-\zeta^{(\mb{k})}_{m_{\rm top}-1}) - n_{\rm{F}} (\zeta^{(\mb{k})}_{m_{\rm top}}) \notag \\
  &=& n_{\rm F} (\w^{(\mb{k})}_{m_{\rm top}})  + \mcal{O}(\e^{ \zeta^{(\mb{k})}_{m_{\rm top}-1}/T}) - \mcal{O}(\e^{-\zeta^{(\mb{k})}_{m_{\rm top}}/T}). 
\end{eqnarray}
Note that the second and the third term in Eq.~(\ref{ntop}) are exponentially small because 
$T \ll - \zeta_{m_{\rm top}-1}^{(\mb{k})}$ and $T \ll \zeta_{m_{\rm top}}^{(\mb{k})}$ [see Eq.~(\ref{eq:low-T})].

Similarly, for momenta $\mb{k}$ at which singularities of $G_{\mb{k}}(\w)$ are given as 
Eq.~(\ref{wfs}) (i.e., at the Fermi surface in the zero-temperature limit) and 
Eq.~(\ref{wbot}) (i.e., above the Fermi surface in the zero-temperature limit), we find that 
\begin{equation}
  \label{nfs}
  n_{\mb{k}} \simeq n_{\rm F} (\w^{(\mb{k})}_{m_{\rm FS}})  + \mcal{O}(\e^{\zeta^{(\mb{k})}_{m_{\rm FS}-1}/T}) - \mcal{O}(\e^{-\zeta^{(\mb{k})}_{m_{\rm FS}}/T}) 
\end{equation}
and
\begin{equation}
  \label{nbot}
  n_{\mb{k}} \simeq n_{\rm F} (\w^{(\mb{k})}_{m_{\rm bot}})  + \mcal{O}(\e^{ \zeta^{(\mb{k})}_{m_{\rm bot}-1}/T}) - \mcal{O}(\e^{-\zeta^{(\mb{k})}_{m_{\rm bot}}/T}),
\end{equation}
respectively. Here, the subscript $m_{\rm FS}$ should be read as a label for an excitation 
on the chemical potential because the Fermi surface is not well defined at finite temperatures. 
Equations~(\ref{ntop})--(\ref{nbot}) clearly show that $n_{\mb{k}}$ defined in Eq.~(\ref{QPT})  
is expressed as the Fermi-Dirac distribution function $n_{\rm F}(\w)$ of the quasiparticle excitation energy 
at momentum $\mb k$. Therefore, we can conclude that $n_{\mb{k}}$ is considered as the 
distribution function of quasiparticles at low temperatures.

Let us now consider $n_{\mb{k}}$ from the analytical aspects of the single-particle Green's function 
$G_{\mb{k}}(z)$. 
As shown in Eq.~(\ref{eq.VL}), $n_{\mb{k}}$ in Eq.~(\ref{QPT}) is also expressed in the contour integral as 
\begin{equation}
  n_{\mb{k}} = \oint_{\Gamma} \frac{\dd z}{2 \pi \imag} n_{\rm F}(z) \frac{\partial \ln G_{\mb{k}}^{-1}(z)}{\partial z}.
  \label{eq:nk_int}
\end{equation}
Explicitly considering the quasiparticle contribution in the single-particle Green's function $G_{\mb{k}}(z)$, 
the Lehmann representation of $G_{\mb{k}}(z)$ [see Eq.~(\ref{lehmann})] is given as   
\begin{equation}
  \label{GFL}
  G_{\mb{k}}(z) 
  = \frac{a_{\mb{k}}}{z - \w_{m_{\rm QP}}^{(\mb{k})}} 
  + \sum_{m(\not = {m_{\rm QP}})}^{P_{\mb{k}}} \frac{|Q_{\mb{k} m}|^2}{z - \w_{m}^{(\mb{k})}},
\end{equation}
where $\w_{m_{\rm QP}}^{(\mb{k})}(=\w_{m_{\rm top}}^{(\mb{k})}, \w_{m_{\rm FS}}^{(\mb{k})}$, or $\w_{m_{\rm bot}}^{(\mb{k})})$ 
is the quasiparticle excitation energy for momentum ${\mb k}$ 
with  the corresponding quasiparticle weight $a_{\mb{k}} = |Q_{\mb{k}{m_{\rm QP}}}|^2 > 0$, 
and the second term in the right-hand side represents the incoherent part of $G_{\mb{k}}(z)$. 

When $z$ is in the vicinity of the quasiparticle excitation energy, i.e., 
$z \simeq \w_{m_{\rm{QP}}}^{(\mb{k})}$,  
the single-particle Green's function is approximated as 
$G_{\mb{k}}(z) \simeq a_{\mb{k}}/(z - \w_{m_{\rm QP}}^{(\mb{k})})$ and thus 
we find that
\begin{equation}
  \frac{\partial \ln G_{\mb{k}}^{-1}(z)}{\partial z} 
  = G_{\mb{k}}(z) \frac{\partial G_{\mb{k}}^{-1}(z)}{\partial z} 
  \simeq \frac{1}{z - \w_{m_{\rm QP}}^{(\mb{k})}}. 
  \label{dlogG}
\end{equation}
This implies that the logarithmic derivative of $G_{\mb{k}}^{-1}(z)$ behaves like a {\it free} 
fermionic single-particle Green's function with the excitation energy $\w_{m_{\rm QP}}^{(\mb{k})}$ 
when $z \simeq \w_{m_{\rm QP}}^{(\mb{k})}$. 
Therefore, the pole of the single-particle Green's function at $\w_{m_{\rm QP}}^{(\mb{k})}$ 
contributes to $n_{\mb{k}}$ in Eq.~(\ref{eq:nk_int}) by $n_{\rm F} (\w_{m_{\rm {QP}}}^{(\mb{k})})$.

The same argument as in Eq.~(\ref{dlogG}) can be applied to each of the remaining $P_{\mb{k}}-1$ poles 
of $G_{\mb{k}}(z)$, which are given in the second term of the right-hand side of Eq.~(\ref{GFL}). 
However, the positive contributions $n(\w_m^{(\mb{k})})$ from these 
poles are mostly canceled by the negative contributions $-n(\zeta_l^{(\mb{k})})$  
from the same number of zeros of $G_{\mb{k}}(z)$ at low temperatures, 
and thereby the net contribution to $n_{\mb{k}}$ from the remaining incoherent 
part is exponentially small. 
We thus again reach the same conclusion that 
\begin{equation}
  n_{\mb{k}} \simeq \oint \frac{\dd z}{2 \pi \imag} n_{\rm F}(z) \frac{1}{z - \w_{m_{\rm QP}}^{(\mb{k})}} = n_{\rm F}(\w_{m_{\rm QP}}^{(\mb{k})}), 
\end{equation}
showing that $n_{\mb{k}}$ is dominated by the lowest-energy single-particle excitation 
$\w_{m_{\rm QP}}^{(\mb{k})}$ 
and the excitation indeed obeys the Fermi-Dirac statistics, as in Eqs.~(\ref{ntop})--(\ref{nbot}). 

The important consequence of this is that 
the Luttinger volume $V_{\rm L} $ in Eq.~(\ref{eq:vl_ft}) provides 
the average number of quasiparticles. 
The Landau's Fermi-liquid theory hypothesizes that the number of particles, $N$, is equal to that of 
quasiparticles~\cite{Luttinger1962}. 
Therefore, 
the argument given here immediately implies that 
this fundamental hypothesis of the Landau's Fermi-liquid theory 
is guaranteed 
when $V_{\rm L} =N$. This is the case when the Luttinger theorem is valid at zero temperature or 
when the particle-hole symmetry is preserved at finite temperatures, as shown in Eq.~(\ref{Vph}).

\begin{center}
  \begin{figure}
    \includegraphics[width=6.8cm]{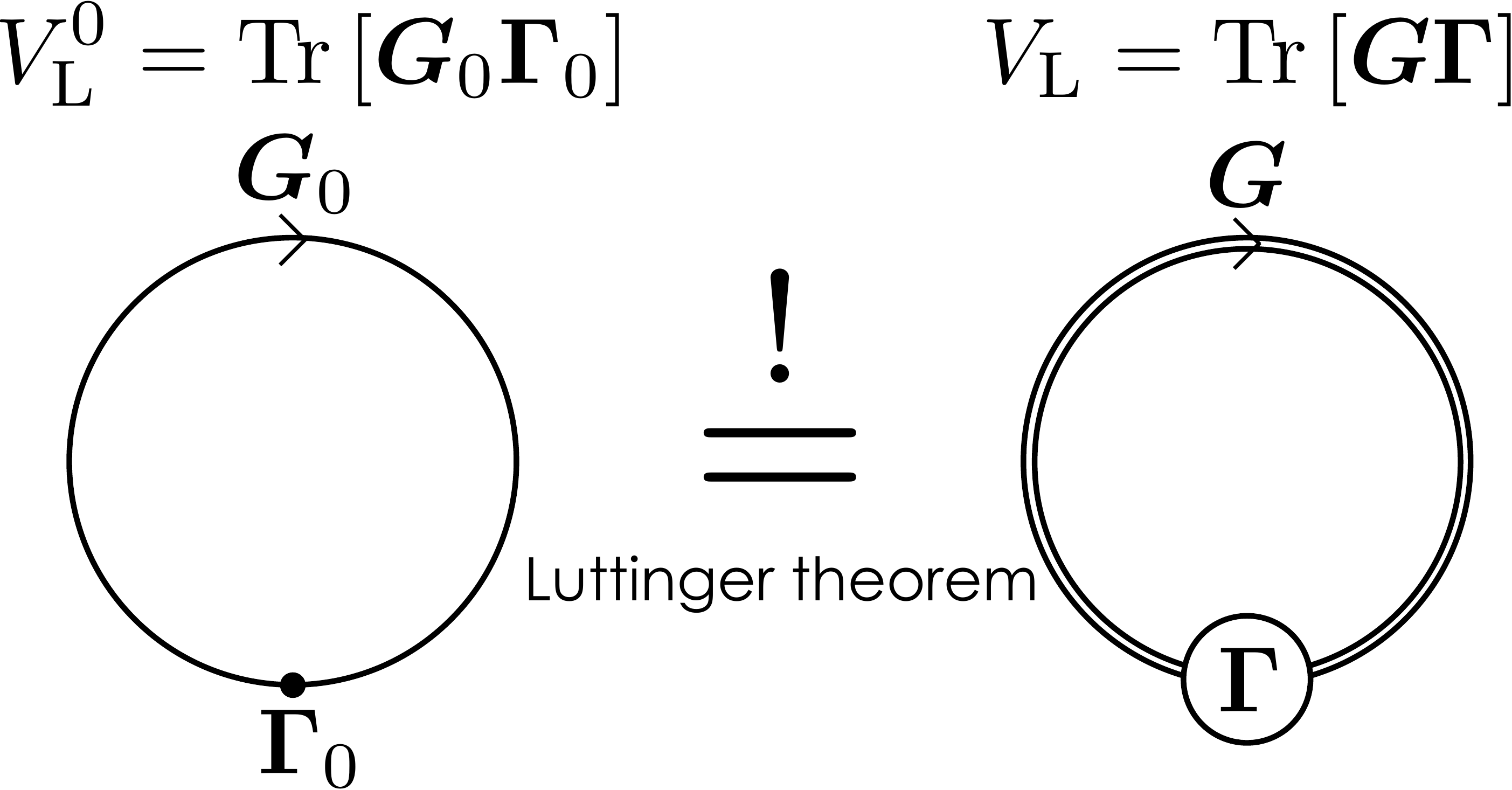}
    \caption{
      Diagrammatic representation for the Luttinger volumes of 
      a noninteracting system $V_{\rm L}^0$ (left) and 
      an interacting system $V_{\rm L}$ (right).  
      Here, $\Tr[\cdots] = T\sum_{\nu=-\infty}^{\infty}\e^{\imag \w_\nu 0^+} \tr[\cdots]$ and 
      $\bs{\Gamma}_{0}=\frac{\partial \bs{G}_{0}^{-1}(z)}{\partial z}=\bs{I}$ (unit matrix). 
      The thin line with an arrow represents $\bs{G}_0$, 
      the dot $\bs{\Gamma}_0$, 
      the double line with an arrow $\bs{G}$, and 
      the circle $\bs{\Gamma}$. 
      The Luttinger theorem equates these two quantities at zero temperature. 
    }
    \label{diagram}
  \end{figure}
\end{center}

Finally, we note that for general complex frequency $z$, 
\begin{equation}
  \frac{\partial \ln G_{\mb{k}}^{-1}(z)}{\partial z} 
  = G_{\mb{k}}(z) \Gamma_{\mb{k}}(z), 
  \label{GGamma}
\end{equation}
where 
$\Gamma_{\mb{k}}(z)=\frac{\partial G_{\mb{k}}^{-1}(z)}{\partial z}=1-\frac{\partial \bs{\Sigma}_{\mb{k}}(z)}{\partial z}$ 
is the scalar vertex function. The comparison with Eq.~(\ref{dlogG}) suggests that 
$\Gamma_{\mb{k}}(z)$ 
enhances the renormalized quasiparticle spectral weight $a_{\mb{k}} (\leqslant 1)$ up to $1$ 
in the interacting single-particle Green's function 
for $z$ near the quasiparticle excitation energy $\w_{m_{\rm QP}}^{(\mb{k})}$. 
Notice also that Eq.~(\ref{GGamma}) allows us for a diagrammatic representation of the Luttinger volume and 
the Luttinger theorem, as shown in Fig.~\ref{diagram}. 

\begin{thebibliography}{99}
\bibitem{note2}
  The same pair of $(r,s)$ but with different order, i.e., $(s,r)$, should be considered 
  separately. 
  In other words, $\sum_{m}\cdots$ in Eq.~(\ref{lehmann}) should be considered 
  simply as $\sum_r\sum_s\cdots$, and thus includes the case when $r=s$.
\bibitem{note}
  It should be noted that zeros of 
  $G_{\alpha \alpha}(z)$ and $R_{\alpha \alpha}(z)$ 
  at infinity ($z=\infty$)
  is not counted in 
  $Z_{\alpha \alpha}$ and $\tilde{Z}_{\alpha \alpha}$, respectively.
\bibitem{note5}
  Note that Eq.~(\ref{Vph}) is still correct even when a singularity appears at zero energy because 
  $n_{\rm F}(0)=1/2$.
\bibitem{note7}
  Note that the chemical potential should be determined by taking the zero-temperature limit becuase 
  generally there is ambiguity in the chemical potential at exactly zero temperature for insulators.
\end{thebibliography}
\end{document}